\documentclass[openacc]{rsproca_arxiv}

\usepackage{lipsum}
\usepackage{capt-of}

\newcommand{\Ebb}{\mathbb{E}}
\newcommand{\Lbb}{\mathbb{L}}
\newcommand{\Nbb}{\mathbb{N}}

\newcommand{\Rbb}{\mathbb{R}}

\newcommand{\Vbb}{\mathbb{V}}

\newcommand{\bbf}{\mathbf{b}}
\newcommand{\Bbf}{\mathbf{B}}

\newcommand{\Cbf}{\mathbf{C}}

\newcommand{\Gbf}{\mathbf{G}}

\newcommand{\Hbf}{\mathbf{H}}

\newcommand{\Ibf}{\mathbf{I}}

\newcommand{\Lbf}{\mathbf{L}}

\newcommand{\rbf}{\mathbf{r}}
\newcommand{\Rbf}{\mathbf{R}}

\newcommand{\Sbf}{\mathbf{S}}
\newcommand{\tbf}{\mathbf{t}}
\newcommand{\Tbf}{\mathbf{T}}
\newcommand{\ubf}{\mathbf{u}}
\newcommand{\Ubf}{\mathbf{U}}
\newcommand{\vbf}{\mathbf{v}}

\newcommand{\wbf}{\mathbf{w}}
\newcommand{\Wbf}{\mathbf{W}}

\newcommand{\Xbf}{\mathbf{X}}
\newcommand{\ybf}{\mathbf{y}}
\newcommand{\Ybf}{\mathbf{Y}}

\newcommand{\ep}{\epsilon}

\newcommand{\CalD}{{\mathcal{D}}}

\newcommand{\CalI}{{\mathcal{I}}}

\newcommand{\CalL}{{\mathcal{L}}}

\newcommand{\CalN}{{\mathcal{N}}}
\newcommand{\CalO}{{\mathcal{O}}}

\renewcommand{\tilde}[1]{\widetilde{#1}}
\newcommand{\argmin}{\text{argmin}}
	
\newcommand{\lan}{\left\langle}
\newcommand{\ran}{\right\rangle}

\newcommand{\wstar}{\wbf^\star}
\newcommand{\what}{{\widehat{\wbf}}}
\newcommand{\nrm}[1]{\left\Vert {#1} \right\Vert}

\makeatletter
\DeclareFontFamily{OMX}{MnSymbolE}{}
\DeclareSymbolFont{MnLargeSymbols}{OMX}{MnSymbolE}{m}{n}
\SetSymbolFont{MnLargeSymbols}{bold}{OMX}{MnSymbolE}{b}{n}
\DeclareFontShape{OMX}{MnSymbolE}{m}{n}{
    <-6>  MnSymbolE5
   <6-7>  MnSymbolE6
   <7-8>  MnSymbolE7
   <8-9>  MnSymbolE8
   <9-10> MnSymbolE9
  <10-12> MnSymbolE10
  <12->   MnSymbolE12
}{}
\DeclareFontShape{OMX}{MnSymbolE}{b}{n}{
    <-6>  MnSymbolE-Bold5
   <6-7>  MnSymbolE-Bold6
   <7-8>  MnSymbolE-Bold7
   <8-9>  MnSymbolE-Bold8
   <9-10> MnSymbolE-Bold9
  <10-12> MnSymbolE-Bold10
  <12->   MnSymbolE-Bold12
}{}

\let\llangle\@undefined
\let\rrangle\@undefined
\DeclareMathDelimiter{\llangle}{\mathopen}%
                     {MnLargeSymbols}{'164}{MnLargeSymbols}{'164}
\DeclareMathDelimiter{\rrangle}{\mathclose}%
                     {MnLargeSymbols}{'171}{MnLargeSymbols}{'171}
\makeatother

\usepackage[utf8]{inputenc}
\usepackage{
algorithmic,
amsmath,
amssymb,
amsthm,
appendix,
bm,
bbm,
caption,
epsf,
enumitem,
empheq,
hyperref,
listings,
mathtools,
mathrsfs,
subcaption,
tikz,
titletoc,
url,
ulem,
}

\usepackage{algorithm2e}[numbered]
\hypersetup{
    colorlinks=true, 
    linktoc=all,     
    linkcolor=black,  
}

\numberwithin{equation}{section}

\graphicspath{{figures/}}

\tikzset{every label/.style={font=\footnotesize,inner sep=1pt}}
\DeclareCaptionFormat{myformat}{#1#2#3\hrulefill}
\usepackage{siunitx}
\usepackage{multicol}

\begin{document}
\title{Weak-Form Inference for Hybrid Dynamical Systems in Ecology}

\author{Daniel Messenger$^{1}$ \\  Greg Dwyer$^{2}$ \\ Vanja Dukic$^{1}$}

\address{$^{1}$Department of Applied Mathematics, University of Colorado, Boulder, CO 80309-0526\\
$^{2}$Department of Ecology \& Evolution, University of Chicago, Chicago, Illinois 60637}

\subject{Biomathematics, Ecology}

\keywords{Multiscale model, hybrid systems, data-driven modeling, system identification, parameter estimation, WSINDy.}

\corres{Daniel A. Messenger\\ \email{daniel.messenger@colorado.edu}}

\begin{abstract}
Species subject to predation and environmental threats commonly exhibit variable periods of population boom and bust over long timescales. Understanding and predicting such behavior, especially given the inherent heterogeneity and stochasticity of exogenous driving factors over short timescales, is an ongoing challenge. A modeling paradigm gaining popularity in the ecological sciences for such multi-scale effects is to couple short-term continuous dynamics to long-term discrete updates.  We develop a data-driven method utilizing weak-form equation learning to extract such hybrid governing equations for population dynamics and to estimate the requisite parameters using sparse intermittent measurements of the discrete and continuous variables. The method produces a set of short-term continuous dynamical system equations parametrized by long-term variables, and long-term discrete equations parametrized by short-term variables, allowing  direct assessment of interdependencies between the two time scales. We demonstrate the utility of the method on a variety of ecological scenarios and provide extensive tests using models previously derived for epizootics experienced by the North American spongy moth ({\it Lymantria dispar dispar}).
\end{abstract}

\maketitle

\begin{multicols}{2}
\section{Introduction}

The ability to estimate model parameters and select appropriate models based on data is of central interest in ecological applications, but is often complicated by the presence of irregularly observed sparse observations \cite{hilborn2013ecoDetective}.  Fortunately, the widespread availability of high-performance computing and the development of sophisticated nonlinear optimization algorithms have made it possible to fit a vast range of nonlinear ecological models to data \cite{bolker2008ecoModelsNData}.  Many applications of model-fitting have focused on human diseases, for which an extensive knowledge base makes it possible to write down at least basic models using only information from the literature \cite{KingIonidesPascualEtAl2008Nature}. A very large set of ecological problems, however, instead involve natural systems about which so little is known that there is often a great deal of uncertainty about the model structure \cite{hilborn2013ecoDetective}. 

In such cases, the standard approach is to write down a set of competing models, each comprising what is believed to be a reasonable model structure, to fit each model separately to the data, and then to use statistical model selection to choose between the models \cite{BurnhamAnderson2007, BurnhamAndersonHuyvaert2011BehavEcolSociobiol, HootenHobbs2015EcolMonogr}.   Because in such cases the models in question are typically based on well-studied ecological models from the literature, the use of model selection results in a model that has a well understood connection to the biology of the system being studied.  Model selection can then offer important insights into the biology of the system \cite{kyle2020stochasticity}.  

A very important disadvantage of the application of model selection to ecological problems, however, is that it typically requires the implementation of complex nonlinear fitting routines in high-performance computing environments, and this implementation must be repeated for each model under consideration \cite{KennedyDukicDwyer2015EnvironEcolStat}.   Model selection in ecology therefore faces daunting computational obstacles when even moderate  numbers of models are considered, which is likely why it has played only a modest role in the development of the field.  

Here we therefore propose a novel alternative approach based on weak-form equation learning \cite{BortzMessengerDukic2023article,MessengerBortz2021SIAMMultiscaleModelSimul, MessengerBortz2021JComputPhys} to accomplish the joint parameter estimation and model selection task, which we will refer to throughout as "model identification". By simultaneously comparing a very large number of models to data at once, our model identification algorithm is able to carry out model selection far more rapidly than the standard information-criterion model selection process.  Moreover, our algorithm produces models that can very accurately reproduce data.   

The models that we consider consist of host-pathogen models that allow for seasonality.  Although classical host-pathogen models typically neglect seasonality \cite{keeling2011modelingInfDiseases}, host-pathogen interactions in nature are often very strongly affected by seasonality \cite{AltizerDobsonHosseiniEtAl2006EcolLett}. Moreover, standard models typically incorporate seasonality in the form of sinusoidal variation in a model parameter, often host reproduction. But this model structure is deeply unrealistic for most host-pathogen systems.

For most host-pathogen systems there is instead typically a period during which pathogen transmission exhausts or nearly exhausts the supply of available hosts.  Transmission then falls to zero or nearly so until host reproduction replenishes the supply of susceptible hosts.  As briefly summarized by Andreasen and Dwyer \cite{AndreasenDwyer2023TheAmericanNaturalist}, this pattern occurs in host-pathogen systems for which hosts are all classes of vertebrates, most if not all terrestrial phyla of invertebrates, and many different plants, while the pathogens include viruses, bacteria and fungi.  Our algorithm is customized to adapt to this specific structure by carrying out model identification using a combination of short-term disease dynamics and long-term population fluctuations.  

We consider in particular the case of baculovirus pathogens of forest insects, which often drive long-period, large-amplitude cycles in the population densities of their insect hosts \cite{DwyerDushoffYee2004Nature}.  Because only larvae are susceptible to baculoviruses, virus epizootics (epidemics in animals) only occur during the larval season.  Larvae that survive to adulthood reproduce, while the virus overwinters on the insect's eggs.  The epizootic is then restarted in the following spring when some larvae become infected as they hatch.   Because many outbreaking host insects have only one generation per year \cite{Hunter1995PopulDynNewApproachesSynth}, and because the larval period typically lasts for only two months \cite{DwyerMihaljevicDukic2022AmNat}, our models consist of continuous-time epizootic models that are nested inside discrete-generation host-reproduction/virus-overwintering models. 
 
These multiscale hybrid models of insect-pathogens were introduced in the ecological literature by \cite{BriggsGodfray1996TheoreticalPopulationBiology}, and have been extended to include evolutionary change \cite{DwyerDushoffElkintonEtAl2000,DwyerMihaljevicDukic2022AmNat,ElderdDushoffDwyer2008TheAmericanNaturalist,fleming2015effects}, variation in the foliage quality of insect host trees \cite{ElderdDwyerDukic2013Epidemics} and weather effects \cite{kyle2020stochasticity}. Hybrid models have been extensively applied to host-pathogen dynamics more generally \cite{SinghEmerick2020preprint,EmerickSingh2016MathematicalBiosciences,EmerickSinghChhetri2020MathematicalBiosciences} and to consumer-resource theory \cite{EskolaGeritz2007BullMathBiol,PachepskyNisbetMurdoch2008Ecology}, sometimes drawing fundamentally different ecological implications compared to purely discrete or purely continuous models. 


\subsection{Inference and learning in dynamical systems}

When the model system equations are of known form, the  unknown system parameters and initial conditions are usually inferred from the data using statistical methods. However, when the model itself is unknown (in addition to having unknown parameters), the problem of statistical inference becomes much more difficult. We discuss both situations next.  


\subsubsection{Parameter Inference} 

In spite of the availability of sophisticated optimization algorithms and high-performance computing, parameter estimation and inference (and inverse problems in general) in non-linear dynamical systems remain a challenging problem  \cite{McGoffMukherjeePillai2015StatSurv}. Many of the existing parameter estimation methods in practice,  such as non-linear least squares and likelihood-based methods, still use optimization (cost) functions which are based on comparisons of data to the solution of the system. This results in iterative schemes that use a forward-solver based optimization routine to minimize the nonlinear-least squares residual at every iteration. Furthermore, the likelihood functions commonly resulting from non-linear  differential equations often exhibit multiple modes and ridges, with  different combinations of parameter values yielding similar state values. In most  scenarios,  converging to globally optimal and unique parameter values can be challenging, requiring long run times and careful diagnostic monitoring \cite{KennedyDukicDwyer2014AmNat, dukic2012, ramsay2007}.

Several alternative approaches have therefore been put forward in an attempt to reduce or altogether avoid using forward-based solvers, with varying levels of success: via likelihood-free sequential Monte Carlo methods \cite{ABC2009}, local approximations \cite{conrad2016}, and manifold-constrained Gaussian processes \cite{YangWongKou2021ProcNatlAcadSci}. Sometimes a purely statistical approach (such as spline and kernel regression methods in \cite{splineODE2018}) is used instead to model a smooth temporal pattern. With the exception of the Gaussian-process based approach in \cite{YangWongKou2021ProcNatlAcadSci}, none of the above methods are guaranteed to solve the underlying dynamical system;  our weak-form approach in contrast guarantees that. Moreover, the Gaussian-process approach in \cite{YangWongKou2021ProcNatlAcadSci} requires Markov chain Monte Carlo sampling to do so and thus it takes notably longer to produce results than the weak form algorithm presented here.


\subsubsection{Model Equation Learning}


As we described earlier, the standard approach to the model learning problems in ecology is to write down a small candidate set of models based on a general understanding of the organisms in question, and then to fit the parameters in each model as if that model was the true model. Statistical model selection is then performed using an information criterion (such as Akaike (AIC), Bayesian (BIC), Watanabe (WAIC), etc.) \cite{Akaike1977Applicationsofstatistics,Schwarz1978AnnStatist,Watanabe2013JMachLearnRes,HootenHobbs2015EcolMonogr,MihaljevicPolivkaMehmelEtAl2020AmNat,DwyerMihaljevicDukic2022AmNat, kyle2020stochasticity}. This process requires a great deal of knowledge of the biology and sizeable computational resources, and is particularly challenging when the models under consideration have either high parameter dimensions or high-dimensional states, or both in the case of hierarchical and random differential equation models \cite{ElderdDwyerDukic2013Epidemics,MihaljevicPolivkaMehmelEtAl2019AmNat}, or in the case of strong seasonality and multiple coupled models, like the discrete-continuous hybrid framework that we consider here.

Our approach is to instead write down very general expressions for the model structure, and then to use data to choose the actual model structure from a large library $\Lbb$ consisting of $|\Lbb|$ possible equation terms (where $|\Lbb|$ in our application is set to 562), resulting in $2^{|\Lbb|}$  possible candidate models for consideration. Such large model spaces cannot be explored in their entirety, but instead rely on guided methods for sequential exploration, such as sequential thresholding \cite{BruntonProctorKutz2016ProcNatlAcadSciUSA,ZhangSchaeffer2019MultiscaleModelSimul}.  The equation learning methodology presented here thus results in the identification of interpretable forms of dominant mechanisms governing a  hybrid discrete-continuous dynamical system for population dynamics, among combinations of hundreds of possible linear and non-linear terms in the library, without the need to separately fit different models to the data.


Previous methods for  model equation learning for hybrid dynamics have generally handled parameter inference only  after the final form of the model has been chosen.  Such methods include combined sparse identification and clustering \cite{ManganAskhamBruntonEtAl2019ProcRSocA,NovelliLenciBelardinelli2022JComputNonlinearDyn}, neural networks \cite{LiXuDuanEtAl2023ChaosInterdiscipJNonlinearSci,BrusaferriMatteucciPortolaniEtAl202020207thIntConfControlDecisInfTechnolCoDIT}, and online learning \cite{LiWuLiu2023PhysRevResearch}.
Ecological hybrid modeling calls for new tools, particularly to address complex dependencies between the discrete and continuous modeling stages.  In addition, in contrast to engineering applications that are the focus of  \cite{ManganAskhamBruntonEtAl2019ProcRSocA},  in many ecological settings the switching times between continuous stages are seasonal and therefore known {\it a priori}. This implies that many existing methods may be unnecessarily complex, providing non-optimal or inefficient estimators for ecological hybrid models. 

Here we develop a data-driven method that performs  weak-form ``equation-learning''  methodology simultaneously with  parameter inference \cite{MessengerBortz2021SIAMMultiscaleModelSimul, MessengerBortz2021JComputPhys, BortzMessengerDukic2023article}. 
For this, we develop a novel weak-form version of the recently developed sparse parametric modeling technique \cite{NicolaouHuoChenEtAl2023PhysRevResearch} and extend it to incorporate  simultaneous parameter inference as in \cite{BortzMessengerDukic2023article}.

Our weak-form method is highly efficient as it does not employ forward-solvers during model identification, requiring on the order of $30$ seconds on a modern laptop to complete, inclusive of the search through a large set of $2^{|\Lbb|}$ models. Our method has relatively low data-requirements and high noise tolerance as compared to other sparse equation learning methods that employ the strong form  (see e.g.\ \cite{MessengerBortz2021SIAMMultiscaleModelSimul} for a comparison). Our method is also orders of magnitude faster than equation learning methods that employ non-linear least squares \cite{BortzMessengerDukic2023article} and orders of magnitude more accurate and robust to observational noise levels than existing dynamics-identification methods (see e.g.\ \cite{BortzMessengerDukic2023article}). We show that a sparse set of intermittent observations 
is sufficient to identify a coupled system of short-term continuous dynamics and long-term discrete dynamics, each a generalization of the annual epidemics and overwintering effects common to seasonal host-pathogen systems. This enables a variety of inference tasks common to ecology to be performed efficiently, such as for example prediction of future boom and bust cycles. In fact, as we will show in the Numerical Experiments, data on 18 nonconsecutive generations can enable accurate prediction for upwards of 40 generations with our method. 

Our method learns a parametrization of the continuous-time model by the discrete-time variables (and vice versa) using the weak-form model learning algorithm (WSINDy) \cite{MessengerBortz2021SIAMMultiscaleModelSimul}, and is therefore  a novel  approach to modeling hybrid systems \cite{GoebelSanfeliceTeel2012,van2007introduction}. Furthermore, learning a system of differential equations using WSINDy, jointly with the recently-developed weak-form parameter inference method WENDy \cite{BortzMessengerDukic2023article}, constitutes a novel approach to model learning.  In addition, both WSINDy and WENDy are novel to ecology applications. WENDy has been shown to lead to a substantial improvement over both traditional nonlinear-least-squares optimization \cite{BortzMessengerDukic2023article} and Bayesian methods \cite{ElderdDukicDwyer2006ProcNatlAcadSci,YangWongKou2021ProcNatlAcadSci}. The resulting improvements in our ability to identify the dominant mechanisms in both short- and long-term modeling components have the potential to improve our understanding of the underlying ecology, to produce better model-based predictions, and to better inform future control strategies. The weak-form equation-learning methods presented here therefore have the capability to allow model identification and inference for a broad class of dynamical systems that express some form of seasonality, and that therefore lie outside of the current capabilities of equation learning.

This article is organized as follows. In Section \ref{sec:motivation} we present more details on the insect-pathogen systems that motivate our seasonal hybrid models. In Section \ref{sec:methods} we describe our proposed model identification method, including a high-level overview of our weak-form equation learning methodology in Section \ref{sec:WFEL}, and a description of the complete algorithm in Section \ref{sec:WFEL_hybrid}. In Section \ref{sec:num_exp} we perform a series of numerical experiments to assess the performance of the method (Sections \ref{sec:longtermforecasting} and \ref{sec:sampling}), and in addition we provide a detailed roadmap of how to interpret, build upon, and utilize the results of the algorithm in practice to make predictions and quantify uncertainty (Section \ref{sec:HITL}). A discussion is provided in Section \ref{sec:discussion}.

\section{Motivation}\label{sec:motivation}
 
In order to motivate the model class relevant to this paper,  in Section \ref{sec:spongy_moth} we review a simplified model for the dynamics of insect-baculovirus interactions, which will form the basis of our numerical experiments in Section \ref{sec:num_exp}. The general class of seasonal discrete-continuous models is presented in Section \ref{sec:general_case}.

\subsection{\raggedright Seasonality in Host-Pathogen Systems}\label{sec:spongy_moth}

Animal and plant host-pathogen systems commonly have a high degree of seasonality that depends on the timing of host reproduction and pathogen transmission (see \cite{AndreasenDwyer2023TheAmericanNaturalist} and the references therein). In many animal hosts, notably invertebrates and amphibians, pathogen transmission can only occur during the larval stage, while in plants host transmission often only occurs during the growing season.  Meanwhile,  transmission in many bird and mammal hosts often ends because of a lack of susceptible hosts.  In either type of host-pathogen interaction, long periods of disease transmission strongly reduce the susceptible host population, after which short periods of host reproduction replenish the susceptible host population. The ubiquitous effect of seasonality in ecological systems demands a robust modeling and inference framework, which we argue is met by hybrid discrete-continuous models and the inference method that we present here.

To introduce the hybrid discrete-continuous modeling framework, and to demonstrate its applicability to strongly seasonal ecological systems, we  consider a model for insect-baculovirus dynamics \cite{DwyerDushoffElkintonEtAl2000}.   As we described earlier, baculoviruses represent a large group of viral pathogens of arthropod hosts \cite{miller1997baculovirusesBook}; in many insects in particular, baculoviruses are transmitted when uninfected larvae that are feeding on the foliage of their host plant accidentally consume infectious occlusion bodies released from the cadavers of hosts killed by the virus \cite{elderdDwyer2019popStructure}.  Baculoviruses must thus kill their host to be transmitted \cite{Dwyer1991Ecology}.  
For many species of forest insects, baculoviruses decimate outbreaking populations \cite{OtvosCunninghamFriskie1987CanEntomol, WoodsElkinton1987JInvertebrPathol}, and therefore appear to drive long-period, large-amplitude predator-prey type cycles in populations of their hosts \cite{Myers1993AmSci}.  The ability of mathematical host-pathogen models to reproduce these cycles provides support for this hypothesis \cite{DwyerDushoffElkintonEtAl2000, DwyerDushoffYee2004Nature}.   

Outbreaking forest insects typically have only one generation per year \cite{Hunter1995PopulDynNewApproachesSynth}.  Because only larvae can become infected, and because host reproduction therefore occurs after each year's viral epizootic is over,  we first construct a continuous-time model of baculovirus dynamics during the larval period, using a modification of a standard SIR model from theoretical epidemiology \cite{keeling2011modelingInfDiseases}.  Let $S_n(t)$ denote the density of susceptible host larvae and $P_n(t)$ the density of baculovirus-infected cadavers in year $n$ (the $n$th generation).  We then have the following short-term epizootic model running from time $t=0$ to $t=T$ (in days, relative to the start of the generation) in terms of ordinary differential equations:
\begin{equation}\label{shorttime}
\begin{dcases}
\dot{S}_n = -\bar{\nu} S_nP_n \left(\frac{S_n(t)}{S_n(0)}\right)^{V}\\
\dot{P}_n = \bar{\nu} S_nP_n \left(\frac{S_n(t)}{S_n(0)}\right)^{V} - \mu P_n
\\ S_n(0) = N_n, \quad P_n(0) = Z_n
\end{dcases} 
\end{equation}
Here $N_n$ and $Z_n$ denote the host and pathogen density at the start of the epizootic in generation $n$, and $\mu$ is the pathogen decay rate, meaning the breakdown rate of infectious larval cadavers on the foliage \cite{FullerElderdDwyer2012TheAmericanNaturalist}.  This model arises via moment closure from the assumption that the host population exhibits heterogeneity in the pathogen's infection rate, following a distribution with mean $\bar{\nu}$ and squared coefficient of variation\footnote{Coefficient of variation is defined as the ratio of the standard deviation to the mean, and serves as a measure of dispersion, or heterogeneity, for a distribution. For a Gamma$(\alpha,\beta)$ density, it is equal to $\alpha^{-1/2}.$} $V$ \cite[Supporting Text]{ElderdDukicDwyer2006ProcNatlAcadSci}.   Moderate values of $V$ produce long-period, large-amplitude population cycles that are in qualitative agreement with observations of population fluctuations in many forest insects   \cite{DwyerDushoffElkintonEtAl2000,DwyerDushoffYee2004Nature}. 
Because larvae by definition do not reproduce, this model describes the change in the host population that results from pathogen infection in the absence of host reproduction.  

Multiple generations are coupled together by the following long-term discrete model:
\begin{equation}\label{longtime}
\begin{dcases}
N_{n+1} = \lambda N_n(1-I_n)\\ 
Z_{n+1} = \phi N_nI_n+\gamma Z_n\\
I_n = 1 - \frac{S_n(T)}{S_n(0)}
\end{dcases}
\end{equation}
These equations describe host reproduction and pathogen overwintering.   The symbol $I_n$ represents the fraction of larvae that die because of pathogen infection during the $n$th generation (recall that the pathogen is fatal.) The density of surviving host larvae given by $S_n(T) = (1-I_n)S_n(0) = (1-I_n)N_n$, gives rise to adults that lay the next generation of eggs, at a fixed rate $\lambda$.  Because we assume that $\lambda$ allows for the non-disease mortality that occurs before and after the epizootic, we refer to $\lambda$ as ``net fecundity''.   The discrete-generation model assumes that the eggs lay dormant throughout the winter and hatch in the spring. The $N_nI_n$ victims of the pathogen become infectious cadavers, which have net over-winter survival rate $\phi$.  Because $\phi$ allows for both pathogen over-wintering and the much higher susceptibility of hatchling larvae, we allow $\phi > 1$.  The parameter $\gamma$ then allows for long-term pathogen survival in the environment.

\subsection{Seasonal Discrete-Continuous Hybrid Models: the General Case}\label{sec:general_case}

The model in \eqref{shorttime} is a specific case of a more general model for seasonal systems that can be cast in the following framework. Letting $X_n = (N_n,Z_n)$ be the discrete-time model states described in \eqref{longtime} and $Y_n(t) = (S_n(t),P_n(t))$ be the continous-time states described in \eqref{shorttime}, the hybrid model combining \eqref{shorttime} and \eqref{longtime} has the following general form:
\begin{align}
\label{Xmap} X_{n+1} &= F(X_n,Y_n(T))\\
\label{Ymap}\dot{Y}_n &= g(Y_n(t),X_n)\\
\label{YICmap} Y_n(0) &= h(X_n) \\ 
\label{XIC}X_0&\in \Rbb^D
\end{align}
for $X_n\in \Rbb^D, Y_n\in \Rbb^d$ and some fixed $T>0$. This motivates our current focus, which is to identify $F$, $g$, and $h$, for the general case \eqref{Xmap}-\eqref{YICmap}, given  measurement data and   $T$. Note we have assumed that $h$ depends only on the discrete (long-time) variable $X_n\in \Rbb^D$, which is primarily for convenience, so that knowledge of $X_0$ is all that is needed to simulate the full system \eqref{Xmap}-\eqref{YICmap} given  $F$, $g$, and $h$. {Our strategy will be to assume that $F$, $g$, and $h$ can all be represented as a linear combination of terms from a library $\Lbb$, and that $\Lbb$ is made sufficiently large and with sufficiently many terms that might be appropriate for the problem at hand.}

While the equations in \eqref{Xmap}-\eqref{XIC} capture a broad class of dynamics, we highlight some interesting extensions. It is possible to replace the dependence of $F$ on $Y_n(T)$ with dependence on all of $Y_n(t),t\in [0,T]$. Further extensions of this model (and \eqref{shorttime} and \eqref{longtime}) include delayed pathogen interactions, genetic drift, and stochasticity \cite{DwyerDushoffElkintonEtAl2000,DwyerMihaljevicDukic2022AmNat,PaezDukicDushoffEtAl2017AmNat}. The method presented here is generally applicable to these extensions, but may require including  additional features in the library. This is certainly the case for  the identification of stochastic hybrid models,  and we aim to pursue that extension in  future work.

We note that simplifying assumptions have been made to replace the short-time model \eqref{Ymap} with a fully discrete map. In \cite{DwyerDushoffElkintonEtAl2000}, the fraction $I_n$ of infected hosts in \eqref{longtime} is approximated directly from $(N_n,Z_n)$, removing the need to simulate the continuous time variables $(S_n,P_n)$. While this is useful for studying long-term properties of the discrete model, the hybrid modeling framework is of more widespread relevance \cite{SinghEmerick2020preprint,EmerickSingh2016MathematicalBiosciences,EmerickSinghChhetri2020MathematicalBiosciences,EskolaGeritz2007BullMathBiol,PachepskyNisbetMurdoch2008Ecology}, and allows for much more flexible short-term dynamics. We defer discovery of purely-discrete surrogate models for \eqref{Xmap}-\eqref{XIC} to future work.

\section{Methods}\label{sec:methods}
The main result of this manuscript is that the discrete-continuous seasonal dynamics \eqref{Xmap}-\eqref{XIC} can be identified from data with moderate measurement noise using the sparse weak-form equation learning approach proposed in this Section. The proposed method is an extension and merger of two existing non-linear dynamics identification algorithms, WSINDy \cite{MessengerBortz2021SIAMMultiscaleModelSimul,MessengerBortz2021JComputPhys} and WENDy \cite{BortzMessengerDukic2023article}. The  method is suitable for incomplete and irregularly collected data, with gaps in observations sometimes stretching over several years, as commonly seen in ecology given the inherent difficulties of collecting field data. 

In what follows, we provide an overview of weak-form inference in Section \ref{sec:WFEL}, including  the WSINDy and WENDy algorithms \cite{MessengerBortz2021SIAMMultiscaleModelSimul,MessengerBortz2021JComputPhys,BortzMessengerDukic2023article}. Specific advancements of these methods for the current setting of inference for hybrid discrete-continuous models of the form \eqref{Xmap}-\eqref{YICmap} is included in Section \ref{sec:WFEL_hybrid}.

\subsection{Weak-form Equation Learning and Parameter Inference}\label{sec:WFEL}

The equations \eqref{Xmap}-\eqref{XIC} are known as the strong form of the differential equations, a form that is characterized by specifying that the equation holds pointwise in time, for  all $t$. In contrast, the {\it weak form} of a differential equation specifies that the equation holds when integrated in time with respect to an arbitrary function referred to as the {\it test function}. 
The crux of weak-form methods relies on choosing the right test functions to achieve a given objective. For example, when the objective is inference from noisy data, smoothness of the test functions will be an important point. We discuss this in detail in the next subsections. 

Weak-form inference methods have origins in parameter estimation dating back to the 1950's 
(see e.g. \cite{Shinbrot1954NACATN3288}). In some disciplines this is known as the modulating function technique \cite{PreisigRippin1993ComputChemEngb}.  Weak form equation learning, which combines parameter estimation with model selection, is a more recent pursuit, and together with works published by authors of the current manuscript \cite{MessengerBortz2022PhysicaDNonlinearPhenomena,MessengerBortz2021JComputPhys,
MessengerBortz2021SIAMMultiscaleModelSimul,MessengerWheelerLiuEtAl2022JRSocInterface,MessengerBurbyBortz2023}, has seen many new developments in recent years 
\cite{
SchaefferMcCalla2017PhysRevE,
WangHuanGarikipati2019ComputerMethodsinAppliedMechanicsandEngineering,
GurevichReinboldGrigoriev2019Chaos,
BertsimasGurnee2022arXiv220600176article,
RussoLaiu2022arXiv220915573article,
WoodallEsparzaGutovaEtAl2023preprinta,
FaselKutzBruntonEtAl2022ProcRSocA,
TangLiaoKuskeEtAl2023JComputPhys,
NicolaouHuoChenEtAl2023PhysRevResearch}. 

Another recent pursuit in the context of parameter estimation is to successively improve weak-form estimates by iteratively incorporating the action of the test function basis $\Phi$ into the covariance structure for the weak form residual. 
This is known as the WENDy method, or Weak-Form Estimation of Nonlinear Dynamics, and has been shown to be advantageous over forward solver-based parameter estimation (such as non-linear least squares) in a variety of ODE models of biological relevance \cite{BortzMessengerDukic2023article}.

In the current work we combine the WENDy and WSINDy methods for the first time, arriving at an efficient inference framework for simultaneously selecting model terms and estimating parameters. Below, we review the WENDy and WSINDy methods.

\paragraph{Strong-form Model Equation Learning.} Given observations $\Ybf = (\ybf_1,\dots,\ybf_M)$ of a system's state variables $y(t)\in \Rbb^d$ where $\ybf_i$ are noisy observations of $y(t_i)$ and $\tbf = (t_0,\dots,t_M)$ is a series of time points\footnote{Although this approach need not be restricted to time-dependent phenomena.}, the goal of {\it equation learning} is to identify an appropriate evolution equation
\begin{equation}\label{eq:general_evoeq}
y_+(t) = F(y(t))
\end{equation}
For discrete-time systems, the left-hand side is defined as  $y_+(t) = y(t+\Delta)$ for a specified time interval $\Delta$; for continuous-time systems $y_+(t) = \frac{d}{dt}y(t)$. Rather than merely approximating the map $F$ from $y$ to $y_+$, equation learning represents $F$ as a human-readable function that can be used to infer the underlying mechanisms driving the system, and thus be used to further our understanding of the ecology of these systems, facilitate  their control, and inform future experimental designs. 

The SINDy equation learning  algorithm \cite{BruntonProctorKutz2016ProcNatlAcadSciUSA} assumes that $F$ admits a sparse expansion
\begin{equation}\label{eq:SINDy}
F_i(y) = \Wbf_{1i}f_1(y)+\dots+\Wbf_{Ji}f_J(y), \quad i=1,\dots,d
\end{equation}
in terms of a {\it trial function library} $\Lbb = (f_j)_{j=1}^J$, with each $f_j\in \Lbb$ having a meaningful  interpretation relevant to the process being modeled. Here $\Wbf=(\Wbf_{ji})_{j=1,i=1}^{J,d}$ is a {\it weight matrix} which is assumed to be sparse, with nonzero coefficient $\Wbf_{ji}$  indicating that $f_j$ is an important mechanism in the dynamics of $y_i$. For instance, if $y = (S,P)$ is a host-pathogen system, the term $f(y) = -PS$ appearing in the $S_+$ equation would indicate that the host population decreases at a rate linear in the pathogen density, whereas the term $-\left(\frac{P}{1+P}\right)S$ would indicate that the rate of decrease levels off at high pathogen density, possibly capturing the effects of crowding or resource scarcity in the pathogen population.

\paragraph{Weak-form Model Equation Learning.} In {\it weak form}  learning, the requirement that the dynamics of $y$ are governed by the so-called {\it strong form} equation \eqref{eq:general_evoeq}, holding at every $t$, is weakened as follows. First, the function $t\to F(y(t))$ is assumed to belong to a space of functions $\cal X$ of  square-integrable  over a time window $[0,T]$. The space $\cal X$ is denoted as:
\[L_2([0,T];\Rbb^d) = \bigg\{g:[0,T]\to \Rbb^d\ \bigg\vert\ \int_0^T \|g(t)\|_2^2dt<\infty\bigg\}\]
 for continuous-time systems, and
\[\ell_2([M];\Rbb^d)=\bigg\{g:[M]\to \Rbb^d\ \bigg\vert\ \sum_{n=0}^{M}\|g(n)\|_2^2<\infty\bigg\}\]
for discrete-time systems, where $[M] = \{0,1,2,\dots,M\}$, and $M = \lfloor T/\Delta\rfloor$. We then let $\cal X^*$ be the {\it dual} of $\cal X$, that is, the space of continuous linear functionals $\varphi: \cal X\to \Rbb$. The {\it weak form} of \eqref{eq:general_evoeq} is then defined as
\begin{equation}\label{eq:weak_evoeq}
\varphi(y_+(\cdot)) = \varphi(F(y(\cdot)) \quad \text{for all}\ \varphi\in X^*.
\end{equation}
When $\cal X$ is $L_2$ (and similarly for $\ell_2$), it holds that
\begin{equation}\label{eq:weakrep}
\varphi(g) = \int_0^T\phi(t)\cdot g(t) dt :=\lan \phi,g\ran
\end{equation} 
for all $\varphi\in \cal X^*$, where $\phi\in \cal X$ is some representer {\it test function}, and $\lan\cdot,\cdot\ran$ is the $L_2$ inner product. We then assume that $F$ admits the same representation as in \ref{eq:SINDy}. 

The weak form \eqref{eq:weak_evoeq} was originally developed to solve partial differential equations with jumps in the  solution or other nonsmooth characteristics \cite{LaxMilgram1954Contributionstothetheoryofpartialdifferentialequations} that do not satisfy the strong-form of the equation. This property makes the weak form beneficial when considering data corrupted by measurement noise, which can be interpreted as a solution to \eqref{eq:weak_evoeq} holding on a suitable {\it subspace} $\Phi\subset \cal X^*$. Inference methods that employ the weak form hinge on selecting an appropriate finite-dimensional subspace $\Phi$, known as the {\it test function basis}, on which to enforce the weak-form \eqref{eq:weak_evoeq}. 

\subsubsection{WSINDy}\label{sec:WSINDy}

We now briefly review the WSINDy algorithm (Weak-form SINDy \cite{MessengerBortz2021SIAMMultiscaleModelSimul,MessengerBortz2021JComputPhys}) and describe its application in continuous and discrete contexts. In both cases, a trial function library $\Lbb = (f_1,\dots,f_J)$ and a test function basis $\Phi = (\phi_1,\dots,\phi_K)$ can be specified by the user. We now describe guidelines for choosing $\Phi$, with specific test functions employed in this work included in Appendix \ref{app:wsindy_eco}.

\paragraph{Continuous Time.} When the data $\Ybf$ are assumed to come from a continuous-time process $y(t)$ which obeys an ordinary differential equation, integration by parts can be leveraged in the weak form to avoid computation of numerical derivatives. For this, the test functions  must at least be continuously differentiable; however, as identified in \cite{MessengerBortz2021SIAMMultiscaleModelSimul}, requiring test functions to have higher-order continuous derivatives can increase accuracy. In many cases though, the choice of $\Phi$ can be informed based on the data $\Ybf$ (see \cite[Appendix A]{MessengerBortz2021JComputPhys}).

\paragraph{Discrete Time.} In discrete time there are no issues with discretizing integrals or numerically approximating derivatives. The test functions can simply be taken as $\delta$-functions,  $\phi_k = \delta_{k\Delta}$, defined by 
\[\lan \delta_{k\Delta}, y\ran = y(k\Delta)\]
for a specified time point $k\Delta$, with $k\in \Nbb$ and $\Delta$ the fixed time interval. This is the strategy employed in the current work, although nothing precludes us from employing $\phi_k$ with a greater support in time. We  discuss this more in Section \ref{sec:discussion}.

\paragraph{Weak-form Linear System.} For given $\Lbb$ and $\Phi$, the {\it weak-form linear system} $(\Gbf,\Bbf)$ is then formed by discretizing the weak form equation \eqref{eq:weak_evoeq} at the given data point values $\Ybf$. The entries of the matrix $\Gbf\in \Rbb^{K\times J}$ are given by 
\begin{equation}\label{eq:Gij}
\Gbf_{kj} = \llangle \phi_k, f_j(\Ybf)\rrangle
\end{equation}
where $\llangle \cdot,\cdot\rrangle$ is a discretized version of the inner product \eqref{eq:weakrep}, taken for continuous systems to be the trapezoidal rule, 
\begin{align*}
&\llangle \phi, f(\Ybf)\rrangle = \\
&\qquad\sum_{i=0}^{M-1} \left(\frac{t_{i+1}-t_i}{2}\right)\left[\phi(t_i)f(\ybf_i)+\phi(t_{i+1})f(\ybf_{i+1})\right]
\end{align*}
In discrete time systems we simply have $\llangle \cdot,\cdot\rrangle = \lan\cdot,\cdot\ran_{\ell_2}$, the  $\ell_2$ inner product. 

In continuous time, the entries of $\Bbf$ are computed using integration by parts for \eqref{eq:weakrep}, yielding
\begin{equation}\label{eq:B_cont}
\Bbf_{ki} = \phi_k\Ybf_i\Big\vert_{t=t_0}^{t=t_M}-\left\llangle \frac{d}{dt}\phi_k,\Ybf_i\right\rrangle
\end{equation} 
where $\Ybf_i$ is the time series of the observed $i$th coordinate of $y\in \Rbb^d$. It is then assumed that $\phi_k(t_0) = \phi_k(t_M)=0$ to remove the boundary terms (first term on the right-hand side of \eqref{eq:B_cont}) from $\Bbf_{ki}$. 

In discrete time, we define $\Ybf^+ = \Tbf_+\Ybf$, where $\Tbf_+$ performs a shift forward in time, $\Ybf^+_{mi} = \ybf_i((m+1)\Delta )$, and similarly $\Tbf_-$ performs a backwards shift, to get

\begin{equation} \label{eq:B_disc}
\begin{split}
\hspace{-0.5cm}\Bbf_{ki} &= \llangle \phi_k, \Ybf_i^+\rrangle = \llangle \Tbf_-\phi_k, \Ybf_i\rrangle\ \ +\\
&\big[\phi_k(M\Delta)\ybf_i((M+1)\Delta)-\phi_k(-\Delta)\ybf_i(0)\big]
\end{split}\hspace{-1cm}
\end{equation}
where again we assume $\phi(-\Delta) = \phi(M\Delta) = 0$ to remove the boundary terms. Placing operators in the $\Bbf$ matrix onto the test functions is central to the WENDy algorithm, outlined below. 

\paragraph{Sparse Regression.} Once $(\Gbf,\Bbf)$ is formed, a sparse regression problem is solved for the set of weights as follows: 
\begin{equation}\label{eq:sparse_reg}
\widehat{\Wbf} = \argmin_{\Wbf}\sum_{i=1}^d\|\Gbf\Wbf_i-\Bbf_i\|^2_2+\lambda^2\|\Wbf_i\|_0
\end{equation}
In SINDy (Sparse Identification of Nonlinear Dynamics \cite{BruntonProctorKutz2016ProcNatlAcadSciUSA}), the STLS algorithm (Sequential Thresholding Least Squares, \cite{BruntonProctorKutz2016ProcNatlAcadSciUSA,ZhangSchaeffer2019MultiscaleModelSimul}) is employed to solve \eqref{eq:sparse_reg} which involves truncation of terms with coefficients $|\Wbf_{ji}|<\lambda$, in other words uniform thresholding according to $\lambda$. In WSINDy, \eqref{eq:sparse_reg} is solved using the MSTLS algorithm (modified STLS), which uses nonuniform thresholding and performs a grid search to select the sparsity threshold $\lambda$. In MSTLS, thresholding intervals $I_{ij}^\lambda$ are defined by
\begin{equation}\label{eq:MSTLS_I}
I_{ji}^\lambda = \left[\lambda\max\left\{1,\frac{\|\Bbf_i\|_2}{\|\Gbf_j\|_2}\right\}, \quad \lambda^{-1}\min\left\{1,\frac{\|\Bbf_i\|_2}{\|\Gbf_j\|_2}\right\}\right]
\end{equation}
and at each step $|\Wbf_{ji}|\notin I_{ji}^\lambda$ is removed. For example, with $\lambda = 10^{-2}$, this says that the overall term magnitude $\|\Gbf_j\Wbf_{ji}\|_2$ as well as the coefficient value $|\Wbf_{ji}|$ are each restricted to lie within 2 orders of magnitude from $\|\Bbf_i\|_2$ and 1, respectively.

\subsubsection{WENDy}\label{sec:WENDy}
The WENDy algorithm \cite{BortzMessengerDukic2023article} is a forward-solver-free method for estimating parameters in non-linear dynamical systems for which the model form is known. The only requirement is that the systems are linear in parameters (as in \eqref{eq:SINDy}), agreeing with the WSINDy construction just described. That is, the sparsity aspect of WSINDy is removed. A very simple approach to solving this problem would be to set $\lambda=0$ in \eqref{eq:sparse_reg}, leading to an ordinary least-squares problem. From a statistical point of view, however, the underlying assumption of least squares is that the $i$th residual vector
\[ \Rbf_i = \Gbf\Wbf_i - \Bbf_i\]
has uncorrelated entries with zero mean. This would be a fundamentally incorrect assumption, however, because $\Gbf$ is in most cases nonlinear in the data $\Ybf$ and because the weak form introduces correlations in time. The idea behind WENDy is to utilize the correlations imparted by the weak form, and to correct for nonlinearities, in order to improve parameter estimates and quantify uncertainty.

In order to incorporate the inherent correlations in $\Rbf$ into an estimate for $\Wbf$, we first define the vectorized quantities $\ybf = \textsf{vec}(\Ybf)$, $\wbf = \textsf{vec}(\Wbf)$ and subsequently the vectorized residual $\rbf(\wbf;\ybf) = \textsf{vec}(\Rbf) \in \Rbb^{Kd}$, with entries 
\[\rbf_{(i-1)K+k}(\wbf;\ybf) = \Rbf_{ki} = \llangle \CalD^*\phi_k, \Ybf_i\rrangle - \llangle \phi_k,\Lbb(\Ybf)\Wbf_i\rrangle\]
Here $\CalD$ denotes either $\frac{d}{dt}$ or the forward shift  operator $\Tbf^{+}$, depending on continuous or discrete systems, and $\CalD^*$ is its {\it adjoint} ($-\frac{d}{dt}$ for continuous systems and $\Tbf^{-}$ for discrete systems). If the samples are perturbed, $\ybf = \ybf^\star+\delta \ybf$, we can propagate the effect of the errors $\delta\ybf$ into the parameter estimates $\widehat{\wbf}$ by Taylor expanding $\rbf$ evaluated at the true weights $\wbf^\star$ around the unperturbed data $\ybf^\star$,
\[\rbf(\wbf^\star;\ybf) \approx \rbf(\wbf^\star;\ybf^\star) + \Lbf(\wbf^\star) \delta \ybf +\CalO(\|\delta \ybf\|^2)\]
where $\Lbf(\wbf^\star)$ is a linear operator defined as the Fr\'echet derivative
\[\Lbf(\wbf^\star)\delta \ybf=\lim_{\ep\to 0}\frac{1}{\ep}\rbf(\wbf^\star;\ybf^\star+\ep\delta\ybf)\bigg\vert_{\ep=0}\]
Moreover, $\Lbf(\wbf^\star)$ is linear in $\wbf^\star$, computed using
\begin{equation}\label{eq:Lfacs}
\Lbf(\wbf^\star) := \Lbf^{(0)}(\Phi)+\sum_{\ell=1}^{Jd}\wbf^\star_{\ell}\Lbf^{(\ell)}(\Phi,\ybf)
\end{equation}
where the factors $\Lbf^{(0)},\dots\Lbf^{(Jd)}$ depend only on $\Phi$, $\Lbb$, and $\ybf$ and can thus can be pre-computed without knowledge of $\wstar$ (see \cite{BortzMessengerDukic2023article} for details). 

When $\delta \ybf$ is normally distributed with mean zero and covariance $\Sigma$, then the residuals $\rbf(\wbf^\star;\ybf)$ are approximately normal with mean-zero and covariance 
\[\Cbf(\wstar) = \Lbf(\wstar)\Sigma\Lbf(\wstar)^T\]
thus an approximate set of weights $\widehat{\wbf}\approx \wbf^\star$ can be obtained by solving the generalized least squares (GLS) problem
\begin{equation}\label{eq:wendy}
\min_\wbf (\Gbf_{_{\tiny \otimes}}\wbf-\bbf)^T\Cbf^{-1}(\wbf)(\Gbf_{_{\tiny \otimes}}\wbf-\bbf)
\end{equation}
where $\bbf = \textsf{vec}(\Bbf)$, $\Gbf_{_{\tiny \otimes}} = \Ibf^{(d)}\otimes \Gbf$ and $\Ibf^{(d)}$ is the identity in $\Rbb^{d\times d}$. However, \eqref{eq:wendy} introduces $\wbf$ dependence into $\Cbf$. In WENDy this is dealt with using iterative reweighted least squares (IRLS), starting from an initial ordinary least squares solve $\wbf^{(0)} = (\Gbf_{_{\tiny \otimes}})^\dagger\bbf$, before updating the covariance $\Cbf(\wbf^{(0)})$ and performing generalized least squares (GLS) for a new set of weights (see lines 7-12 of Algorithm \ref{alg:MSTLS_WENDy}).  

\subsubsection{WSINDy-WENDy} 
In this work, we combine the WSINDy and WENDy approaches. The former is employed to perform efficient model selection, while WENDy is used to estimate the selected  model's parameters more accurately and more precisely \cite{BortzMessengerDukic2023article}. 

\paragraph{MSTLS-WENDy.} The first step in this combination is the new sparse regression algorithm \texttt{MSTLS-WENDy}, provided in Algorithm \ref{alg:MSTLS_WENDy}, whereby WENDy replaces the usual ordinary least squares steps in MSTLS. Hyperparameters for \texttt{MSTLS-WENDy} include WENDy hyperparameters $\tau$ and \texttt{MaxIts}, which represent the stopping tolerance and maximum number of WENDy iterations, respectively, as well as the candidate thresholds $\pmb{\lambda}$ used in the MSTLS grid search.

\paragraph{Library Incrementation.} To increase efficiency, we sequentially increment the library $\Lbb$ using covariance information from WENDy to estimate the expected norm of the residual $\rbf$. In the current setting, parametrized libraries can quickly grow to exceed computational resources. To circumvent this, we consider a nested sequence of libraries $\Lbb^{(0)}\subset \Lbb^{(1)}\subset \cdots$ together with an incrementation operator $\Lbb_+$ defined by
\begin{equation}\label{eq:lib_inc}
\Lbb_+(\Lbb^{(j)}) = \Lbb^{(j+1)}
\end{equation}
We then define an incrementation criteria based on the outputs $(\what^{(\ell)}$, $\widehat{\Cbf}^{(\ell)}, \widehat{\Sbf}^{(\ell)})$ and weak-form matrix $\Gbf_{_{\tiny \otimes}}^{(\ell)}$ of \texttt{MSTLS-WENDy} with library $\Lbb^{(\ell)}$ as follows. The WENDy  residual $\widehat{\rbf}^{(\ell)} = \Gbf_{_{\tiny \otimes}}^{(\ell)}\what^{(\ell)}-\bbf$ is approximately normally distributed \cite{BortzMessengerDukic2023article}:
\[\widehat{\rbf}^{(\ell)} \sim \CalN(\textbf{0},\widehat{\Cbf}^{(\ell)})\]
Similarly, we have that the mean squared residual is approximately normally distributed\footnote{This holds for large $Kd$, where $K$ is the number of test functions and $d$ is the number of equations. In the examples below, $Kd=84|\CalI|$ for the continuous dynamics and $Kd=2|\CalI|$ for the discrete and initial conditions maps, where $|\CalI|$ is the number of observed generations.}:
\[r^{(\ell)} = \frac{1}{Kd}\sum_{k=1}^{Kd}(\widehat{\rbf}^{(\ell)}_k)^2\sim \CalN(\hat{M}^{(\ell)},\hat{V}^{(\ell)})\] 
where $\hat{M}^{(\ell)}$ and $\hat{V}^{(\ell)}$ are the sample mean and covariance of the diagonal of $\widehat{\Cbf}^{(\ell)}.$ A criterion to increment the library to $\Lbb^{(\ell+1)}$, for a given confidence level $c\in (0,1)$, is then  
\begin{equation} \label{eq:LibInc}
\begin{split}
&\texttt{LibIncFun}(\what^{(\ell)},\widehat{\Cbf}^{(\ell)},c)\\
&:=\Big\{ r^{(\ell)} 
> \max\big(\hat{M}^{(\ell)} + \Psi^{-1}(c)\sqrt{\hat{V}^{(\ell)}},\ \texttt{tol}\|\bbf\|_\text{rms}^2\big)\Big\}
\end{split}
\end{equation}
where $\Psi$ is the standard normal CDF. A value of $\texttt{LibIncFun}(\what^{(\ell)},\widehat{\Cbf}^{(\ell)},c)=\texttt{True}$, indicates that $\widehat{\rbf}^{(\ell)}$ 
is outside of the confidence interval defined by $c$, which indicates that a better model fit might be warranted. In the event that the covariance is grossly underestimated, the method defaults to a simple residual-based criteria given by $r^{(\ell)} > \texttt{tol}\|\bbf\|_\text{rms}^2$, where $\texttt{tol}=0.01$ in the numerical experiments below. 

\subsection{Weak-form sparse identification for Seasonal Discrete-Continuous Models}\label{sec:WFEL_hybrid}

Our proposed approach to identify dynamical systems of the general seasonal hybrid form \eqref{YICmap}-\eqref{Xmap}, which we will refer to as the \texttt{WSINDy-Eco} algorithm, is presented in Algorithm \ref{alg:full}. Given data of the form $\Xbf = (X_n,X_{n+1})_{n\in\CalI}$ and $\Ybf = (Y_n(\tbf_n))_{n\in \CalI}$ for a finite set of generations $\CalI$ and timegrids $(\tbf_n)_{n\in \CalI}$, \texttt{WSINDy-Eco} outputs weight vectors $(\what^\text{IC},\what^Y,\what^X)$ along with libraries $(\Lbb^\text{IC}, \Lbb^Y, \Lbb^X)$ to define the learned model, as well as estimates $(\widehat{\Cbf}^\text{IC},\widehat{\Cbf}^Y,\widehat{\Cbf}^X)$ for the covariances of the residuals for each submodel, and parameter covariance estimates $(\widehat{\Sbf}^\text{IC},\widehat{\Sbf}^Y,\widehat{\Sbf}^X)$. Crucially, the parameter covariances allow for uncertainty quantification in a variety of predictive scenarios, as described in Sections \ref{sec:UQ} and \ref{sec:HITL}. Choices of hyperparameters for \texttt{WSINDy-Eco} are presented in Appendix \ref{sec:parMSTLSWENDy}.

A main component \texttt{WSINDy-Eco} is the application of \texttt{MSTLS-WENDy} to parametrized libraries, dubbed \texttt{MSTLS-WENDy-Par}, described below in Section \ref{sec:parMSTLSWENDy} with pseudocode given in Algorithm \ref{alg:MSTLS-WENDy-Par} in the Appendix. \texttt{MSTLS-WENDy-Par} utilizes recently developed methods for identifying parametric dependencies in PDEs \cite{NicolaouHuoChenEtAl2023PhysRevResearch}. However, our approach employs parametrized libraries to learn dependencies between the discrete and continuous variables in each of the submodels \eqref{YICmap}-\eqref{Xmap}. 

In order to learn the discrete dynamics \eqref{Xmap}, an estimate of the continuous variables at time $T$, $\widehat{Y}_n(T)$, is required, provided $Y_n(T)$ is not directly observed (which is often the case and is assumed here). In \texttt{WSINDy-Eco} this is handled by simulating the learned continuous dynamics given by $(\what^Y,\Lbb^Y)$ using the learned initial conditions map given by $(\what^\text{IC},\Lbb^\text{IC})$, to obtain state estimates $\widehat{\Ybf}_T = (\widehat{Y}_n(T))_{n\in \CalI}$. These estimates are then used in conjunction with the observed data $(Y_n(\tbf_n))_{n\in \CalI}$ to inform a covariance estimate $\widehat{\Sigma}^{\widehat{\Ybf}_T}$ for use in the final \texttt{MSTLS-WENDy-Par} stage to approximate the discrete dynamics. This process is encompassed in the algorithm \texttt{ForwardSim} (Algorithm \ref{alg:ForwardSim}). This is the only stage of \texttt{WSINDy-Eco} requiring forward solves (and is unique to the hybrid model framework); in particular, no forward solves are needed during model identification.

\subsubsection{Parametric MSTLS-WENDy}\label{sec:parMSTLSWENDy}
Each of the three submodels \eqref{YICmap}-\eqref{Xmap} can be viewed as a parametrized equation, with state variables  and parameters, of the form 
\begin{equation}\label{eq:u_sp}
\CalD u^{(s)} = f(u^{(s)};u^{(p)})
\end{equation}
In the initial conditions map \eqref{YICmap}, $u^{(s)} = Y_n(0)$, $u^{(p)} = X_n$, and $\CalD$ is simply the identity. For the continuous-time map \eqref{Ymap}, $u^{(s)} = Y_n(t)$, $u^{(p)} = X_n$, and $\CalD = \frac{d}{dt}$. Finally, for the discrete-time map \eqref{Xmap} $u^{(s)} = X_n$, $u^{(p)} = Y_n(T)$, and $\CalD = \Tbf_+$. In \cite{NicolaouHuoChenEtAl2023PhysRevResearch} a sparse identification method is introduced to identify such parametric dependencies in dynamical systems, by building a tensorized library
\[\Lbb_{(s\otimes p)} = \Lbb_{(s)}\otimes \Lbb_{(p)}\]
constructed from libraries $\Lbb_{(s)}$ and $\Lbb_{(p)}$ in the state and parametric variables, respectively. Applying \texttt{MSTLS-WENDy} to this setting is straightforward, assuming data $\Ubf^{(s)}, \Ubf^{(p)}$ is available on $u^{(s)}$ and $u^{(p)}$, and data covariance matrices $\Sigma^{(s)},\Sigma^{(p)}$ are available such that approximately 
\[\text{Cov}(\textsf{vec}(\Ubf^{(s)})) = \Sigma^{(s)}, \quad \text{Cov}(\textsf{vec}(\Ubf^{(p)})) = \Sigma^{(p)}\] 
The concatenated data $\ubf = (\textsf{vec}(\Ubf^{(s)})^T,\textsf{vec}(\Ubf^{(p)})^T)^T$ then  approximately satisfies
\begin{equation}\label{eq:full_cov}
\Sigma^{(s\otimes p)} :=\text{Cov}(\ubf) = \begin{pmatrix} \Sigma^{(s)} & 0 \\ 0 & \Sigma^{(p)}\end{pmatrix}
\end{equation}

\SetKwComment{Comment}{/* }{ */}
\begin{algorithm*}
\caption{\label{alg:full}\texttt{WSINDy-Eco}. Weak-form model identification for seasonal hybrid discrete-continuous models  \eqref{Xmap}-\eqref{YICmap}.}
	\SetKwInOut{Input}{input} \SetKwInOut{Output}{output} 
	\Input{Data $\Xbf = (X_n,X_{n+1})_{n\in\CalI}$, $\Ybf = (Y_n(\tbf_n))_{n\in \CalI}$, $(\tbf_n)_{n\in \CalI}$; Data covariances $\Sigma^\Ybf,\Sigma^\Xbf$; Initial libraries $\Lbb^\text{IC}_{(p)}$, $\Lbb^Y_{(s)}$, $\Lbb^Y_{(p)}$, $\Lbb^X_{(s)}$, $\Lbb^X_{(p)}$; Library Incrementers $\Lbb_+^\text{IC}$,$\Lbb_+^Y$,$\Lbb_+^X$; test functions $\Phi^\text{IC},\Phi^\text{Y},\Phi^\text{X}$; candidate thresholds $\pmb{\lambda}$; WENDy stopping tolerance $\tau>0$, maximum WENDy iterations \texttt{MaxIts}, library incrementation confidence level $c$, simulation timepoints $t_\text{sim}$ with $T\in t_\text{sim}$}
    \Output{Parameter Estimates $(\what^\text{IC},\what^Y,\what^X)$, Residual Covariance Estimates $(\widehat{\Cbf}^\text{IC},\widehat{\Cbf}^Y,\widehat{\Cbf}^X)$, Parameter Covariance Estimates $(\widehat{\Sbf}^\text{IC},\widehat{\Sbf}^Y,\widehat{\Sbf}^X)$, Final Libraries $\Lbb^\text{IC}, \Lbb^Y, \Lbb^X$} 
$\Xbf_0 \gets (X_n)_{n\in \CalI}$\; 
$\Xbf_1 \gets (X_{n+1})_{n\in \CalI}$\;
$\Ybf_0 \gets (Y_{n}(\tbf_n(0)))_{n\in \CalI}$\;
$\Sigma^Y_0\gets \Sigma^\Ybf\big\vert_{\Ybf_0}$ \Comment*[r]{marginal covariance of $\Ybf_0$}
$\Sigma^X_0\gets \Sigma^\Xbf\big\vert_{\Xbf_0}$ \Comment*[r]{marginal covariance of $\Xbf_0$}
$(\what^\text{IC},\widehat{\Cbf}^\text{IC},\widehat{\Sbf}^\text{IC}) \gets \texttt{MSTLS-WENDy-Par}\Big(\Ibf_d,(\Ybf_0,\Xbf_0),(\Sigma^Y_0,\Sigma^X_0),([],\Lbb^\text{IC}_{(p)}),\Lbb_+^\text{IC},\Phi^\text{IC},\pmb{\lambda},\tau,\texttt{MaxIts},c\Big)$\;
\Comment*[r]{get IC map \eqref{YICmap}}
$(\what^Y,\widehat{\Cbf}^Y,\widehat{\Sbf}^Y) \gets \texttt{MSTLS-WENDy-Par}\Big(\frac{d}{dt},(\Ybf,\Xbf_0),(\Sigma^\Ybf,\Sigma^X_0),(\Lbb^Y_{(s)},\Lbb^Y_{(p)}),\Lbb_+^Y,\Phi^Y,\pmb{\lambda},\tau,\texttt{MaxIts},c\Big)$\;\Comment*[r]{get dynamics \eqref{Ymap}}
$(\widehat{\Ybf}_T,\widehat{\Sigma}^{\widehat{\Ybf}_T}) \gets \texttt{ForwardSim}\Big(\Xbf_0,\Ybf,\Sigma^\Ybf,\what^\text{IC},\Lbb^\text{IC},\what^Y,\Lbb^Y,t_\text{sim}\Big)$ \Comment*[r]{get time-$T$ estimates}
$(\what^X,\widehat{\Cbf}^X,\widehat{\Sbf}^X) \gets \texttt{MSTLS-WENDy-Par}\Big(\Tbf_+,\Xbf,\widehat{\Ybf}_T,\Sigma^\Xbf,\widehat{\Sigma}^{\widehat{\Ybf}_T},\Lbb^X_{(s)},\Lbb^X_{(p)},\Lbb_+^X,\Phi^X,\pmb{\lambda},\tau,\texttt{MaxIts},c\Big)$\;\Comment*[r]{get dynamics \eqref{Xmap}}
\end{algorithm*}

\section{Numerical Experiments}\label{sec:num_exp}

We now examine the performance of \texttt{WSINDy-Eco} applied to seasonal host-pathogen dynamics. To simplify exposition of the inference method, we restrict our examples to the epizootic model \eqref{shorttime}-\eqref{longtime}, which captures the key components of a seasonal hybrid model. Below we describe the data generation process, several performance metrics, and methods of uncertainty quantification, before diving into results concerning long-term prediction (Sec.\ \ref{sec:longtermforecasting}). We show that the method  predicts accurately over many cycles of the continuous-time variables (despite requiring accurate updates and re-initializations in terms of the discrete variables at every generation $n$). We then describe a human-in-the-loop modeling framework that leverages the uncertainty information provided by the WENDy regression approach in Section \ref{sec:HITL}. Finally, we evaluate the performance of the method under different sampling strategies in Section \ref{sec:sampling}, revealing that peak sampling (data collection during outbreaks, as is customary in field studies), provides a clear advantage over random sampling in terms of predictive accuracy. 

\subsection{Data Generation Process} 

We perform simulations of the epizootic model  \eqref{longtime} with parameters $\lambda=5.5$, $\phi=7$, $\gamma = 0.3$, $V=0.5$, $\bar{\nu} = 1$, and $\mu= 0.01$, based on previous results from the literature and field studies \cite{DwyerDushoffElkintonEtAl2000,ElderdDushoffDwyer2008TheAmericanNaturalist,PaezDukicDushoffEtAl2017AmNat}, and initial conditions $(N_0,Z_0)=(0.00713,0.0183)$ leading to the true continuous dynamics
\begin{align}
\label{eq:ICres}&\begin{dcases}
S_n(0)= (1.0)N_n\\
P_n(0) = (1.0)Z_n
\end{dcases}  \\
\label{eq:Yres}&\begin{dcases}
\dot{S}_n(t) = -(1.0)S_n(t)P_n(t)\left(\frac{S_n(t)}{N_n}\right)^{V}\\
\dot{P}_n(t) = (1.0)S_n(t)P_n(t)\left(\frac{S_n(t)}{N_n}\right)^{V}-(0.01)P_n(t)\\
\end{dcases} \hspace{-1cm}\\
\intertext{and the discrete update}
\label{eq:Xres}&\begin{dcases}
N_{n+1}= (5.5)S_n(T)\\
Z_{n+1} = (7.0)N_n-(7.0)S_n(T)+(0.3)Z_n
\end{dcases}
\end{align}
Here, a duration of $T=56$ days is chosen to correspond to 
an 8-week epizootic period of the spongy moth  \cite{PaezDukicDushoffEtAl2017AmNat}. See Figures \ref{fig:example_snrY01} and \ref{fig:example_snrY05} for the  host-pathogen dynamics that result from this model (dashed green lines in the top row plots).

The method introduced here requires data $\Xbf = (X_n,X_{n+1})_{n\in\CalI}$ on the discrete variables and data $\Ybf = (Y_n(\tbf_n))_{n\in \CalI}$ on the continuous variables, over a finite set of generations $\CalI$ and timegrids $\{\tbf_n\}_{n\in \CalI}$. To test our method we simulate the model \eqref{eq:ICres}-\eqref{eq:Xres} for $n_{\max}=80$ generations, with $\CalI$ either sampled uniformly at random from the first 40 generations or clustered near peaks of the host density $N_n$. We fix the timegrids $\tbf_n  = \tbf = (0,\Delta t,\dots, M\Delta t)$ across all generations and we use the realistic regime of one sample per day for 42 days (42 samples per generation), that is 
\begin{equation}\label{eq:sampling}
\Delta t=1, \quad  M\Delta t = (0.75) T = 42
\end{equation}
We study the performance of the method while varying  $\CalI$ and the noise level $\sigma_{NR}$ (defined below).

To emulate the effects of measurement noise, we apply log-normal noise to the continuous variables $(S,P)$, and leave the discrete variables $(N,Z)$ noise-free. This is reasonable because the continuous variables are typically population counts gathered in the field to represent the full population, while the discrete variables are aggregate quantities observed on a yearly basis. In addition, the presence of noise in both discrete and continuous variables would make the joint learning of states, parameters and the initial conditions difficult due to identifiability tradeoffs (see the Discussion Section for more information). We use log-normal noise to preserve non-negativity of the population densities $S$ and $P$. That is, for each $n\in \CalI$, $t\in \tbf_n$, and $Y\in \{S,P\}$, we set 
\[Y_n(t) = Y^\star_n(t)\exp(\mu(t)+\sigma(t)\ep_t)\] 
for $\ep_t \sim \CalN(0,1)$ i.i.d.\ and $\mu(t)$ and $\sigma(t)$ chosen such that $\Ebb[Y_n(t)] = Y_n^\star(t)$ and
$\Vbb[Y_n(t)] = \|Y^\star_n(\tbf_n)\|_\text{rms}^2\sigma_{NR}^2$. Here,  $Y_n^\star(t)$ are the simulated (true, up to the numerical solver accuracy) solution values. We examine performance of the method under different {\it noise ratios} $\sigma_{NR}$, which approximately measures the relative error between the clean and noisy data, since
\begin{equation}
\sigma_{NR} \approx \frac{\|Y_n^\star(\tbf_n)-Y_n(\tbf_n)\|_\text{rms}}{\|Y^\star_n(\tbf_n)\|_\text{rms}}
\end{equation} 

\subsection{Performance Measures}

We are primarily interested in the ability of the proposed method to identify the correct model and accurately predict the time series of population densities under different sampling and noise regimes. We focus on performance as a function of the number of  generations observed $|\CalI|$ and the noise level $\sigma_{NR}$. 

To measure the ability of the algorithm to correctly identify terms having nonzero coefficients in the true model \eqref{eq:ICres}-\eqref{eq:Xres}, we use  the \textit{true positive ratio} (or the Jaccard Index) as in \cite{LagergrenNardiniLavigneEtAl2020ProcRSocA} defined by 
\begin{equation}\label{eq:tpr}
\text{TPR}(\what) = \frac{\text{TP}}{\text{TP}+\text{FN}+\text{FP}}
\end{equation}
where TP is the number of correctly identified nonzero coefficients, FN is the number of coefficients falsely identified as zero, and FP is the number of coefficients falsely identified as nonzero. Identification of the correct model results in a TPR of 1, while identification of half of the correct terms and no incorrect additional terms results in TPR of 0.5. This is applied to each of the vectors $(\what^\text{IC},\what^Y,\what^X)$ separately.

To assess the accuracy of the recovered coefficients we measure the 
relative magnitude of the $\ell^2$ distance in parameter space using 
\begin{equation}\label{eq:E2}
E_2(\what) := \frac{\|\what-\wstar\|_{2}}{\|\wstar\|_{2}}
\end{equation}
where $\|\what-\wstar\|_{2}$ 
corresponds to the sum of square differences between the estimated parameters and the true parameter values used to simulate the data, $\sum_{i} (\widehat w_i - w_i^*)^2$. This sum is then normalized by the total norm of the true parameter vector, $\sum_{i} (w_i^*)^2$ to provide the metric of relative parameter accuracy. 
Results for $E_2$ applied to $\what^X$ are depicted in Figure \ref{fig:sampling_strategies}. 

To measure prediction accuracy, we record the number of accurately predicted generations of the discrete variables, since this requires that all three submodels (initial conditions map, continuous dynamics, and discrete dynamics) have been accurately captured. Let $\Delta_n(\what)$ denote the relative cumulative error in the discrete variables up to generation $n$, given by 
\begin{equation}\label{p12}
\Delta_n(\what) :=  \sqrt{\frac{\sum_{n'=0}^n\nrm{X^\star_{n'}-\hat{X}_{n'}}^2_2}{\sum_{n'=0}^n\nrm{X^\star_{n'}}^2_2}}
\end{equation} 
where $\hat{X}$ represents the learned model output with coefficients $\what$ and $X^\star$ the corresponding true model output, both initialized at the same discrete variables $X_0$. In words, $\Delta_n(\what)$ measures the goodness of fit between the learned trajectory and the noise-free true trajectory up to generation $n$, each simulated from the same initial conditions $X_0$, in the relative root-mean-squared sense. We report the number of generations that $\Delta_n$ remains below a certain tolerance, 
\begin{equation}\label{eq:Ttol}
n_{\text{tol}}(\what) := \max\{n\leq n_{\max}\ :\ \Delta_n(\what)\leq \text{tol}\}
\end{equation} 
We set $\text{tol} = 0.5$ throughout, which we empirically observe corresponds to the generation $n_{\text{tol}}$ at which the learned model visually departs from the ground truth. For example, Figure \ref{fig:example_snrY05} depicts an example dataset that achieved $n_{0.5}(\what) = 32$, indicating that the learned model is able to forecast accurately (with a relative cumulative error not exceeding $50\%$) for 32 generations, given only $(N_0,Z_0)$, the values of the discrete variables at generation $n=0$.

To generate statistics for the metrics \eqref{eq:tpr}, \eqref{eq:E2}, and \eqref{eq:Ttol}, for each noise level $\sigma_{NR}$ and number of observed generations $|\CalI|$ the algorithm was run on $R$ independently-generated datasets, each with randomly generated noise. For peak sampling, the generation set $\CalI$ is fixed for each $|\CalI|$ and $R=500$, while for random sampling we sample $\CalI$ uniformly at random from $\{1,\dots,40\}$, and we use $R=2000$. These values of $R$ were sufficient to yield stable statistics, with larger $R$ needed for random sampling due to the fact that both the noise and the generation set $\CalI$ are random. 

\subsection{Uncertainty Quantification (UQ)}\label{sec:UQ}

\begin{figure*}
\begin{center}
\begin{tabular}{@{}c@{}}
\includegraphics[trim={0 0 0 0},clip,width=0.8\textwidth]{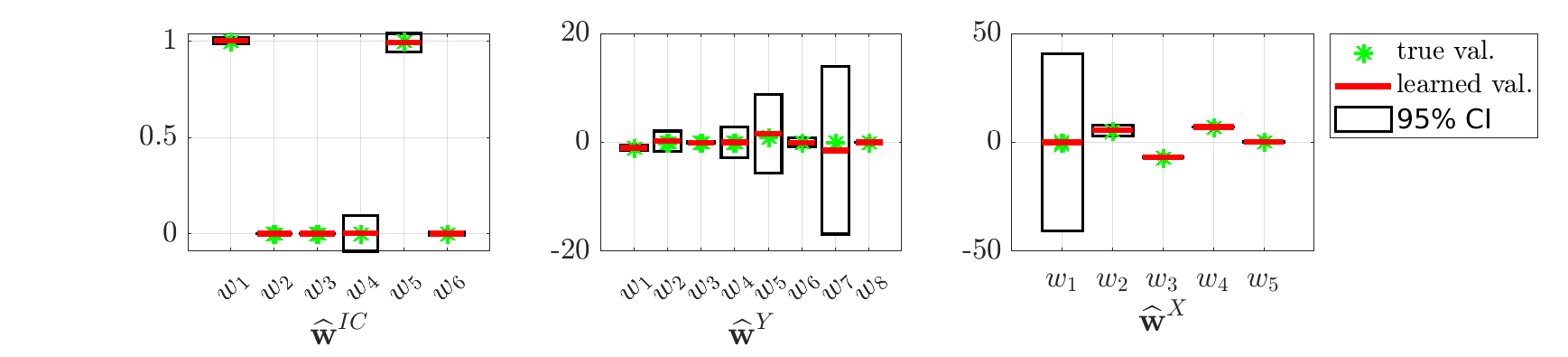}
\\
\includegraphics[trim={40 0 0 30},clip,width=0.65\textwidth]{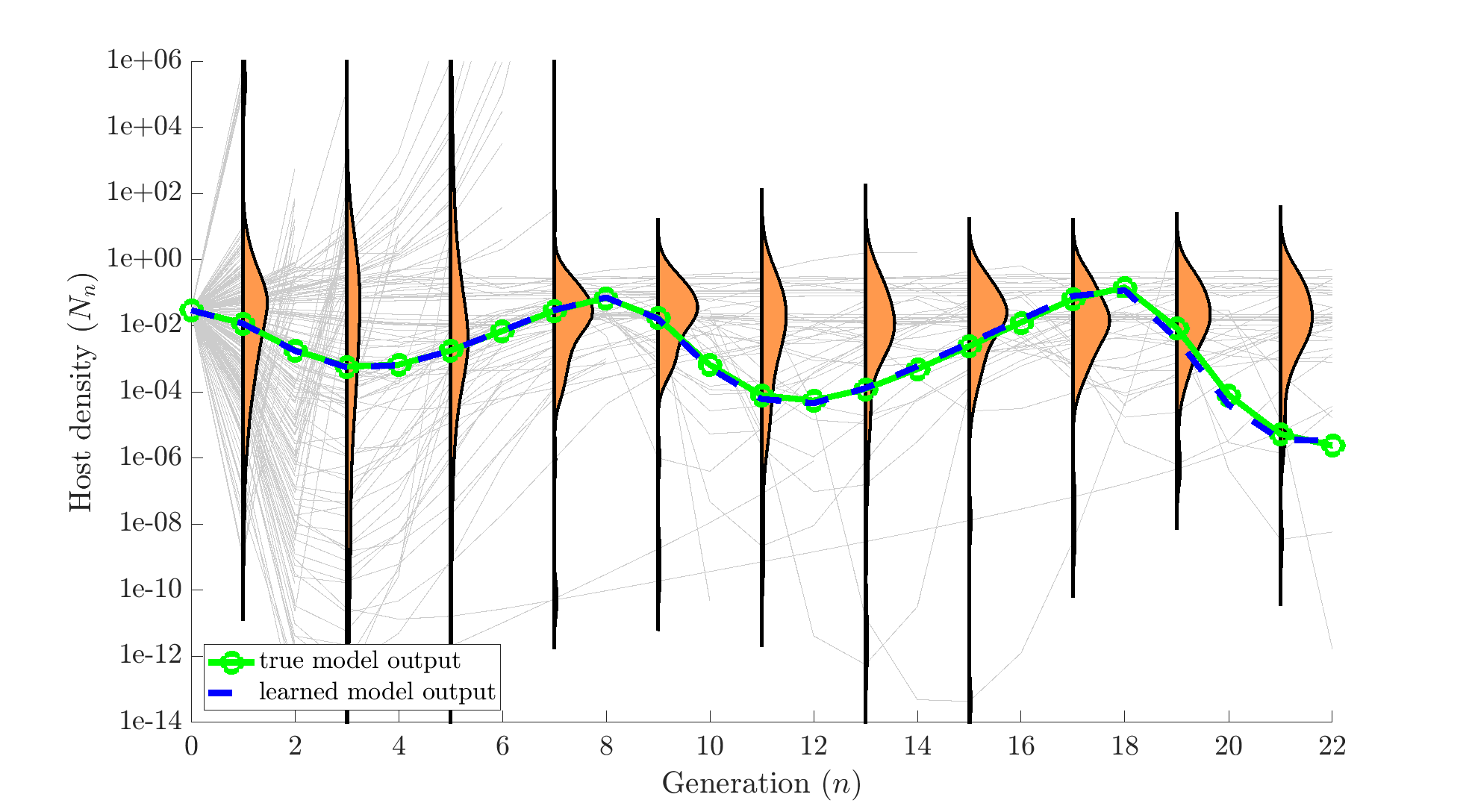}
\end{tabular}
\end{center}
\caption{{\bf UQ under misidentified model, pre-manual model pruning}. In the case of a misidentified model (also from data with $5\%$ noise), the learned model output still fits the data well (blue vs green curves). However, uncertainty in trajectories is large, with different trajectories showing wildly different behavior. Information provided by WENDy, combined with domain knowledge, can be used to manually rule out terms that do not seem biologically realistic; the next figure displays the results obtained from re-solving the system after eliminating $\{w_2,w_3,w_4,w_6\}$ from $\what^\text{IC}$, $\{w_4,w_6,w_7\}$ from $\what^Y$, and $w_1$ from $\what^X$, due to  disproportionately smaller magnitudes and/or  larger confidence intervals (top plots).}
\label{fig:UQ_uncorr_red}
\end{figure*}

\begin{figure*}
\begin{center}
\begin{tabular}{@{}c@{}}
\includegraphics[trim={0 0 0 0},clip,width=0.8\textwidth]{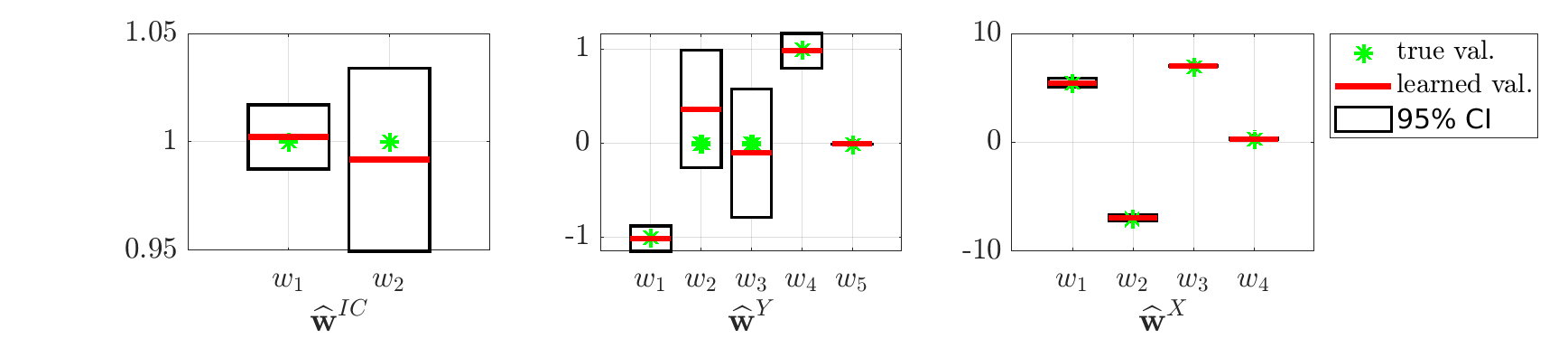}
\\
\includegraphics[trim={40 00 0 30},clip,width=0.65\textwidth]{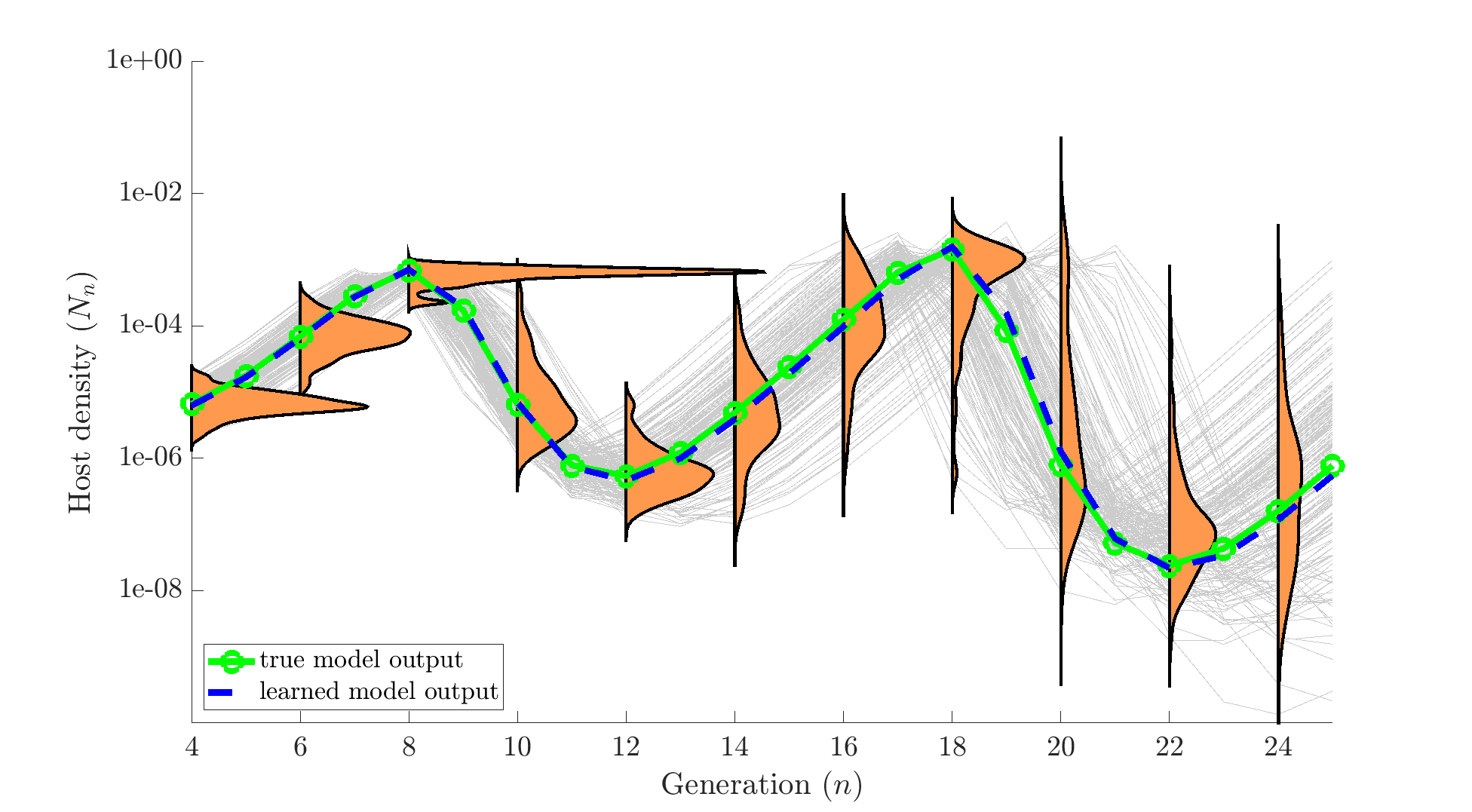}
\end{tabular}
\end{center}
\caption{{\bf UQ for the misidentified model described in Figure \ref{fig:UQ_uncorr_red}, post model pruning}. After having eliminated $\{w_2,w_3,w_4,w_6\}$ from $\what^\text{IC}$, $\{w_4,w_6,w_7\}$ from $\what^Y$, and $w_1$ from $\what^X$ (see Fig.\ \ref{fig:UQ_uncorr_red}) and resolving the system using WENDy, a clearer trajectory pattern is revealed with tighter spread of oscillations. In particular, the next host population peak is predicted to occur at $n=8$ with a high degree of certainty.
}
\label{fig:UQ_corr_red}
\end{figure*}

\texttt{WSINDy-Eco} outputs a distributional model with asymptotically multivariate normal parameters
\[ \wbf^\text{IC} \hspace{-0.1cm}\sim \CalN(\what^\text{IC},\widehat{\Sbf}^\text{IC}),\ \wbf^Y
\hspace{-0.1cm}\sim \CalN(\what^Y,\widehat{\Sbf}^Y),\ 
\wbf^X
 \hspace{-0.1cm}\sim \CalN(\what^X,\widehat{\Sbf}^X)\] 
Collectively, we will refer to the full model output as $\wbf
\sim \CalN(\what,\widehat{\Sbf})$. The simplest uncertainty metric we can readily quantify are the confidence intervals around each parameter, obtained from $\textsf{diag}(\widehat{\Sbf})$. We will, however, also estimate the uncertainty in predictions using parametric bootstrap sampling:  in Section \ref{sec:longtermforecasting}  we quantify uncertainty in population peak times and amplitudes, and  in Section \ref{sec:HITL} we present a human-in-the-loop modeling approach using various UQ metrics. In each case we make 200 independent parameter draws $\wbf^{(\ell)}\sim \CalN(\what,\widehat{\Sbf})$, $\ell=1,\dots,200$. Then, for each draw, the resulting model is constructed using the output libraries $\Lbb^\text{IC}, \Lbb^Y, \Lbb^X$ and simulated from the same observed initial conditions. At each generation $n=0,\dots,80$ we compute histograms for the discrete host densities $\{N_n^{(\ell)}\}_{\ell=1}^{200}$ to visualize the growth of uncertainty over time (Figures \ref{fig:UQ_corr}-\ref{fig:UQ_corr_red}). Similarly, peaks $P_1^{(\ell)}, P_2^{(\ell)},\dots$ 
are computed for each draw $\ell=1,\dots,200$, and  confidence regions for each peak location and  magnitude are found based on  marginal sample order statistics, from 2.5th quantile to 97.5th quantile.
(see Figures \ref{fig:example_snrY01}, \ref{fig:example_snrY05}). 

\subsection{Long-term Prediction with UQ}\label{sec:longtermforecasting}
\begin{figure*}
\begin{center}
\begin{tabular}{@{}c@{}|c@{}}
\includegraphics[trim={23 0 50 25},clip,width=0.49\textwidth]{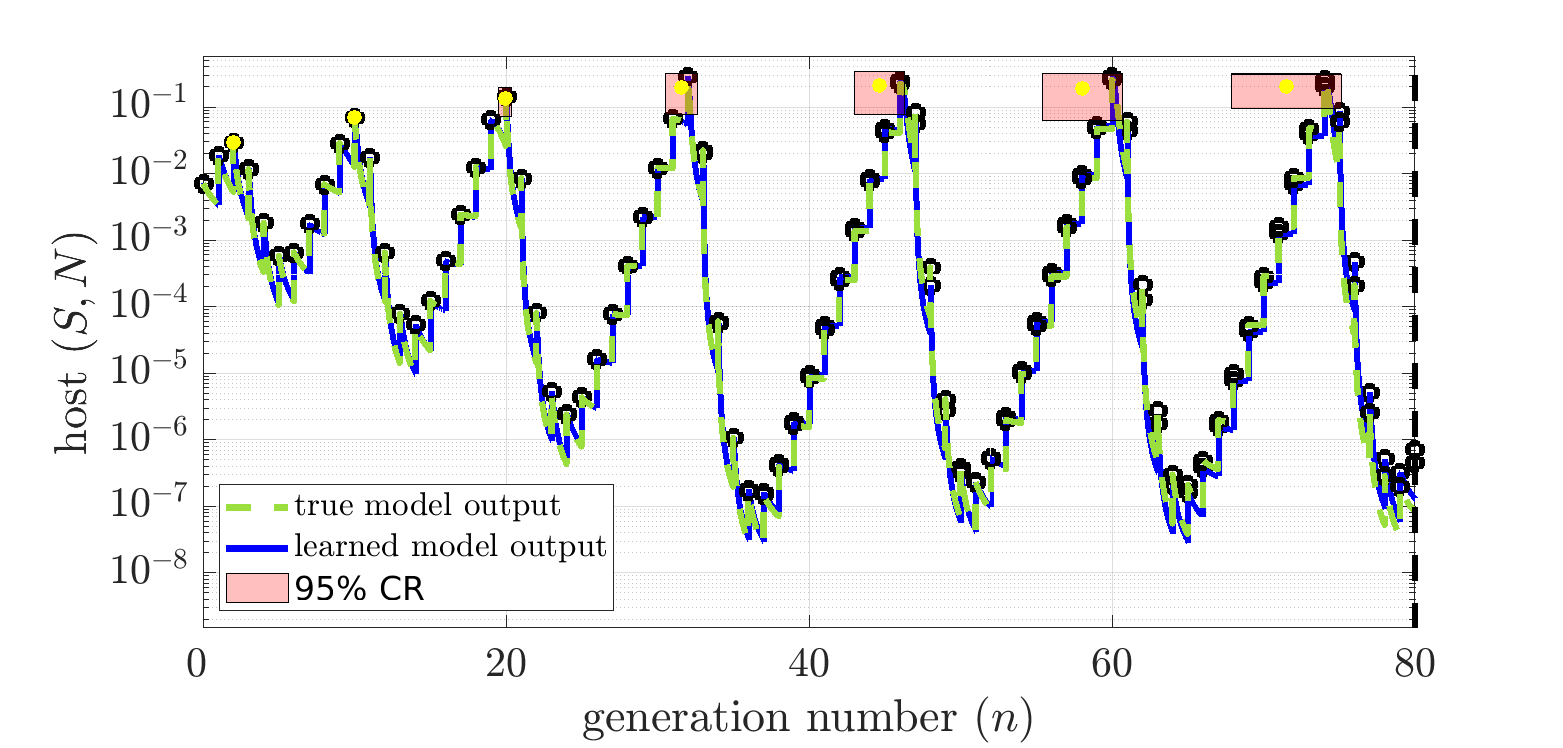} & 
\includegraphics[trim={23 0 50 25},clip,width=0.49\textwidth]{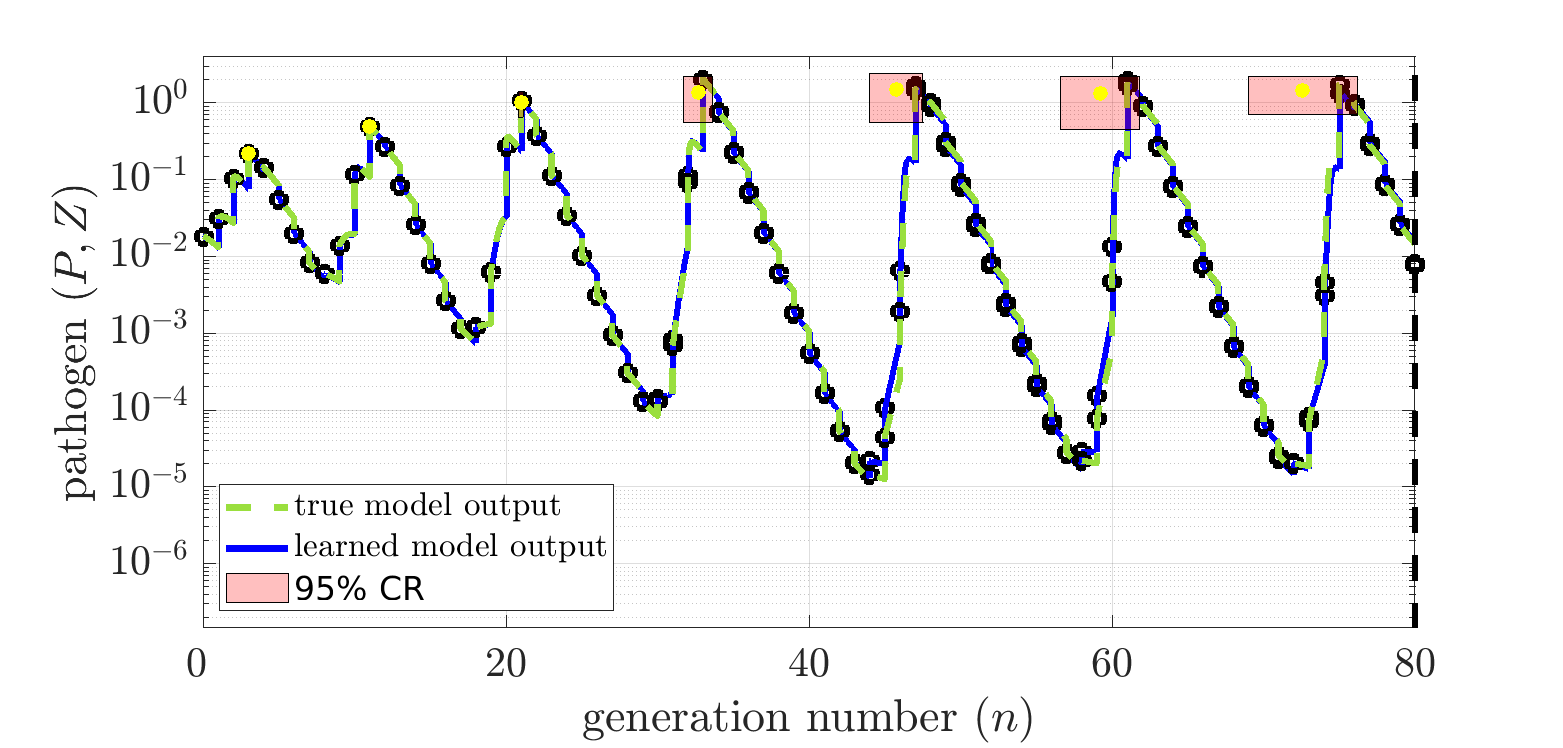} \\ 
\includegraphics[trim={0 0 0 0},clip,width=0.24\textwidth]{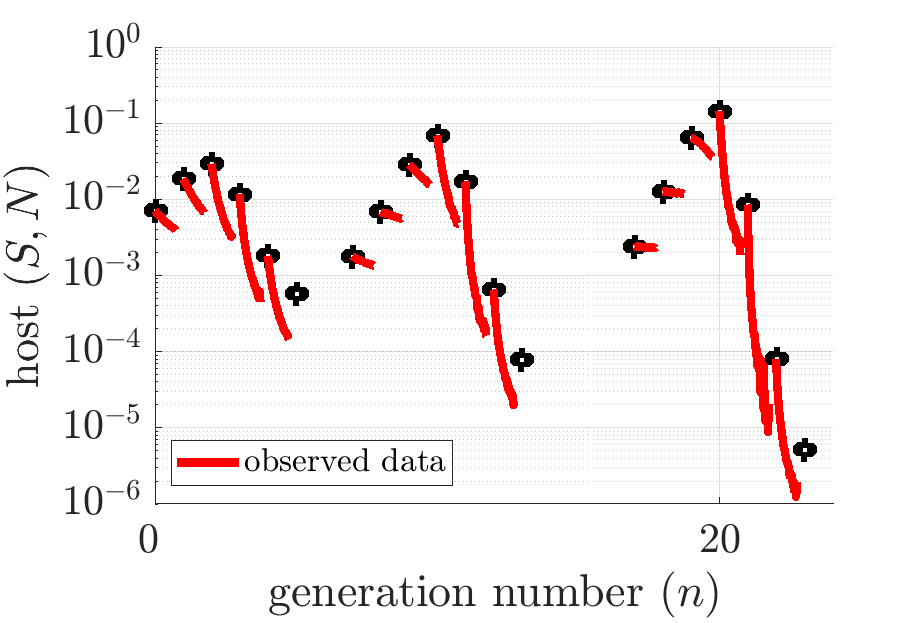}
\includegraphics[trim={0 0 0 0},clip,width=0.24\textwidth]{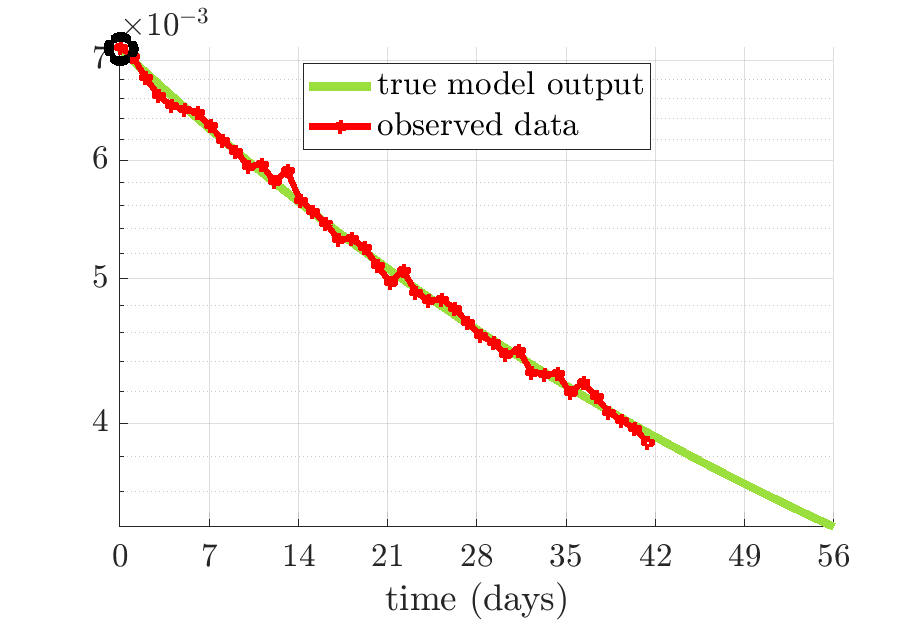} &
\includegraphics[trim={0 0 0 0},clip,width=0.24\textwidth]{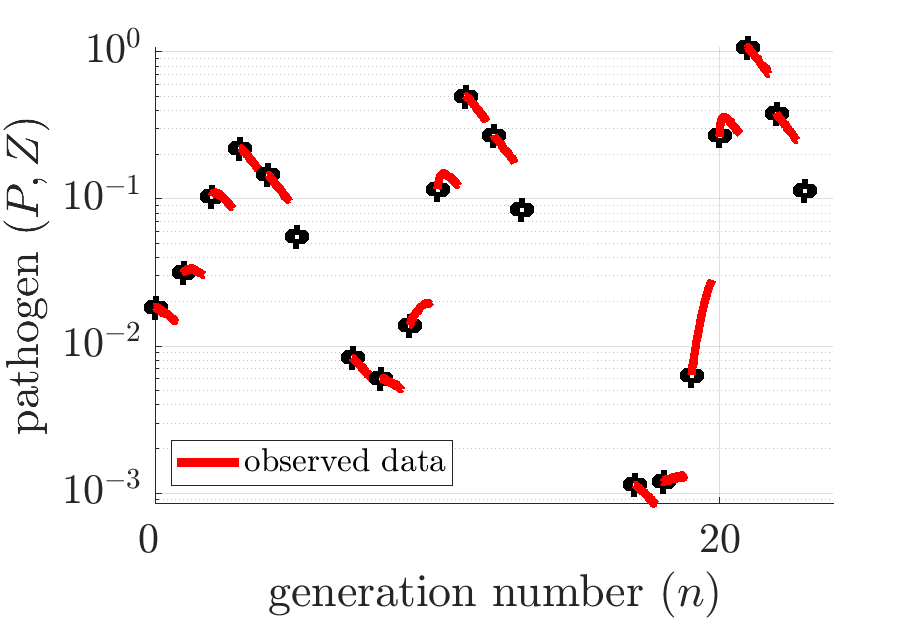} 
\includegraphics[trim={0 0 0 0},clip,width=0.24\textwidth]{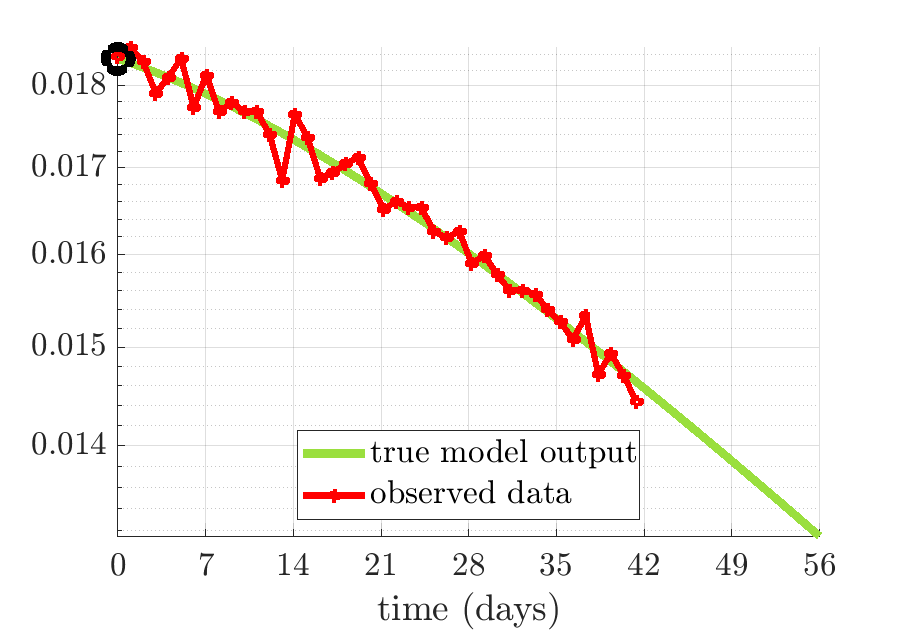}
\end{tabular}
\end{center}
\caption{{\bf Example: $1\%$ noise, accurate predictions up to 80 generations}. Left and right columns visualize performance of WSINDy-ECO in predicting the host and pathogen densities, respectively. The top plots display learned model outputs overlapping the true model output for all 80 generations ($n_\text{0.5}(\what)=80$, black dashed lines), given $|\CalI|=18$ samples from the first 3 host density peaks. The lower left plots in each column display the observed data, with black dots for discrete $(N_n,Z_n)$ variables and red lines for continuous $(S_n,P_n)$ variables over each generation. The modest 1\% noise level in $(S_n,P_n)$ is visualized for the $n=0$ generation in the bottom right plots of each column. By sampling the output parameters $\what$ using the learned covariance $\widehat{\Sbf}$ and simulating the resulting models, $95\%$ confidence regions can be ascribed to the peak locations and heights (red rectangles, with width and height given by $95\%$ confidence intervals around mean peak location and height, respectively, with yellow dots for means), which in each case contain the true peaks and amplitudes. Note that although the parameter distribution is Gaussian $(\wbf\sim \CalN(\what,\widehat{\Sbf}))$, the uncertainty does not propagate linearly through the nonlinear dynamical system, resulting in nonsymmetric confidence regions (hence yellow dots do not align with peaks).}
\label{fig:example_snrY01}
\end{figure*}

\begin{figure*}
\begin{center}
\begin{tabular}{@{}c@{}|c@{}}
\includegraphics[trim={23 0 50 25},clip,width=0.49\textwidth]{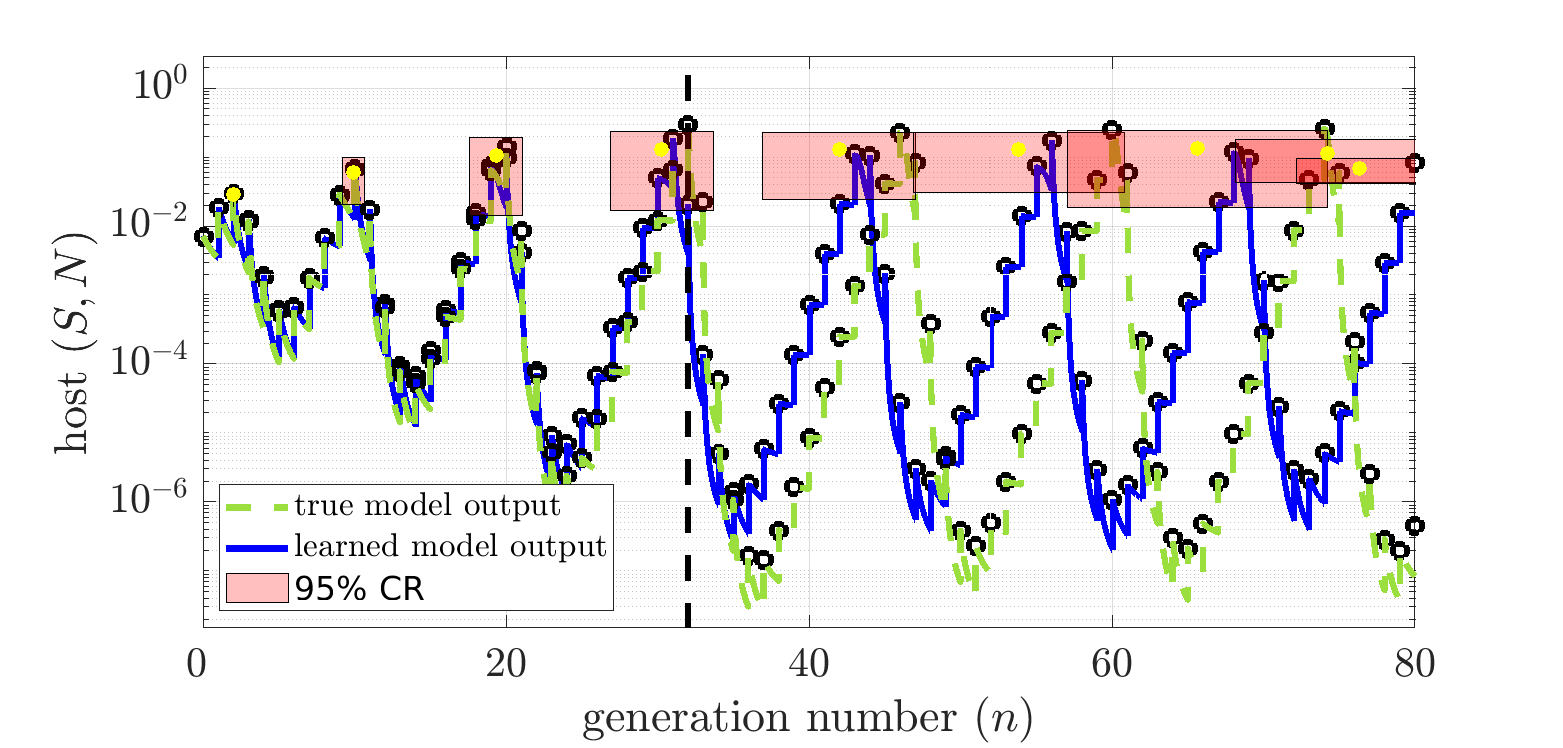} & 
\includegraphics[trim={23 0 50 25},clip,width=0.49\textwidth]{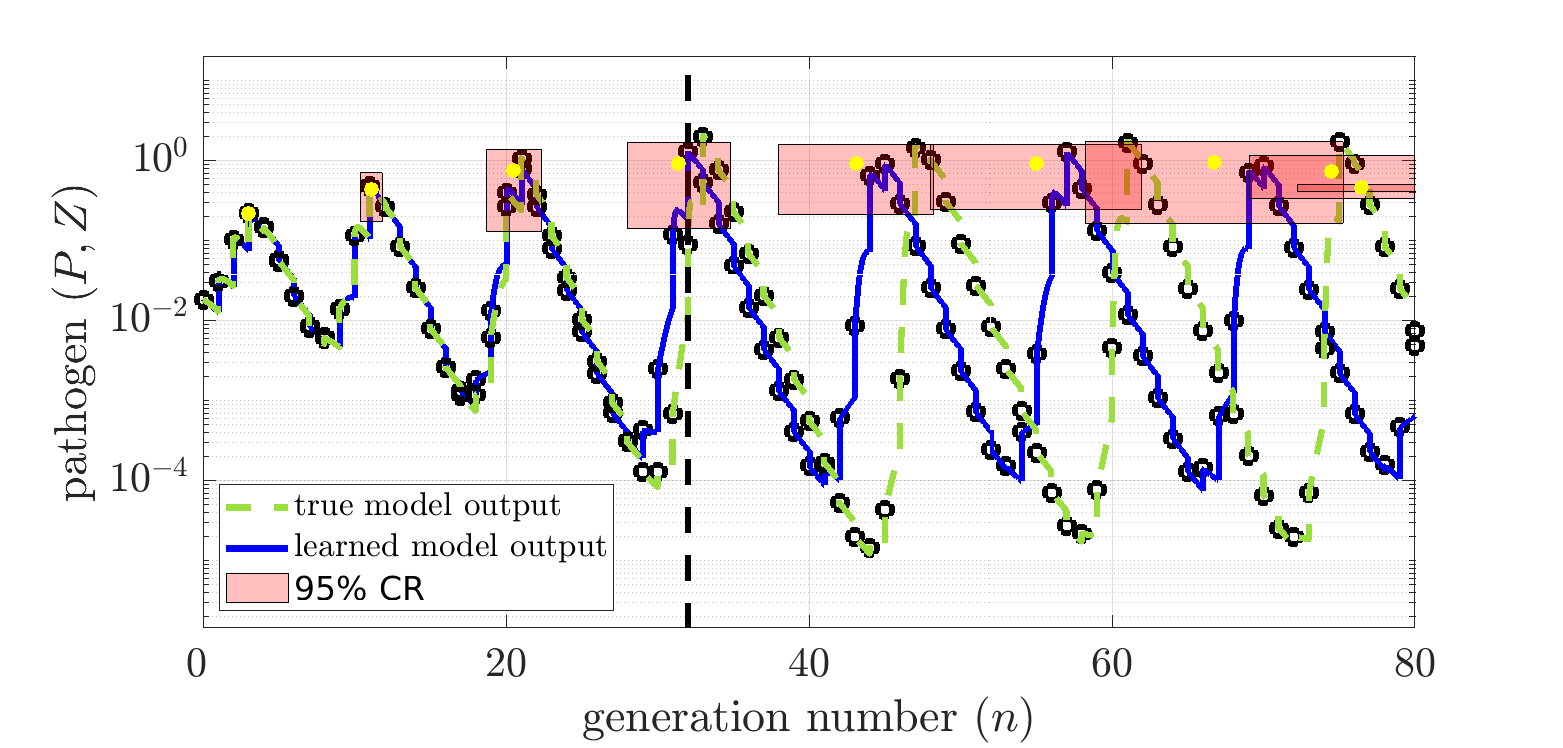} \\ 
\includegraphics[trim={0 0 0 0},clip,width=0.24\textwidth]{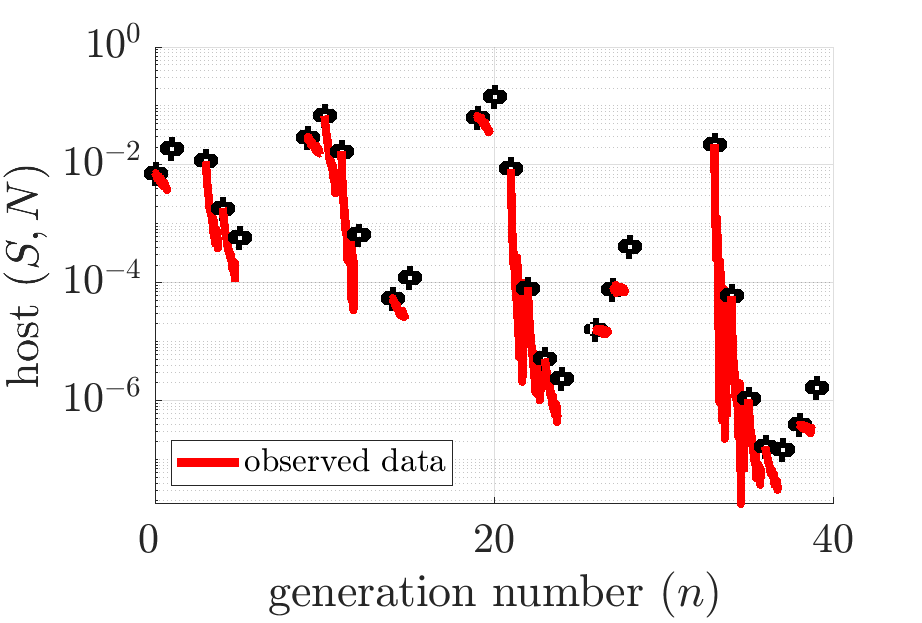}
\includegraphics[trim={0 0 0 0},clip,width=0.24\textwidth]{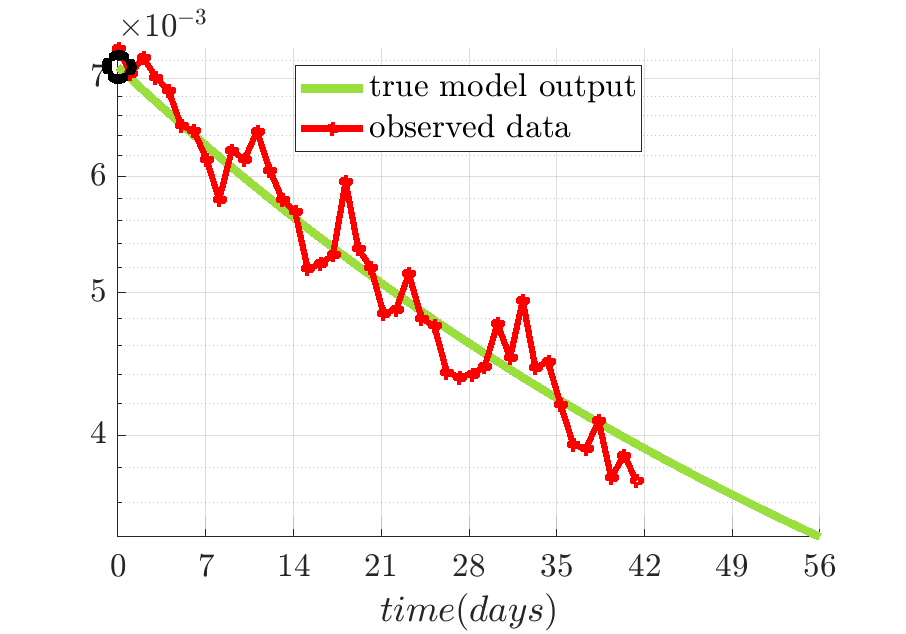} &
\includegraphics[trim={0 0 0 0},clip,width=0.24\textwidth]{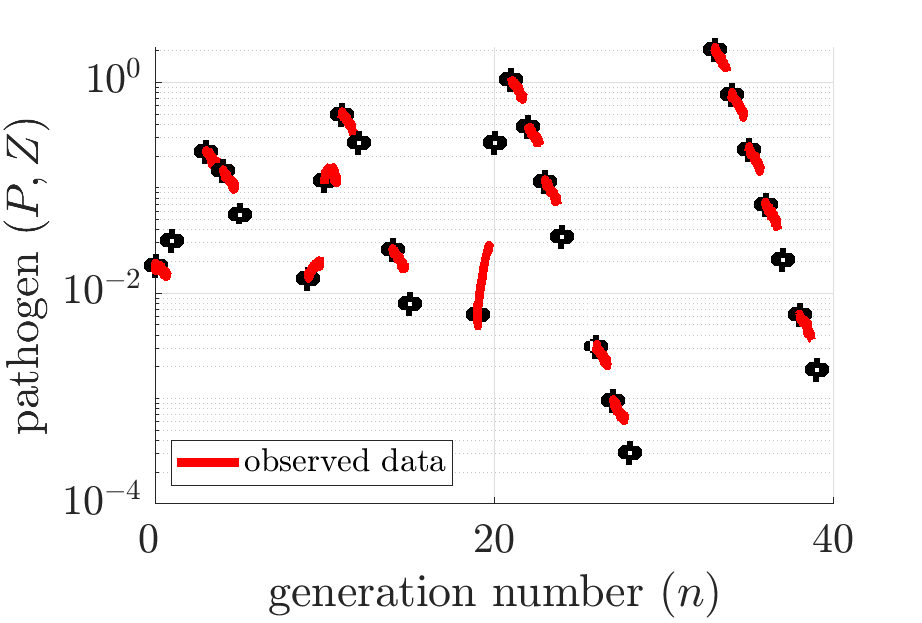} 
\includegraphics[trim={0 0 0 0},clip,width=0.24\textwidth]{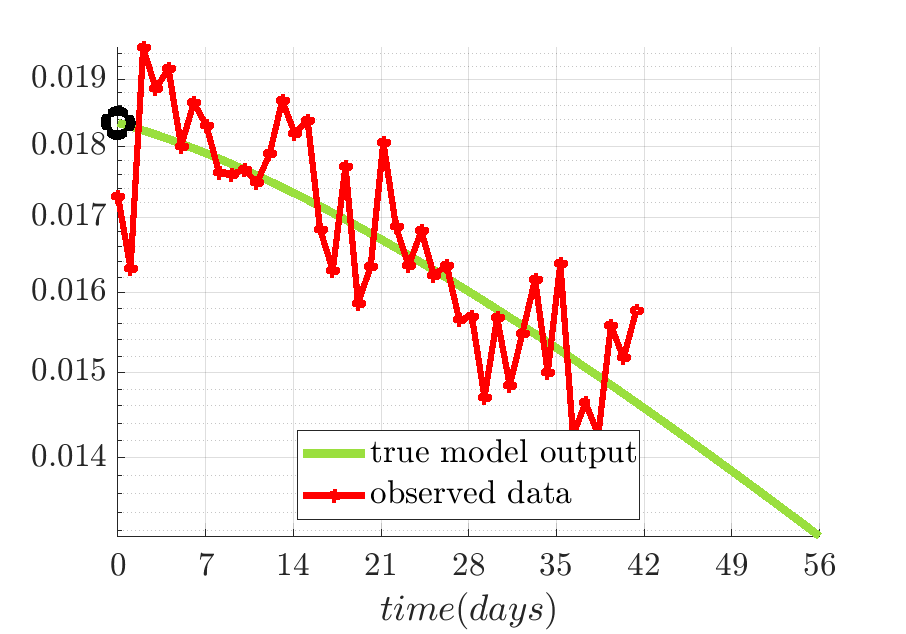}
\end{tabular}
\end{center}
\caption{{\bf Example: $5\%$ noise, accurate predictions up to 32 generations}. Similar to Figure \ref{fig:example_snrY05}, only now with $5\%$ noise in $(S,P)$ (see lower right plots in each column), and $|\CalI|=18$ randomly sampled generations. The larger noise level limits accurate predictions to $n_\text{0.5}(\what)=32$ generations. The 95\% confidence regions contain the true peaks up to $n_\text{0.5}(\what)$, after which the 95\% CI for peak location still contains the true peak, but the true heights fall narrowly outside of the 95\% CI.}
\label{fig:example_snrY05}
\end{figure*}

Figures \ref{fig:example_snrY01} and \ref{fig:example_snrY05} display typical results for $1\%$ and $5\%$ noise, respectively ($\sigma_{NR}=0.01,0.05$). In the $1\%$ noise case, each of the submodels \eqref{eq:ICres}-\eqref{eq:Xres} is identified correctly $(\text{TPR}(\what^\text{IC})= \text{TPR}(\what^Y)=\text{TPR}(\what^X)=1)$, with  coefficient relative errors $E_2(\what^\text{IC}) = 7.8\times 10^{-4}$, $E_2(\what^Y) = 1.9\times10^{-3}$, and $E_2(\what^X) = 1.9\times10^{-5}$. The algorithm runs in 12 seconds on a modern laptop, and allows for accurate prediction up to 80 generations (the maximum simulated). By sampling model parameters using the parameter covariance $\widehat{\Sbf}$, as described in Section \ref{sec:UQ}, we can compute confidence intervals for the host and pathogen population peak locations and amplitudes, which in Figure \ref{fig:example_snrY01} are seen to contain the true peaks for all predicted cycles. For example, the fifth peak is estimated to occur approximately 44 years in the future with a 1 year margin.  

The effects of larger measurement noise are depicted in Figure \ref{fig:example_snrY05}, showing  exemplary results for $5\%$ noise in the $(S_n,P_n)$ variables (see the bottom right subplots of each column for noise visualizations). The algorithm in this case is still able to correctly identify all submodels \eqref{eq:ICres}-\eqref{eq:Xres}, yet the coefficient errors $E_2(\what^\text{IC}) = 1.9\times 10^{-2}$, $E_2(\what^Y) = 9.4\times10^{-3}$, and $E_2(\what^X) = 1.1\times10^{-2}$ are now large enough to restrict the long-term predictive abilities of the learned model. Still, the learned model is able to capture the underlying true model output for 32 generations, after which the behavior is qualitatively correct, namely a stable cycle appears to have been reached. The peak location and amplitude confidence intervals in this case communicate that later population booms (peaks 5 and later) cannot be accurately predicted; however, we anticipate that in many cases only information on the next 1-3 peaks in is expected. 

\begin{figure*}
\begin{center}
\begin{tabular}{@{}c@{}c@{}}
Ordinary Least Squares & WENDy \\ 
\includegraphics[trim={0 0 0 0},clip,width=0.4\textwidth]{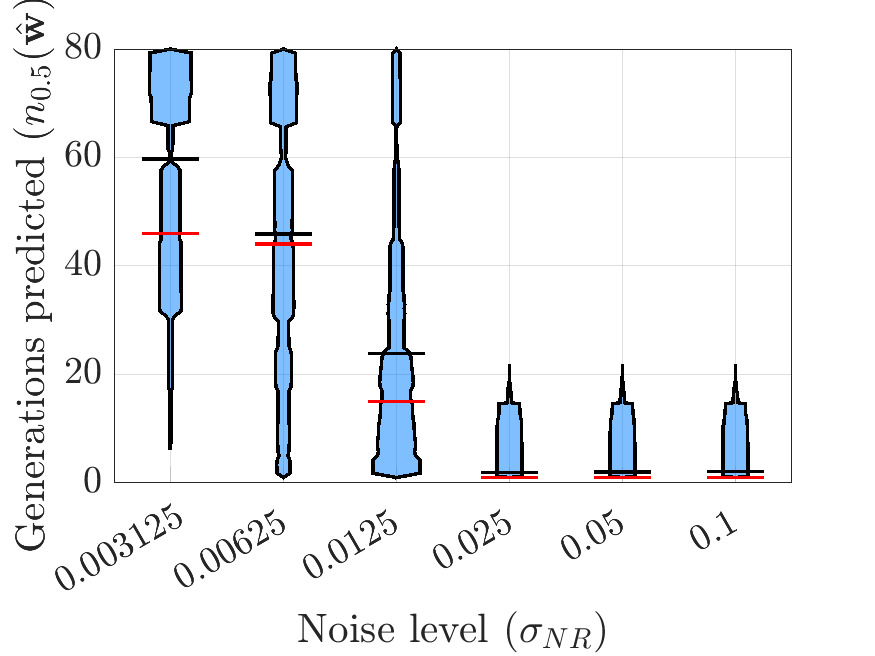}
& 
\includegraphics[trim={0 0 0 0},clip,width=0.4\textwidth]{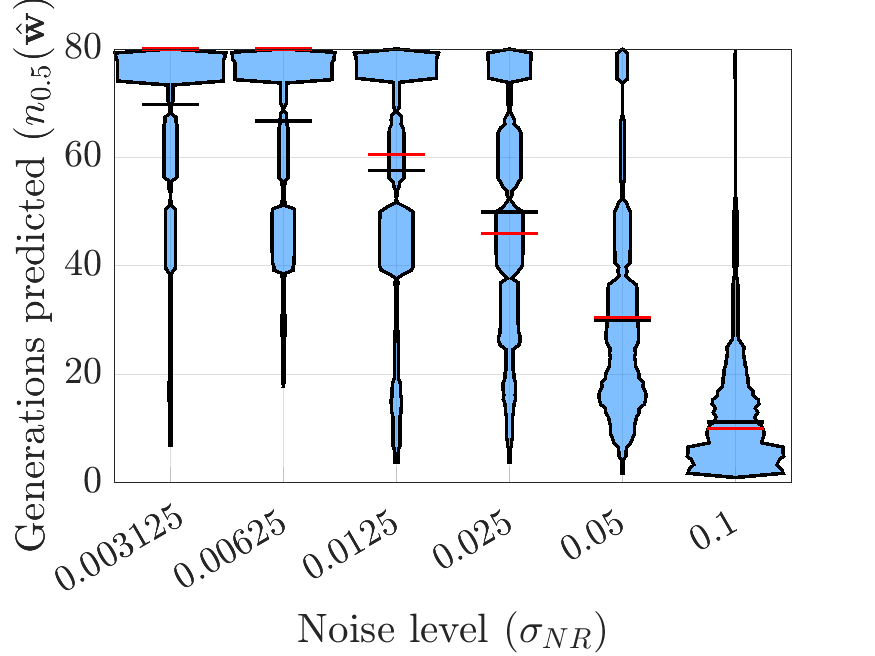}
\end{tabular}
\end{center}
\caption{{\bf Advantages of WENDy regression in long-term prediction}. Both plots depict the performance of \texttt{WSINDy-Eco} with data sampled from the first five host peaks (see Fig. \ref{fig:sampling_strategies} for sampling related performance), with ordinary least squares (OLS) regression on the left (\texttt{MaxIts}=0) and WENDy regression on the right (\texttt{MaxIts}=5). Employing WENDy increases predictive capabilities from 20 to 60 generations on average for data with $1.25\%$ noise in the continuous variables. OLS does not allow for effective model learning at higher noise levels, leading to models that cannot predict well forward in time, while WENDy is still able to predict 45, 32, and 10 generations forward at $2.5\%$, $5\%$, and $10\%$ noise, respectively.}
\label{fig:wendy_advantage}
\end{figure*}

Finally, we report the overall advantages of the WENDy regression approach in affording long-term prediction in Figure \ref{fig:wendy_advantage}. Results are shown for sampling data from the first five host population peaks (see Section \ref{sec:sampling} for an in-depth treatment of sampling strategies). Simply using ordinary least squares (OLS, left plot), obtained by setting \texttt{MaxIts}=0 in \texttt{WSINDy-Eco}, results in a lack of robust predictions when noise in the $(S_n,P_n)$ variables exceeds $1.25\%$. On the other hand, accurately accounting for the errors-in-variables nature of the problem using WENDy with just 5 iterations (\texttt{MaxIts}=5) enables accurate prediction for at least 10 generations on average for noise levels up to $10\%$, with 5\% noise resulting in accurate predictions up to 32 generations, and 2.5\% noise yielding accurate prediction up to 45 generations.  

\begin{figure*}
\begin{center}
\begin{tabular}{@{}c@{}}
\includegraphics[trim={0 0 0 0},clip,width=0.8\textwidth]{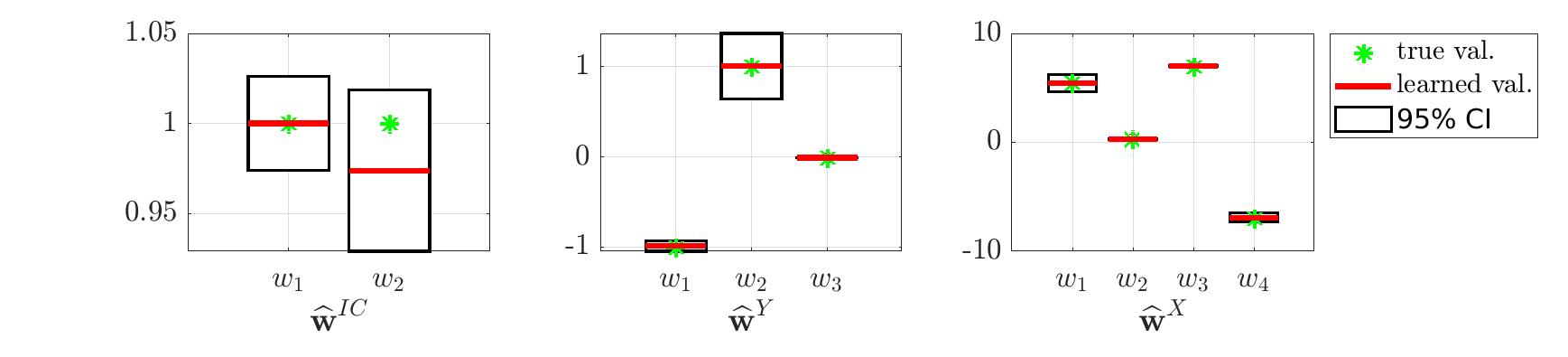}
\\
\includegraphics[trim={40 0 0 30},clip,width=0.7\textwidth]{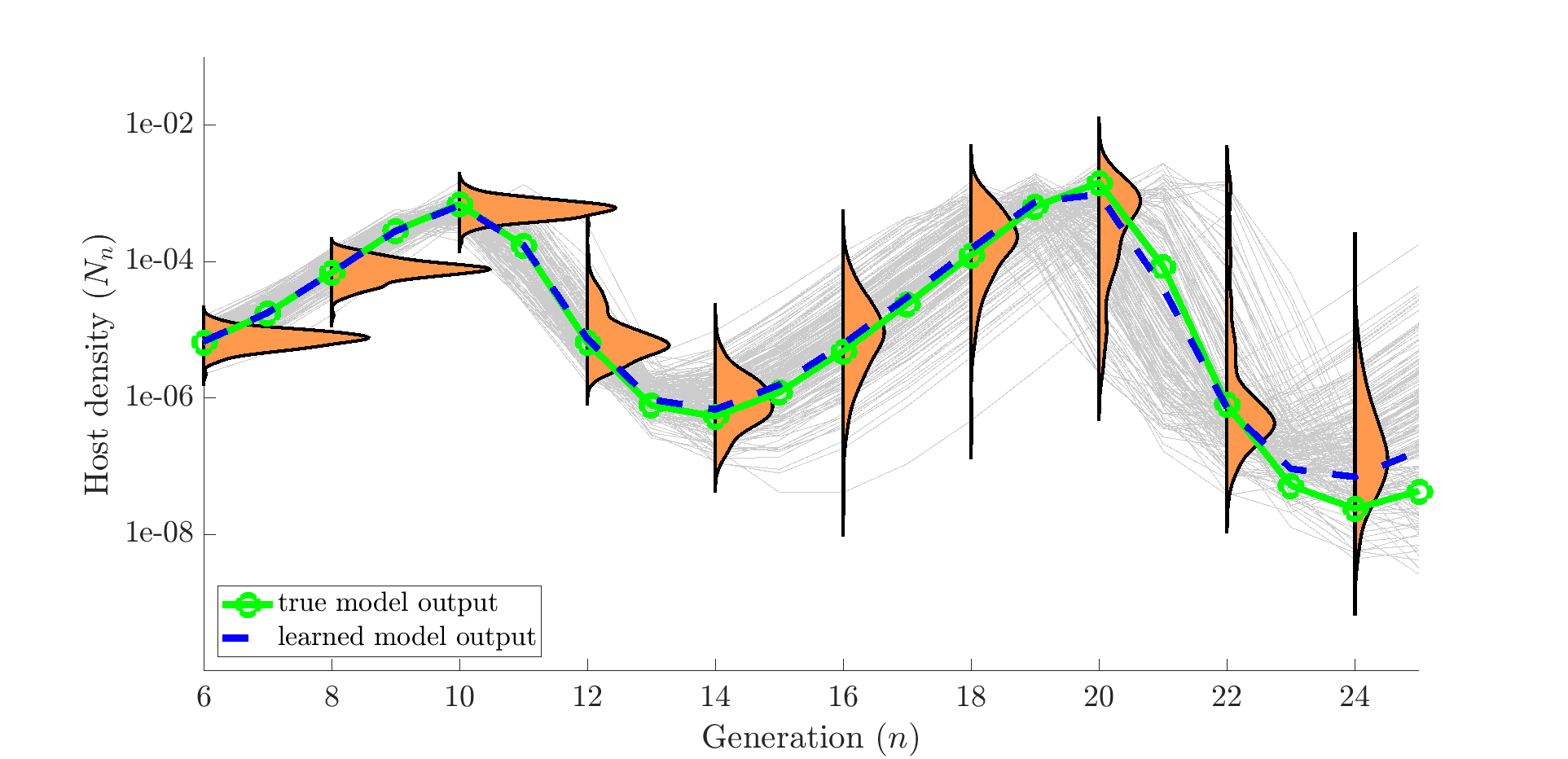}
\end{tabular}
\end{center}
\caption{{\bf UQ under correct model identification}. In this example with 5\% noise (same data as in Figure \ref{fig:example_snrY05}), the learned model is able to capture the data well (blue vs.\ green curve). The added uncertainty quantification allows the \texttt{WSINDy-Eco} approach to provide a view into the validity and usefullness of the model. Confidence intervals around learned parameter values (top plots) each contain the true value and show model parameter uncertainty, indicating low uncertainty (narrow intervals) and good identification. Similarly,  model trajectories (gray curves) obtained via parametric bootstrap samples from the WENDy parameter distribution, provide visualization of the uncertainty in state trajectories themselves. They also allow quantification of the spread of predictive values  and  propagation of uncertainty into forward predictions (orange kernel density estimates) at any point in time. The next figure displays corresponding results for the case of a misidentified model.}
\label{fig:UQ_corr}
\end{figure*}

An additional way to quantify uncertainty is to understand how parameter uncertainty propagates into dynamical system trajectories. To see this, we sample model parameter sets (using parametric bootstrap as explained above), and plot the solution trajectories for each parameter set in Figure \ref{fig:UQ_corr}. This  allows us to assess prediction uncertainty in the state trajectories themselves, which are non-linear transformations of the parameters (and parameter uncertainty). Histograms of solution samples at each generation $n$ (orange kernel density estimates superimposed on the cloud of trajectories) reveal the spread of state values in the host population at each particular time slice. In this case, with the true model having been identified by the algorithm, the sampled state trajectories remain relatively tight and close to the true model trajectory. It is worth noting however that the state trajectories start to show reduced cohesiveness and heavier tails as the uncertainty increases the further into the future we are trying to predict.

\subsection{Human-in-the-loop Modeling}\label{sec:HITL}

A desirable quality of equation learning combined with covariance information (as afforded by the current method) is that models can be vetted by experts, through direct examination of the learned model terms, their respective coefficients, and their corresponding covariance structure and confidence intervals. For example, Figure \ref{fig:UQ_corr} displays  confidence intervals for the model described in Figure \ref{fig:example_snrY05}, and shows that in this case each confidence interval contains the true value of the parameter (green star). Note that normally we would expect 95\% of our confidence intervals to contain the true values over repeated sampling. When confidence intervals are large however, and include 0, this allows for a discussion about whether those terms should be excluded from the final model, as long as multiple testing strategies are employed.

In practice, however, there is no ground truth model, and it is quite possible that the learned model could be improved upon by further inspection from a researcher with expertise in the ecological systems being studied. It is also possible that low data quality and high noise can lead to suboptimal sparse recovery (see Section \ref{sec:sampling}, Sampling Strategies). In particular, if the model goodness of fit  is not diminished in any meaningful way with the {\it removal} of terms, then by Occam's razor  the simpler model should be used. In the current framework, this can be tested immediately. For example, Figure \ref{fig:UQ_uncorr_red} displays the  parameter confidence intervals (top) and forward simulation uncertainty (bottom) corresponding to the misidentified model in equation \eqref{eq:misspec}, with spurious terms noted in red (see \eqref{eq:ICres}-\eqref{eq:Xres}). The coefficient values and their confidence intervals are listed in Table \ref{tab:coeff_table}. At first glance, the terms look reasonable: the initial conditions map is linear, the continuous dynamics show competitive interactions (all terms with $S_nP_n$) with decay in the pathogen density $P_n$, while the discrete dynamics replace the host update $N_{n+1}$ with a linear combination of discrete and continuous host variables.  
\begin{table*}
\begin{equation}
\label{eq:misspec} 
\hspace{-0.3cm}\begin{dcases}
S_n(0)= w_1^{IC}N_n \\
\hspace{0.7cm}\textcolor{red}{+w_2^{IC}Z_n+w_3^{IC}}\\
P_n(0) = w_4^{IC}Z_n \\
\hspace{0.7cm}\textcolor{red}{+w_5^{IC}N_n+w_6^{IC}}
\end{dcases} \hspace{0.5cm}
\begin{dcases}
\dot{S}_n(t) = S_n(t)P_n(t)\bigg[w_1^{Y}\left(\frac{S_n(t)}{N_n}\right)^V\hspace{-1cm}\\  \qquad \textcolor{red}{+w_2^{Y}(S_n(t))^V +w_3^{Y}(1)+w_4^{Y}S_n(t)(N_n)^{-V}}\bigg]\\
\dot{P}_n(t) = S_n(t)P_n(t)\bigg[w_5^{Y}\left(\frac{S_n(t)}{N_n}\right)^V\textcolor{red}{+w_6^{Y}(N_n)^{-V}}\hspace{-1cm}\\
\qquad \textcolor{red}{+w_7^{Y}S_n(t)(N_n)^{-V}}\bigg]+w_8^{Y}P_n(t)\\
\end{dcases}\hspace{0.4cm}
\begin{dcases}
N_{n+1}= \textcolor{red}{w_1^{X}N_n}+w_2^{X}S_n(T) \\
Z_{n+1} = w_3^{X}S_n(T) + w_4^{X}N_n \\
\hspace{1.3cm}+ w_5^{X}Z_n.
\end{dcases}
\end{equation}
\end{table*}

\begin{table*}
\begin{center}
\begin{tabular}{@{}l@{}}
\begin{tabular}{p{4cm}|p{1.2cm}p{1.2cm}p{1.2cm}p{1.2cm}p{1.2cm}p{1.2cm}}

\hline\hline
&&&&&& \\ Coefficient &$w^{IC}_1$&$w^{IC}_2$&$w^{IC}_3$&$w^{IC}_4$&$w^{IC}_5$&$w^{IC}_6$\\
Term & $N_n$ & $Z_n$ & 1 & $Z_n$ & $N_n$ & 1 \\  
&&&&&& \\ \hline
{\bf True model} (eq. \eqref{eq:ICres}-\eqref{eq:Xres})  & 1 & 0 & 0 & 1 & 0 & 0\\
{\bf Misidentified model} (Fig.\ \ref{fig:UQ_uncorr_red})  & $1.00$ & $-\num{4.41e-5}$ & $-\num{2.10e-7}$ & 0.99 & $\num{1.17e-3}$ & $-\num{6.77e-5}$\\
\hspace{0.75cm} (95\% CI) & $\pm\num{0.017}$ & $\pm\num{5.3e-4}$ & $\pm\num{6.2e-4}$ & $\pm\num{0.092}$& $\pm\num{0.049}$& $\pm\num{0.01}$ \\
{\bf Corrected model} (Fig.\ \ref{fig:UQ_corr_red}) & 1.00 & 0 & 0 & 0 & 0.99 & 0 \\
\hspace{0.75cm} (95\% CI)& $\pm\num{0.015}$ & 0 & 0 & 0 & $\pm\num{0.042}$& 0 \\
\end{tabular}
\\ \hline\hline
\begin{tabular}{p{4cm}|p{1.2cm}p{1.2cm}p{1.2cm}p{1.2cm}p{1.2cm}p{1.2cm}p{1cm}p{1cm}}
&&&&&&&& \\
Coefficient&$w^Y_1$&$w^Y_2$&$w^Y_3$&$w^Y_4$&$w^Y_5$&$w^Y_6$&$w^Y_7$&$w^Y_8$ \\ 
Term & $\frac{S^{1+V}_n}{N_n^V}P_n$ & $S_n^{1+V}P_n$ & $S_nP_n$ & $\frac{S^2_n}{N_n^V}P_n$ & $\frac{S^{1+V}_n}{N_n^V}P_n$ & $\frac{S_n}{N_n^V}P_n$ & $\frac{S^2_n}{N_n^V}P_n$ & $P_n$\\ &&&&&&&& \\ \hline
{\bf True model} (eq. \eqref{eq:ICres}-\eqref{eq:Xres}) &-1 & 0  & 0  & 0 &1 & 0 & 0 & -0.01 \\
{\bf Misidentified model} (Fig.\ \ref{fig:UQ_uncorr_red}) & 0.99 & 0.22 & $-\num{0.041}$ & $-0.01$ & $1.53$ & $-0.047$ & $-1.49$ &$-\num{9.7e-3}$ \\
\hspace{0.75cm} (95\% CI)& $\pm\num{0.48}$& $\pm\num{1.8}$& $\pm\num{0.18}$& $\pm\num{2.9}$& $\pm\num{7.2}$& $\pm\num{0.8}$& $\pm\num{15}$& $\pm\num{1.8e-3}$\\
{\bf Corrected model} (Fig.\ \ref{fig:UQ_corr_red}) & -1.02 & 0.36 & -0.12 & 0 & 0.98 & 0 & 0 & -0.01\\
\hspace{0.75cm} (95\% CI)& $\pm\num{0.13}$& $\pm\num{0.62}$& $\pm\num{0.68}$& 0 & $\pm\num{0.19}$ & 0 & 0 & $\pm\num{1.1e-3}$\\
\end{tabular}
\\ \hline\hline
\begin{tabular}{p{4cm}|p{1.2cm}p{1.2cm}p{1.2cm}p{1.2cm}p{1.2cm}}
&&&&& \\
Coefficient&$w^{X}_1$ &$w^{X}_2$ &$w^{X}_3$ &$w^{X}_4$ &$w^{X}_5$ \\
Term & $N_n$ & $S_n(T)$ & $S_n(T)$ & $N_n$ & $Z_n$ \\ 
&&&&& \\\hline
{\bf True model} (eq. \eqref{eq:ICres}-\eqref{eq:Xres}) & 0 & 5.5 & -7 & 7 & 0.3 \\
{\bf Misidentified model} (Fig.\ \ref{fig:UQ_uncorr_red}) & -\num{1.1e-3} & 5.42 & -6.98 & 7.00 & 0.3 \\
\hspace{0.75cm} (95\% CI)& $\pm\num{41}$& $\pm\num{2.4}$& $\pm\num{0.3}$& $\pm\num{0.037}$& $\pm\num{0.11}$ \\
{\bf Corrected model} (Fig.\ \ref{fig:UQ_corr_red}) & 0 & 5.45 & -6.93 & 7.00 & 0.3 \\
\hspace{0.75cm} (95\% CI)& 0 & $\pm\num{0.42}$& $\pm\num{0.23}$& $\pm\num{0.024}$& $\pm\num{0.088}$ \\
\end{tabular}
\end{tabular}
\end{center}
\caption{Comparison of true and learned model coefficients, together with 95\% confidence intervals, for the misidentified model in equations \eqref{eq:misspec} (visualized in Figure \ref{fig:UQ_uncorr_red}) and the corrected model obtained from eliminating $\{w_2,w_3,w_4,w_6\}$ from $\what^\text{IC}$, $\{w_4,w_6,w_7\}$ from $\what^Y$, and $w_1$ from $\what^X$ (visualized in Figure \ref{fig:UQ_corr_red}).}
\label{tab:coeff_table}
\end{table*}

To investigate the validity of this model, in the light of uncertainty in the dynamics the model describes, we simulate an ensemble of solution trajectories corresponding to a set of sampled parameters (sampled using parametric bootstrap  as described earlier), as shown in Figure \ref{fig:UQ_uncorr_red} (bottom).  Qualitatively,  we see that a significant portion of trajectories diverge or stabilize early on. While such behavior may be biologically realistic and valid (for instance  host population  blow-up could correspond to pathogen extinction),  the observed oscillations provide some scrutiny of those mechanisms. Investigation of the coefficients $\what$ and their respective confidence intervals (Table~\ref{tab:coeff_table})   provides support for model modification by removal of terms. Confidence intervals around $w_2,w_3,w_4,w_6$ of $\what^{IC}$, $w_4,w_6,w_7$ of $\what^{Y}$, and $w_1$ of $\what^{X}$ indicate that the sign of the respective coefficients cannot be reasonably inferred, and thus a value of zero cannot be ruled out. In addition,
the magnitudes of  $w_2,w_3,w_6$ in $\what^{IC}$ are  of several orders smaller than the remaining coefficients in $\what^{IC}$, which further advocates that their mechanisms might be insignificant in the seasonal reinitialization of host and pathogen densities. 

We can therefore ``manually'' set each of these coefficients to zero, and re-solve the differential equation system with the remaining terms using just the WENDy algorithm. The result is depicted in Figure \ref{fig:UQ_corr_red}. Although this system still contains two spurious model terms, the ensemble of trajectories (bottom row) now better communicates the spread of oscillations in trajectories. For example, one can infer from the sharply peaked density that the next population boom will likely occur at generation $n=8$ with high confidence.  Finally, to decide if we can formally adopt the pruned (reduced) model as the replacement for the original (full) model in equation \eqref{eq:misspec}, we would need to formally compare the full and reduced models using a model selection criterion such as for example AIC \cite{Akaike1974IEEETransAutomControl,HootenHobbs2015EcolMonogr}. In our case, the reduced model does indeed achieve a lower AIC score  of $771.14$ (using Eqs. \eqref{eq:AIC},\eqref{eq:loglike}), compared to the full model's AIC score of $782.59$, and thus the reduced model should be selected.

 Overall, this process reveals a procedure by which expert knowledge can be used to modify equations and quickly assess hypotheses using the modified equations, as well as assess the uncertainty in the learned model and its predictions. We reiterate that the entire learning algorithm, including assessment and modification of the learned model as just described, occurs with no more than several minutes of computation time on a modern laptop. 

\subsection{Sampling Strategies}\label{sec:sampling}

Data collection (sampling) strategies are an important consideration in ecological applications,  due in part to limited resources and challenges associated with field work. 
Specific questions may include when to sample, how many samples to collect, and during which years to collect samples. This Section   discusses quantitative performance results from two different collection strategies, random sampling and peak sampling.  The random sampling strategy involves sampling the generation set $\CalI$ uniformly at random from the first 40 generations, with $|\CalI| \in \{12,16,20\}$. Peak sampling, which is the strategy typically employed in studies of outbreaking organisms, means sampling generations around host population peaks. Letting $n_i$ denote the location of the $i$th host population peak, we include $\{n_i-2,n_i-1,n_i,n_i+1\}\subset\in \CalI$ for each of the sampled peaks. We consider  peak sampling from the first 3, 4, or 5 peaks, and perform a direct comparison with the same corresponding randomly sampled number of generations $|\CalI|$. 

\begin{figure*}
\begin{center}
\begin{tabular}{@{}c@{}c@{}c@{}}
\includegraphics[trim={0 0 0 0},clip,width=0.33\textwidth]{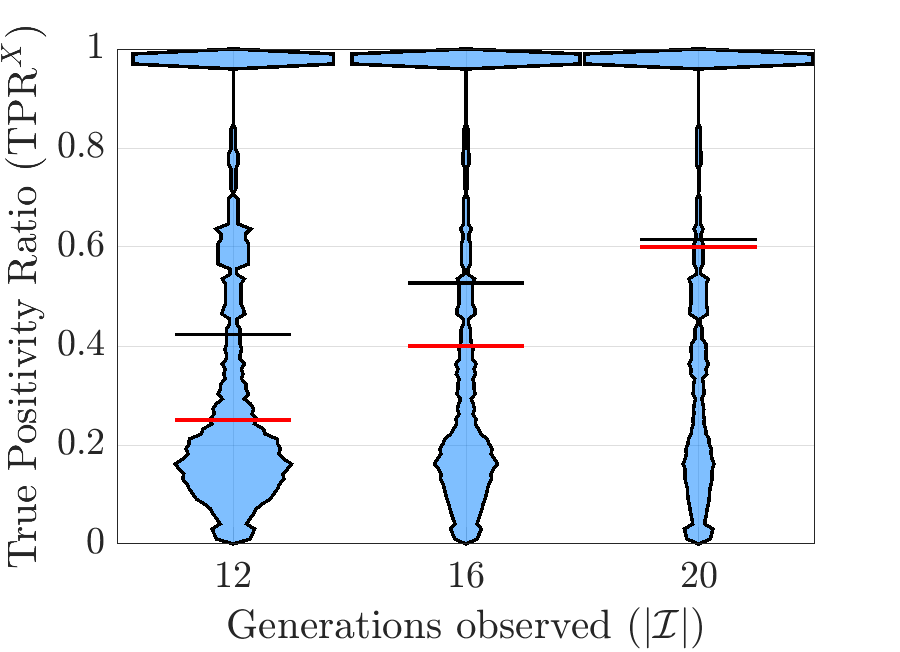}&
\includegraphics[trim={0 0 0 0},clip,width=0.33\textwidth]{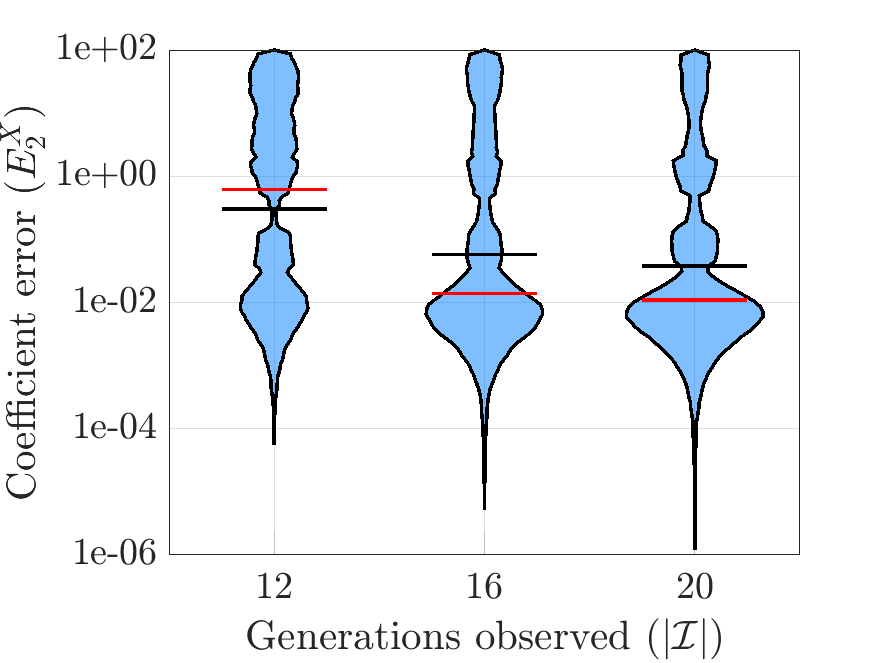}&
\includegraphics[trim={0 0 0 0},clip,width=0.33\textwidth]{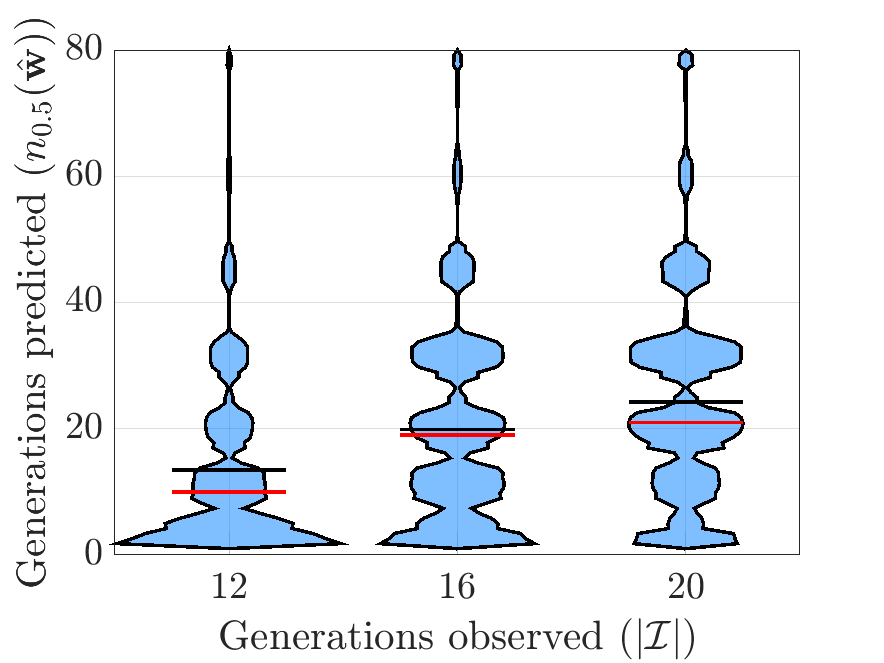} \\
\includegraphics[trim={0 0 0 0},clip,width=0.33\textwidth]{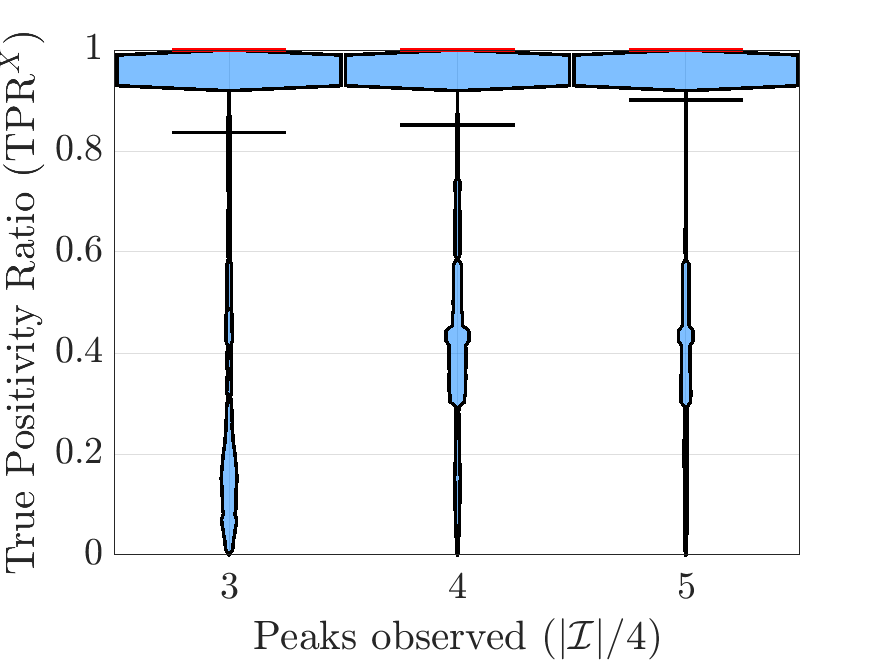}&
\includegraphics[trim={0 0 0 0},clip,width=0.33\textwidth]{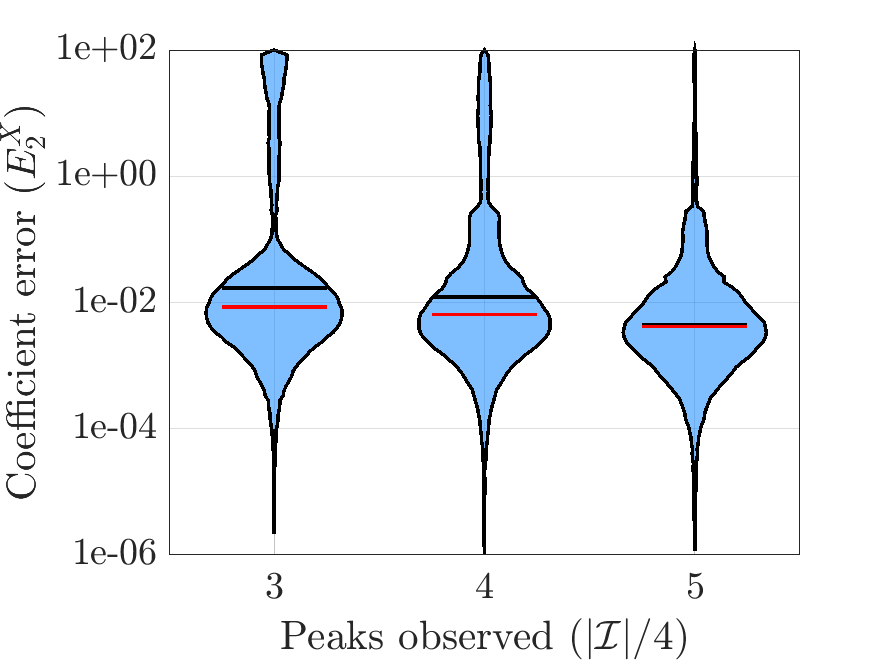}&
\includegraphics[trim={0 0 0 0},clip,width=0.33\textwidth]{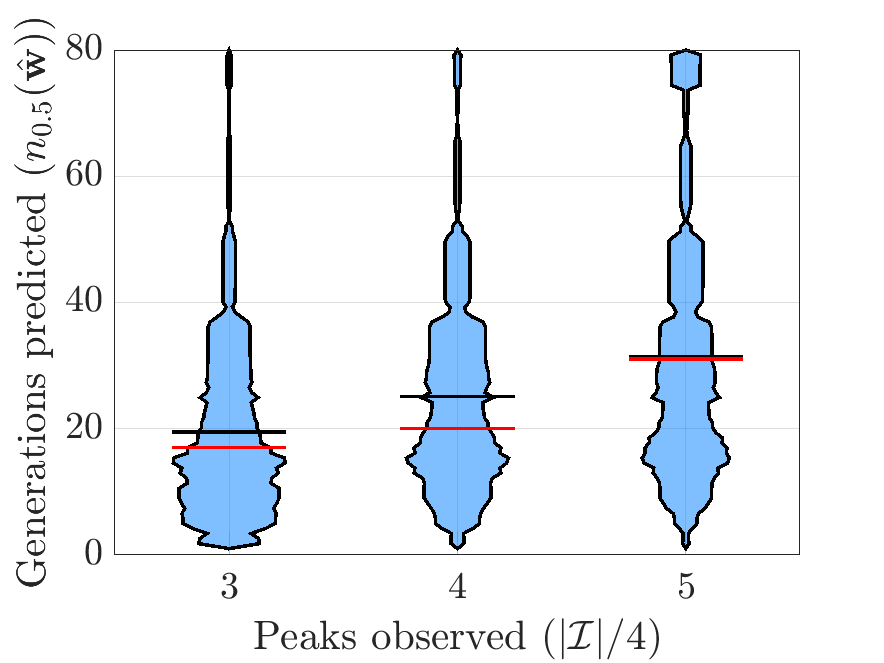}
\end{tabular}
\end{center}
\caption{{\bf Violin plots comparing aggregate performance for random and peak sampling strategies}. Left to right column: TPR$^X$ (model recovery accuracy for the discrete dynamics), $E_2^X$ (coefficient accuracy for discrete dynamics), and $n_{0.5}$ (prediction length) for data with $5\%$ noise in the continuous variables (eq.\ \eqref{eq:tpr}-\eqref{eq:Ttol}). Top and bottom rows display results for random and peak sampling, respectively. In each violin plot the sample mean and median are displayed with black and red lines, respectively. For visualization, values with $E_2^X>100$ have been removed, representing approximately $40\%$, $36\%$, and $19\%$ of values for random sampling 
and $6.2\%$, $1.8\%$, and $2.6\%$ of values for peak sampling. It can be observed from the left plots that sampling 3 peaks only is sufficient to identify the dynamics in the vast majority of cases with average TPR above 0.8, whereas random sampling average TPR is below 0.6 even with the equivalent of 5 peaks worth of data. Correspondingly, the right plots show that peak sampling with only 3 peaks doubles the accurate prediction length to 20 generations on average, compared to the equivalent data quantity random sampling with 12 generations.}
\label{fig:sampling_strategies}
\end{figure*}

Figure \ref{fig:sampling_strategies} top row presents violin plots summarizing the method performance  when using random sampling, as  data quantity (the number of generations sampled) increases, over 2000 simulated datasets for each $|\CalI|$. Similarly, the bottom row of Figure~\ref{fig:sampling_strategies} presents violin plots summarizing the method performance  when using peak sampling, as  data quantity (the number of peaks sampled, with each peak having 4 generations) increases, over 500 simulated datasets for each number of peaks ($|\CalI|/4$). The first column of the figure shows TPR (defined in Eq~\ref{eq:tpr}), with higher values conveying better performance at learning the correct model. The middle column of Figure~\ref{fig:sampling_strategies} presents coefficient relative error (defined in Eq~\ref{eq:E2}), with higher values corresponding to worse performance. Finally, the right column of Figure~\ref{fig:sampling_strategies} presents the number of generations accurately predicted  (metric defined in Eq~\ref{eq:Ttol}), with higher values corresponding to greater number of accurately predicted generations and better method performance.

In the low peak-sampling regime, when only 3 peaks are observed, the learned models on average still robustly identify the discrete map ($\text{TPR}^X\approx 1$, bottom left plot), compared to random sampling, which yields $\text{TPR}^X\approx 0.3$ on average (top left). Recall that correct identification of the discrete map relies on accurate identification of both the initial conditions and the continuous map (see Algorithm \ref{alg:full}), making TPR$^X$ a good indicator of accurate identification of the full model. This directly correlates with the observed prediction accuracy, as measured by $n_{0.5}(\what)$ (right plots), whereby 3 peaks is sufficient to predict up to 20 generations (with a long tail extending up to 80 generations), nearly twice that of random sampling. This is also reflected in the coefficient accuracy (middle plots), from which an improvement by two orders of magnitude (on average) is observed using peak sampling. 

When 5 peaks are observed, results improve slightly, in particular $n_{0.5}(\what)$ on average increases to $\approx 30$ for peak sampling and $\approx 20$ for random sampling. However, random sampling still trails behind peak sampling with TPR$^X\approx 0.6$ compared to TPR$^X\approx 1$ for peak sampling, along with larger coefficient errors $E^X_2$.  Additional performance metrics for the initial conditions and continuous dynamics are shown in Figure \ref{fig:sampling_strategies_app} of the Appendix, showing similar trends. 


In summary, the bottleneck in predictive accuracy of the learned model is found to be identification of the correct model terms (e.g.\ as measured by the TPR). Peak sampling provides a clear advantage by offering sufficient data quality for the sparse regression routines to be effective even when few total generations are observed, indicating that host peak samples are more informative for long-term prediction. Overall however, the proposed approach is able to work effectively with intermittent generation samples, and data gaps, especially when peak sampling is employed, despite the log-normal noise model falling outside of the normality assumptions inherent to regression via WENDy thanks to the consistency properties of the embedded GLS routine which depend on only the first two moments \cite{BollerslevWooldridge1992EconometricReviews,Wooldridge2024,GourierouxMonfortTrognon1984Econometrica}.


\section{Discussion}\label{sec:discussion}

\paragraph{Summary.} We present a general framework for rapidly learning hybrid discrete-continuous dynamics driven by strong seasonality with relevance to ecological systems. The resulting \texttt{WSINDy-Eco} algorithm runs in seconds on a modern laptop. Sparse model selection identifies dominant mechanisms in the dynamics which can easily be interpreted by the explicit model expressions, including complex coupling between the discrete and continuous variables. The method is well-suited for a variety of sampling designs, including data with irregular sampling characterized by missing data for several generations. Distributional information about the  WENDy parameter estimators afforded by the methodology \cite{BortzMessengerDukic2023article} enables uncertainty quantification which can be used to provide confidence around model predictions, and to easily modify learned models, as described in Section \ref{sec:HITL}. Moreover, the framework is not restricted to models in ecology, but can easily be applied to general hybrid systems featuring coupled discrete and continuous variables. 

\paragraph{Extensions.} A major benefit of the general weak formulation and library-based model learning is that libraries and test functions can also be tailored to suit different problems. Even though the library employed in this work is large, the terms in it were realistic for the problem at hand. For example, when we consider genetic drift and stochasticity in future work, to complement previous ecological studies \cite{DwyerDushoffElkintonEtAl2000,DwyerMihaljevicDukic2022AmNat,PaezDukicDushoffEtAl2017AmNat}, we will append additional terms to the library. Similarly, different test functions for discrete-time inference may be used, reflecting   different  correlation structures if so desired.

\paragraph{Data quality and sparse regression.} The data we considered here is meant to match realistic field studies, often limited to sampling a few times per week, often with 10-20 generations of data on hand \cite{skaller1985patternsGM11yrs, williams1991densityGM, ostfeld1996densityGM, maclauchlan2009DFTMinBC}. Results did not vary significantly  when the number of inter-generational samples $M$ and sampling rate $\Delta t$ were varied outside of the fixed values \eqref{eq:sampling}, with $\Delta t M\in(0.5T,T)$ and $M\geq 24$. Furthermore, we note that for the system \eqref{eq:ICres}-\eqref{eq:Xres}, the coarsest sampling needed to identify the true model under zero-noise conditions is $M=8$ samples per generation, corresponding to roughly 1 sample per week, over $9$ (not necessarily consecutive) generations, which leads to all parameters estimated to within $0.08\%$ relative error.  

As in general we will not know a priori what the noise distribution is, we have presented a general framework that we tested on a misspecified likelihood by treating log-normal noise using a GLS estimator within WENDy. Instead of tayloring to a specific likelihood (e.g.\ log-normal), which would require additional knowledge, we have relied on consistency properties of GLS, which depend only on correct specification of the first two moments, which our algorithm was able to handle well.

In the low data / high noise limit, sparse regression is not expected to return perfect results, as the sparsity threshold adapts to the current noise level (See Appendix \ref{app:wsindy_eco} and Figure \ref{fig:mstlsloss} for visualization of sparsity threshold selection under different noise levels). In Figure \ref{fig:sampling_strategies_10percent} of the Appendix, we include a treatment with 10\% noise, displaying limitations in forward prediction accuracy. This is due to the unavoidable high coefficient errors imparted by large noise and low sampling rates, which  limits the selectivity of sparse regression routines\footnote{We remark that in \cite{BortzMessengerDukic2023article} it is reported that 10\% noise and coarse sampling (e.g.\ 64 points in time) leads to 10\% relative errors in the coefficients on average, which is expected to lead to errors in the sparse identification.}. In such cases, making use of the uncertainty as detailed in Section \ref{sec:HITL} is expected to faithfully quantify predictive loss.

In such data-poor regimes, there are several ways to improve results, such as limiting the number of terms in the model libraries to reduce the size of the model space, or structuring the candidate models such that certain terms are favored (e.g.\ including them directly in the model a priori). For example, if intraspecific competition is known to be important, density-dependent survival or reproduction terms can easily be added to the models.

We also note that measurement noise in the discrete ($X$) variables has not been considered, as it is more realistic that measurement errors are found in the continuous variables (which in reality are typically stochastic). A trade-off exists between noise in continuous and discrete variables, and indeed the present sampling regime is sensitive to noise in the discrete variables, due to the low number of observations, and the fact that we are also estimating initial conditions for the continuous $(Y)$ variables at each generation (noise in the discrete variables is amplified by the noise in the initial conditions). Data with $|\CalI|=12$ generation pairs corresponds to linear systems with only 12 rows when identifying the initial conditions map and discrete dynamics, which is often insufficient to accurately select terms in the presence of noise in both $X$ and $Y$. 
This could be alleviated by the introduction of a Bayesian framework, and through the use of literature and experiment-derived prior knowledge as a way to regularize the estimation and provide tighter inference. We  defer the Bayesian extensions to future work.

\end{multicols}

\newpage
\section*{Appendix}

\appendix

\section{Supplementary Info: Sec.\ \ref{sec:methods}}

\subsection{Additional Algorithms}

\paragraph{\texttt{MSTLS-WENDy}.}Algorithm \ref{alg:MSTLS_WENDy} contains pseudocode for the \texttt{MSTLS-WENDy} algorithm, a sparse regression routine for dynamical system identification which accounts for the errors-in-variables nature of the problem. 

\paragraph{\texttt{MSTLS-WENDy-Par}.}Algorithm \ref{alg:MSTLS-WENDy-Par} contains pseudo-code for the \texttt{MSTLS-WENDy-Par} algorithm, which adapts \texttt{MSTLS-WENDy} to the setting of equations with both state and parametric variables.  

\paragraph{\texttt{ForwardSim}.}In the cases of interest, the continuous variable at time $T$, $Y_n(T)$ is not observed and must be simulated using the learned system. This is carried out for $n\in \CalI$ using Runge-Kutta 45 routine in Matlab (and easily parallelized over $n\in \CalI$). This is detailed in Algorithm \ref{alg:ForwardSim}, which includes a heuristic computation for the covariance $\widehat{\Sigma}^{\widehat{\Ybf}_T}$ of the resulting estimates $\widehat{\Ybf}_T$.

\SetKwComment{Comment}{/* }{ */}
\begin{algorithm*}[ht]
\caption{\label{alg:MSTLS_WENDy}\texttt{MSTLS-WENDy}. Sparse regression for errors-in-variables problems. Note this assumes the necessary vectorization operations have been performed, e.g. $\Gbf = \Gbf_{_{\tiny \otimes}}$ in \eqref{eq:wendy}. In line 8, the hard thresholding operator $H_S$ sets to zero the entries that are not in $S$.}
	\SetKwInOut{Input}{input} \SetKwInOut{Output}{output} 
	\Input{Linear system $(\bbf,\Gbf)\in \Rbb^{K}\times\Rbb^{K\times J}$, data covariance $\Sigma\in\Rbb^{M\times M}$, library covariance factors $(\Lbf_j)_{j=0}^J\subset \Rbb^{K\times M}$, WENDy stopping tolerance $\tau>0$, maximum WENDy iterations \texttt{MaxIts}, candidate sparsity thresholds $\pmb{\lambda}\in \Rbb^L$}
	\Output{Parameter Estimate $\widehat{\wbf}$, Residual Covariance Estimate $\widehat{\Cbf}$, Parameter Covariance Estimate $\widehat{\Sbf}$} 
	\BlankLine
        $\wbf^{(0)} \gets (\Gbf^T\Gbf)^{-1}\Gbf^T\bbf$ \Comment*[r]{Initial guess}
        $\Wbf \gets \textbf{0}\in\Rbb^{J\times L}$\;
        $\Hbf \gets \textbf{0}\in\Rbb^{K\times K\times L}$\;
        $\widehat{\Sbf}\gets \textbf{0}\in\Rbb^{J\times J}$\;
        \For{$\ell=1,\dots,L$}{
        $\lambda = \pmb{\lambda}_\ell$\;
        Define thresholding intervals $(I_j^\lambda)_{j=1}^J$ using \eqref{eq:MSTLS_I}\;
        $n \gets 0$\;
        \While{$n<J$}{
            $S^{(n)} \gets \{j\ :\ |\wbf_j^{(n)}|\in I_j\}$ \Comment*[r]{Get sparsity pattern}
            $\vbf^{(0)} \gets H_{S^{(n)}}(\wbf^{(n)})$  \Comment*[r]{Initialize WENDy}
            $m\gets 0$\;
            $\texttt{check}\gets \texttt{True}$\;
            \While{\texttt{check}}{        
                $\Lbf^{(m)}\gets \Lbf_0+\sum_{j\in S^{(n)}}\vbf^{(m)}_j\Lbf_j$ \Comment*[r]{Form covariance factor}
                $\Cbf^{(m)} \gets \Lbf^{(m)}\Sigma(\Lbf^{(m)})^T$ \Comment*[r]{Build covariance matrix}
                $\vbf^{(m+1)}_{S^{(n)}} \gets \left(\Gbf_{S^{(n)}}^T(\Cbf^{(m)})^{-1}\Gbf_{S^{(n)}}\right)^{-1}\Gbf_{S^{(n)}}^T(\Cbf^{(m)})^{-1}\bbf$\Comment*[r]{Solve GLS problem}
                $\texttt{check}\gets \{m<\texttt{MaxIts}\  \&\ \|\vbf^{(m+1)}-\vbf^{(m)}\|_2/\|\vbf^{(m)}\|_2>\tau\}$\;
                $m\gets m+1$\;
            }
            $\wbf^{(n+1)}\gets \vbf^{(m)}$\;
            $n\gets n+1$\;
        }
        $\Wbf(:,\ell) = \wbf^{(n)}$\;
        $\Hbf(:,:,\ell) = \Cbf^{(m-1)}$\;
        }
        $\hat{\ell} \gets \argmin_{1\leq \ell\leq L}\CalL(\pmb{\lambda}_\ell)$ \Comment*[r]{Select sparsity threshold (see \cite[MSTLS]{MessengerBortz2021JComputPhys})}
        $\what \gets \Wbf(:,\hat{\ell})$\;
        $S \gets \{j\ :\ \what_j\neq 0\}$\;
        $\widehat{\Cbf} \gets \Hbf(:,:,\hat{\ell})$\;
        $\widehat{\Sbf}_{S,S} \gets (\Gbf_S^T\Gbf_S)^{-1}\Gbf_S^T)\ 
    \widehat{\Cbf}\ (\Gbf_S(\Gbf_S^T\Gbf_S)^{-1})$\;
\end{algorithm*}

\SetKwComment{Comment}{/* }{ */}
\begin{algorithm*}
\caption{\label{alg:MSTLS-WENDy-Par} \texttt{MSTLS-WENDy-Par}. Adaptation of \texttt{MSTLS-WENDy} to data with state and parametric dependencies.}
	\SetKwInOut{Input}{input} \SetKwInOut{Output}{output} 
	\Input{Left-hand side operator $\CalD$, Data $(\Ubf^{(s)},\Ubf^{(p)})$, Data covariances $(\Sigma^{(s)},\Sigma^{(p)})$, Initial libraries $(\Lbb_{(s)}$, $\Lbb_{(p)})$, library incrementer $\Lbb_+$, test functions $\Phi$, candidate thresholds $\pmb{\lambda}$, WENDy stopping tolerance $\tau>0$, maximum WENDy iterations \texttt{MaxIts}, library increment confidence level $c$}
	\Output{Parameter Estimate $\widehat{\wbf}$, Residual Covariance Estimate $\widehat{\Cbf}$, Parameter Covariance Estimate $\widehat{\Sbf}$}
Build $\bbf=\textsf{vec}(\Bbf)$ from $\CalD$, $\Phi$, and $\Ubf^{(s)}$ according to \eqref{eq:u_sp} and \eqref{eq:B_cont} or \eqref{eq:B_disc}\;
Build $\Sigma^{(s\otimes p)}$ according to \eqref{eq:full_cov}\;
$\texttt{check} \gets \texttt{True}$\;
\While{\texttt{check}}{
$\Lbb_{(p)}\gets\Lbb_+(\Lbb_{(p)})$ \Comment*[r]{Increment library (eq.\ \eqref{eq:lib_inc})}
$\Lbb_{(s\otimes p)}\gets \Lbb_{(s)}\otimes \Lbb_{(p)}$\;
Build $\Gbf_{_{\tiny \otimes}},(\Lbf_j)_{j=1}^J$ from $\Lbb_{(s\otimes p)}$ (eqs.\ \eqref{eq:Gij}, \eqref{eq:Lfacs})\;
$(\what,\widehat{\Cbf},\widehat{\Sbf})$ = \texttt{MSTLS-WENDy}$\left(\bbf,\Gbf_{_{\tiny \otimes}},\Sigma^{s\otimes p},(\Lbf_j)\right)$\;
$\texttt{check} \gets \texttt{LibIncFun}(\what,\widehat{\Cbf},c)$ (eq. \eqref{eq:LibInc}) \;
 }
\end{algorithm*}

\SetKwComment{Comment}{/* }{ */}
\begin{algorithm*}
\caption{\label{alg:ForwardSim}\texttt{ForwardSim}. From learned dynamics estimate continuous variables at time $T$ together with covariance estimates.}
	\SetKwInOut{Input}{input} \SetKwInOut{Output}{output} 
	\Input{Discrete initial conditions $\Xbf_0 = (X_n)_{n\in\CalI}$; Observed continuous dynamics $(\Ybf,\tbf)$; Continuous dynamics covariance $\Sigma^\Ybf$; Initial conditions weights $\what^\text{IC}$; Initial conditions library $ \Lbb^\text{IC}$; Continuous dynamics weights $\what^Y$; Continuous dynamics library $ \Lbb^Y$; Simulation interval $t_\text{sim}$; Additional simulation arguments \texttt{Args}}
	\Output{Estimated final time values $\widehat{\Ybf}_T$, Estimated final time covariance $\widehat{\Sigma}^{\widehat{\Ybf}_T}$} 
$h(X)\gets \Lbb^\text{IC}(X)\what^\text{IC}$ \Comment*[r]{Form initial conditions map}
$g(Y;X)\gets \Lbb^\text{Y}(Y,X)\what^Y$ \Comment*[r]{Form continuous dynamics map}
$\widehat{\Sigma}^{\widehat{\Ybf}_T} \gets \mathbf{0} \in \Rbb^{|\CalI|\times |\CalI|}$\;
\For{$n\in \CalI$}{
$Y_n(0) \gets h(X_n)$ \Comment*[r]{Initialize continuous dynamics}
$g_n(Y) \gets g(Y;X_n)$ \Comment*[r]{Parametrize continuous dynamics map}
$\hat{Y}_n(t_\text{sim}) \gets \texttt{RK45}(g_n,Y_n(0),t_\text{sim},\texttt{Args})$ \Comment*[r]{Simulate with (e.g.) Runge-Kutta 45}
$\widehat{\Ybf}_T \gets (\hat{Y}_n(T))_{n\in \CalI}$ \Comment*[r]{Extract final time estimate}
$\tilde{Y}_n \gets \texttt{interp}(t_\text{sim},\hat{Y}_n(t_\text{sim}),\tbf)$ \Comment*[r]{Interpolate simulated data onto observed grid}
$\widehat{\Sigma}^{\widehat{\Ybf}_T}_{n,n} = \texttt{Var}({\tilde{Y}_n - Y_n})$ \Comment*[r]{Compute simulated data sample variance}
}
\end{algorithm*}

\subsection{\texttt{WSINDy-Eco} Hyperparameters}\label{app:wsindy_eco}

Throughout we set the WENDy iteration tolerance to $\tau = 10^{-4}$, typically reached in under 10 iterations, and we use \texttt{MaxIts} $= 5$. The library, test function, and data covariance estimation choices are described below, together with a discussion about the sparsity parameters.

\paragraph{Trial function libraries.}In this work, we let each library $\Lbb$ be of power-law type, with functions $f_j(u) = u_1^{p_1}\cdots u_d^{p_d}$ for $p_i\in \Rbb$. These trial functions are chosen for their ability to model any dynamical system if $p_i$ is chosen large enough, as well as their computational efficiency, and their widespread appearance in models of natural phenomena. The parameter libraries $\Lbb^\text{IC}_{(p)},\Lbb^X_{(p)}$ are initialized to only contain the constant function $\{1\}$, while $\Lbb^Y_{(p)}$ is initialized to contain the terms $\{1,N_n^{-V}\}$. They are then incremented by increasing the total polynomial degree by 1 in each incrementation step, up to a total degree of 4. The state libraries $\Lbb^Y_{(s)},\Lbb^X_{(s)}$ are fixed with total degrees 3 and 2, that is, one degree greater than that appearing in the true model (or what we think the true model is), in addition the term $S_n^{1+V}$ is added to $\Lbb^Y_{(s)}$. Altogether this leads to a total of 562 possible terms across all equations in all submodels.

\paragraph{Test function basis.}For the continuous dynamics we employ piece-wise polynomial test functions of the form 
\[\phi_k(t) = \max\left\{0,1-\left(\frac{t-t_k}{m_t\Delta t}\right)^2\right\}^\eta\]
which are supported on the time intervals $(t_k-m_t\Delta t, t_k+m_t\Delta t)$ where $\Delta t$ is the within-epidemic sampling rate and $m_t$ is the {\it test function radius}. The power $\eta$ controls the highest order of differentiability, and thus the degree of accuracy of numerical integration in the absence of noise. The radius $m_t$ is largely responsible for variance reduction and mitigating high-frequency noise. The parameters $\eta,m_t$ are chosen from the observed data $\Ybf$ using the method detailed in \cite[Appendix A]{MessengerBortz2021JComputPhys}, which interprets the weak-form  as an approximate low-pass filter. 

\paragraph{Data Covariance Estimation.}The covariance matrix $\Sigma^\Ybf$ for the observed continuous data $\Ybf = (Y_n(\tbf_n))_{n\in \CalI}$ is taken to be diagonal and is estimated by assuming that each value in the time series $Y_n(\tbf)$ has a constant variance $\sigma_n$, which we approximate using a high-order filter (see \cite[Appendix G]{messenger2022asymptotic}). The covariance matrix $\Sigma^\text{IC}$ for the initial condition data is then taken to be diagonal with variance $\sigma_n$ for $Y_n(0)$ and variance 0 for $X_n$. For the covariance matrix $\Sigma^\Xbf$ of the data $(X_{n+1},X_n,Y_n(T))$, we assume that $Y_n(T)$ is random but $X_n,X_{n+1}$ are non-random (have zero variance). We set the variance estimate for each $Y_n(T)$ to be the empirical variance of  the observed data $Y_n(\tbf_n)$ around computed values $\widehat{Y}_n(\tbf_n)$, as described in Algorithm \ref{alg:ForwardSim}.

\begin{figure*}
\begin{center}
\begin{tabular}{@{}c@{}c@{}c@{}}
\includegraphics[trim={0 0 0 0},clip,width=0.33\textwidth]{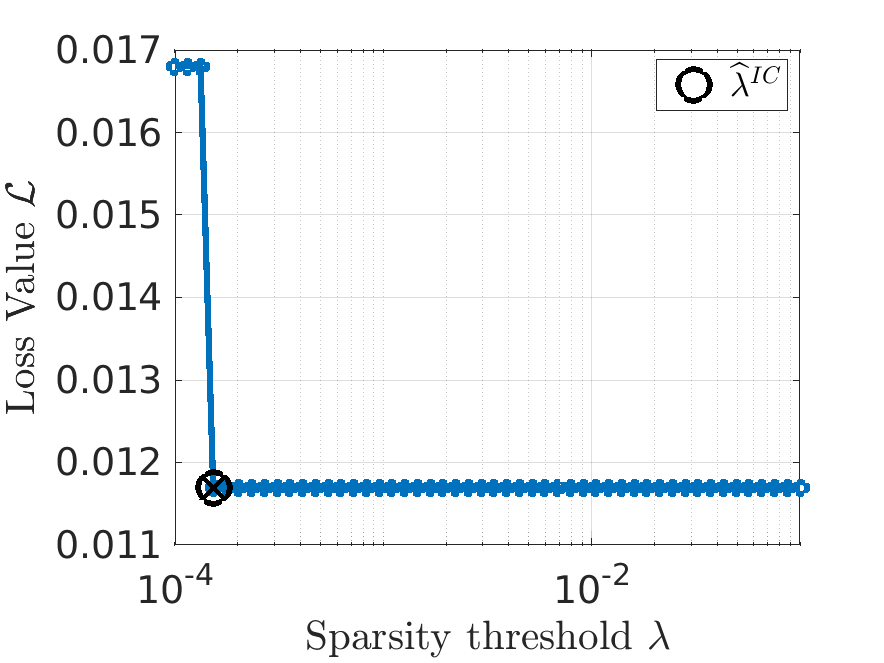}&
\includegraphics[trim={0 0 0 0},clip,width=0.33\textwidth]{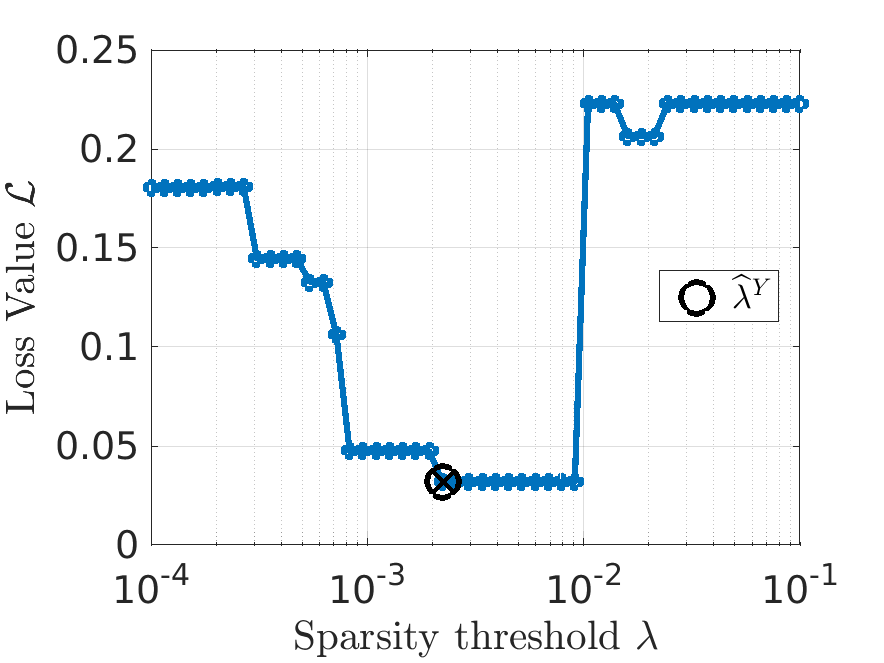}&
\includegraphics[trim={0 0 0 0},clip,width=0.33\textwidth]{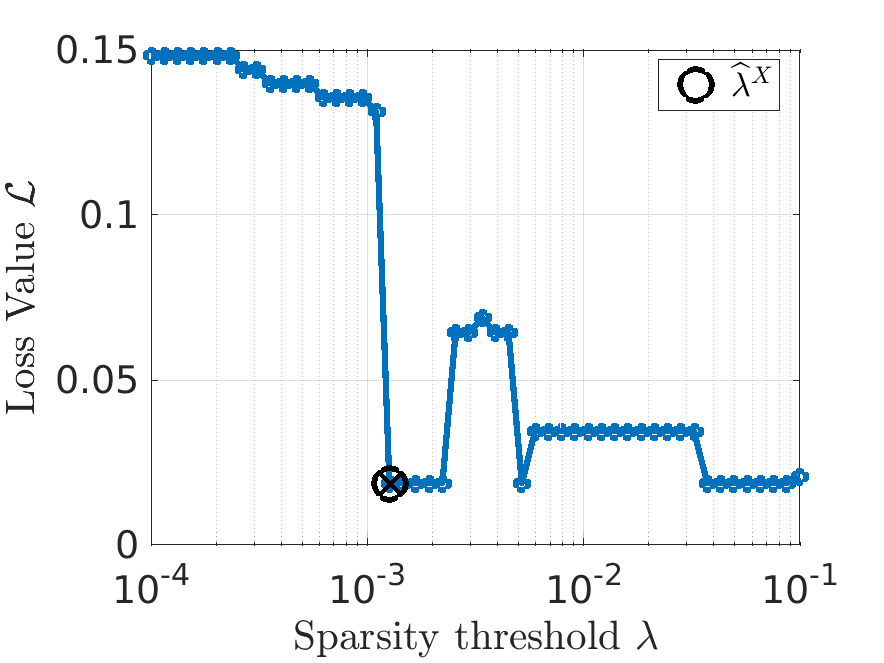}\\
\includegraphics[trim={0 0 0 0},clip,width=0.33\textwidth]{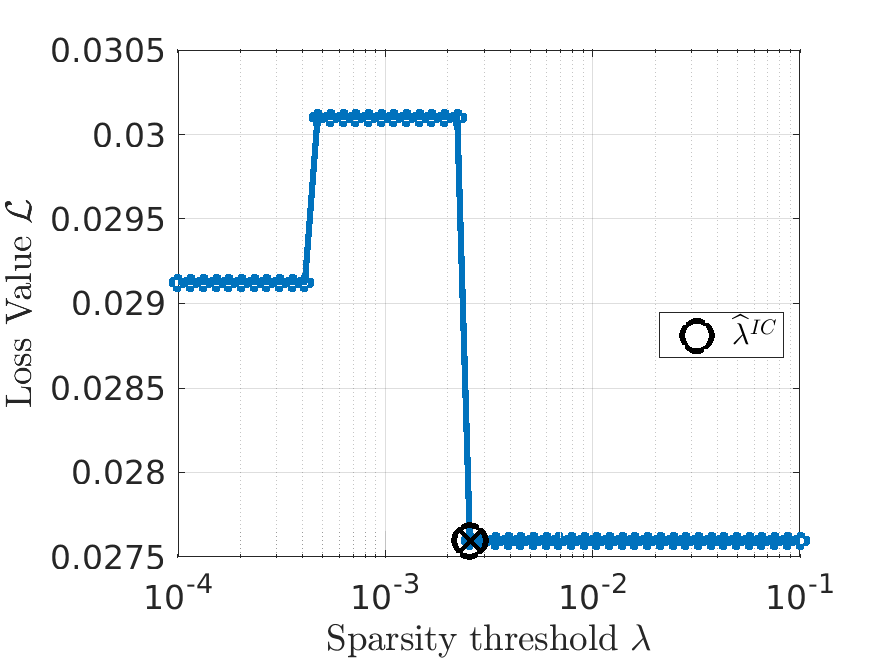}&
\includegraphics[trim={0 0 0 0},clip,width=0.33\textwidth]{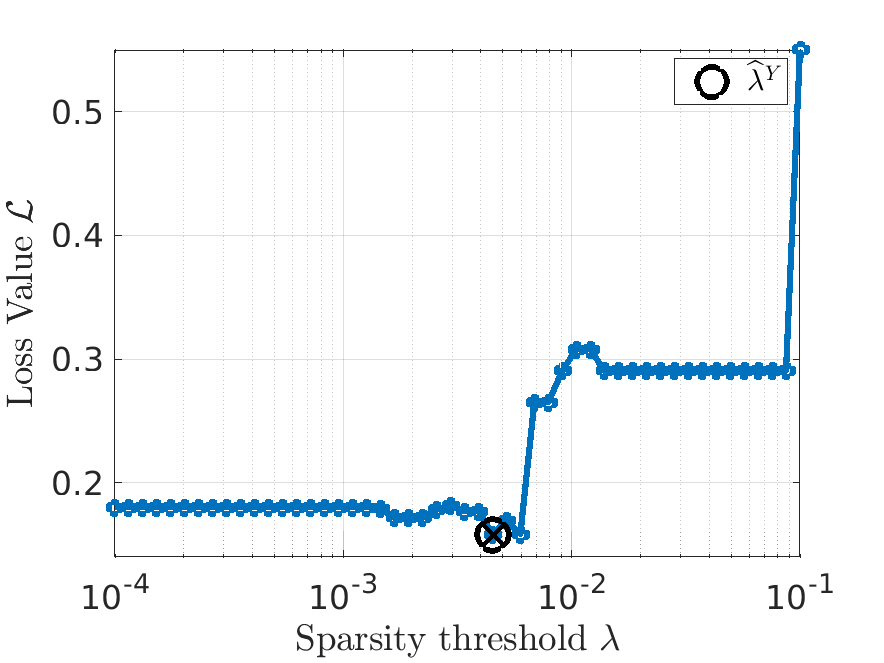}&
\includegraphics[trim={0 0 0 0},clip,width=0.33\textwidth]{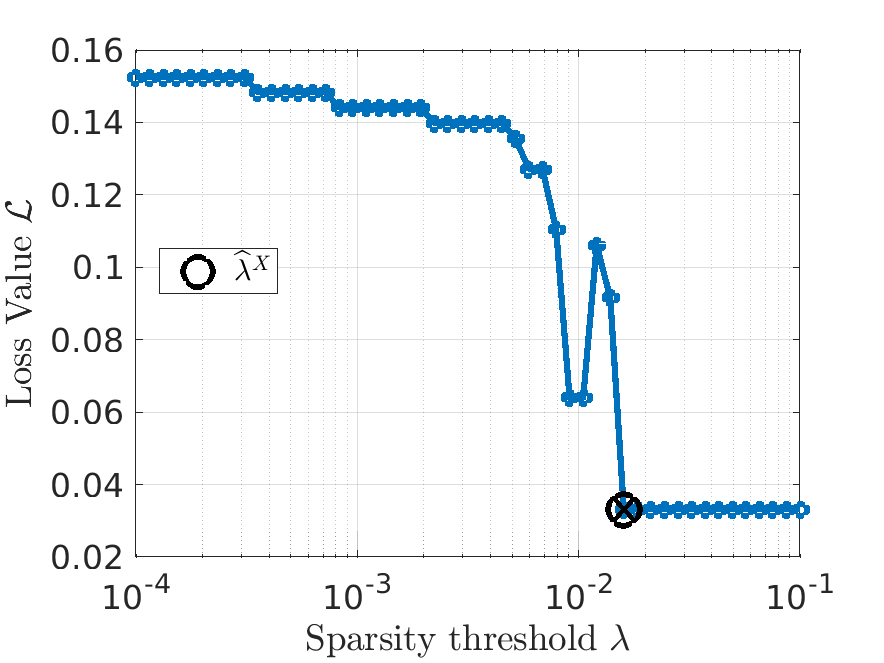}
\end{tabular}
\end{center}
\caption{{\bf Visualization of sparsity threshold selection in MSTLS}. Left to right: selection of thresholds for the initial conditions map, continuous dynamics, and discrete dynamics. Top and bottom: 1\% noise and 5\% noise. There is a clear relationship between the selected threshold and the level of noise in the data, with higher noise levels corresponding to the need for (and the selection of) a larger threshold $\widehat{\lambda}$, and consequently less complex models. In the case of the continuous dynamics (middle plots) the loss function becomes flatter, indicating that there are potentially multiple  models describing the data equally well. In these examples all submodels were identified correctly with the exception of the continuous dynamics at $5\%$ noise, yet the recovered model is accurate enough to enable identification of the discrete dynamics (recall this depends on the outputs of the simulated learned continuous dynamics).}
\label{fig:mstlsloss}
\end{figure*}

\paragraph{Sparsity Thresholds.} We use equally log-spaced candidate thresholds $\log_{10}\pmb{\lambda} = \texttt{linspace}(-4,0,50)$, seen to work well in previous studies utilizing MSTLS \cite{MessengerBortz2021SIAMMultiscaleModelSimul, MessengerBortz2021JComputPhys}. The ultimate sparsity threshold $\lambda$ used in the each final model is determined by the MSTLS loss function \cite{MessengerBortz2021JComputPhys}.  We note that there is a clear relationship between the selected threshold and the level of noise in the data, with higher noise levels corresponding to the need for (and the selection of) larger thresholds, and consequently less complex models. Figure \ref{fig:mstlsloss} displays this trend for two exemplary cases, using peak sampling with 3 peaks at 1\% and 5\% noise respectively. The selected threshold is seen to adapt to the noise level, however at larger noise it may be challenging to select the correct model, as many models fair equally well according to the given loss (bottom middle plot).


\section{Supplementary Info: Sec.\ \ref{sec:HITL}}
The Akaike Information Criteria (AIC) is defined is general by \cite{Akaike1974IEEETransAutomControl,HootenHobbs2015EcolMonogr}
\begin{equation}\label{eq:AIC}
\text{AIC} = 2k - 2\log(\widehat{L})
\end{equation}
where $k$ is the number of non-zero entries of the model parameter vector $\what$ and $L=L(\wbf)$ is a suitable likelihood function, and $\widehat{L} = L(\what)$ is its maximum, attained at the parameters $\what$. In our case, the negative log-likelihood function is simply given by the sum of each generalized least-squares residual,
\begin{equation}\label{eq:loglike}
\begin{split}
-\log(L(\what)) = \frac{1}{2}\Big[(\Gbf^\text{IC}&\what^\text{IC}-\bbf^\text{IC})^T(\widehat{\Cbf}^\text{IC})^{-1}(\Gbf^\text{IC}\what^\text{IC}-\bbf^\text{IC}) \\
&+(\Gbf^Y\what^Y-\bbf^Y)^T(\widehat{\Cbf}^Y)^{-1}(\Gbf^Y\what^Y-\bbf^Y)\\
&\qquad+(\Gbf^X\what^X-\bbf^X)^T(\widehat{\Cbf}^X)^{-1}(\Gbf^X\what^X-\bbf^X)\Big]
\end{split}
\end{equation}
with superscripts $\{\text{IC},X,Y\}$ denoting the submodel which the quantity corresponds to.  Similarly, the number of non-zero model parameters is $k = \|\what\|_0 = \|\what^\text{IC}\|_0+\|\what^Y\|_0+\|\what^X\|_0$.

\section{Supplementary Info: Sec.\ \ref{sec:sampling}}
\begin{figure*}
\begin{center}
\begin{tabular}{@{}c@{}c@{}c@{}c@{}}
\includegraphics[trim={0 0 0 0},clip,width=0.25\textwidth]{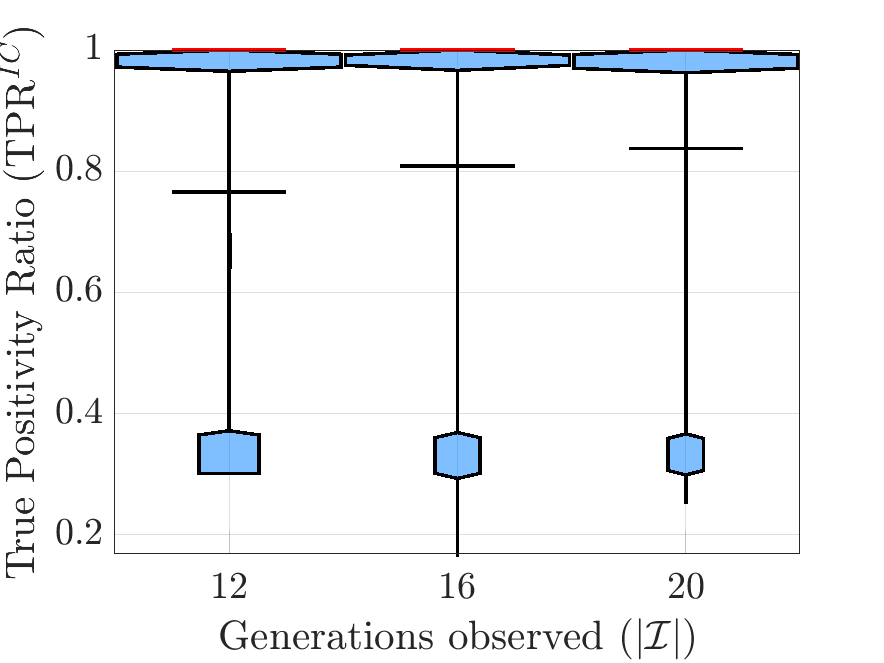}&
\includegraphics[trim={0 0 0 0},clip,width=0.25\textwidth]{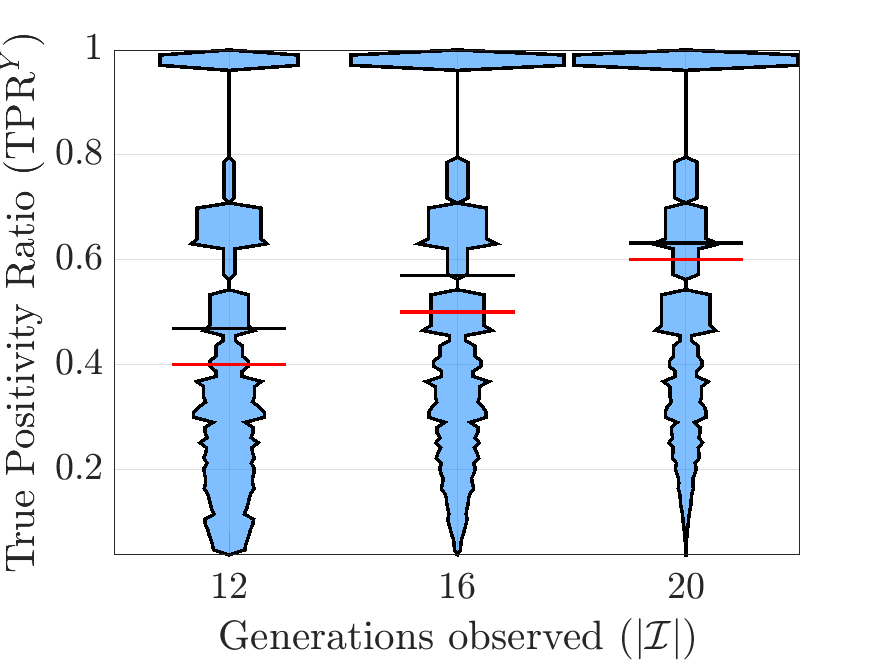}&
\includegraphics[trim={0 0 0 0},clip,width=0.25\textwidth]{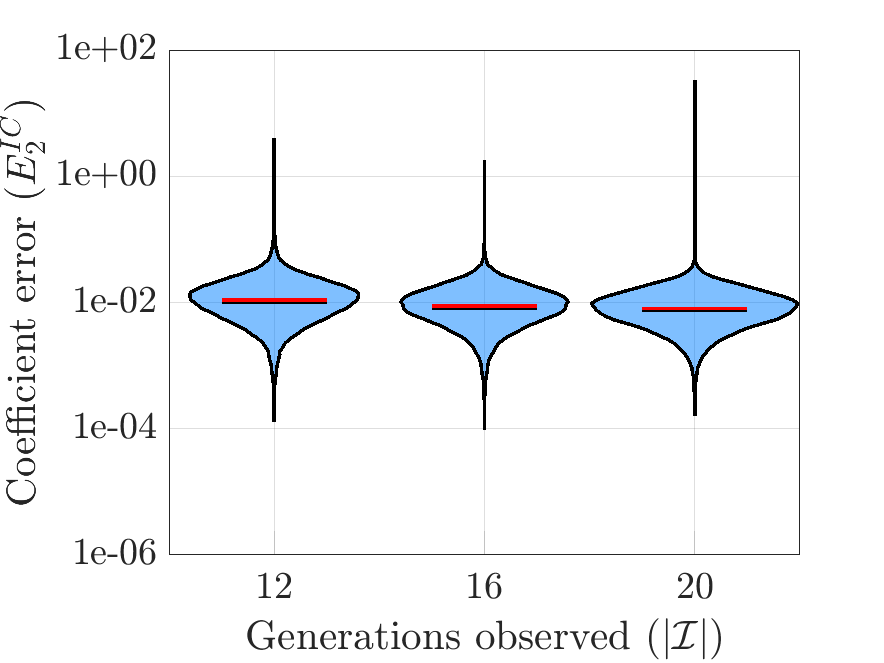}&
\includegraphics[trim={0 0 0 0},clip,width=0.25\textwidth]{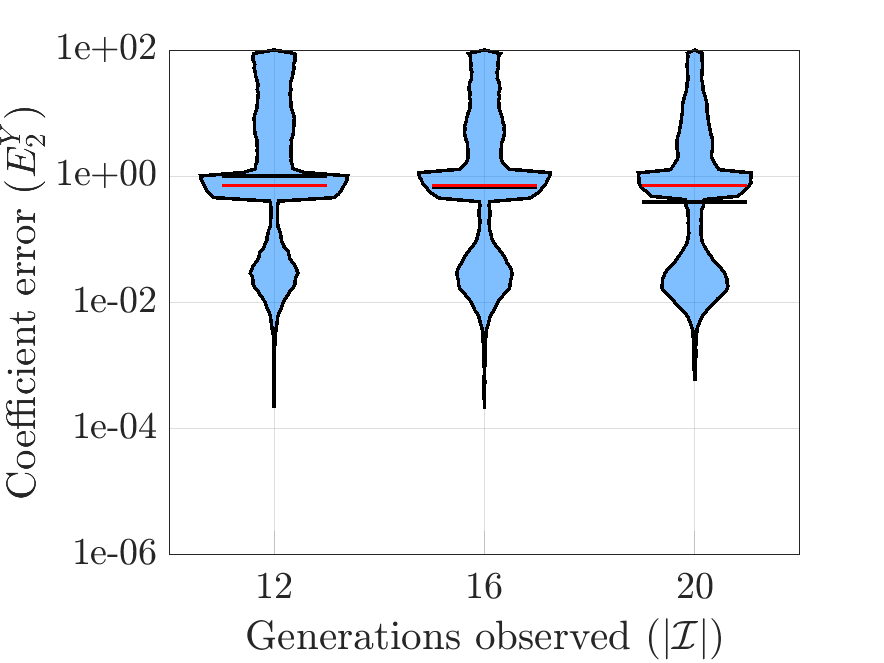}\\
\includegraphics[trim={0 0 0 0},clip,width=0.25\textwidth]{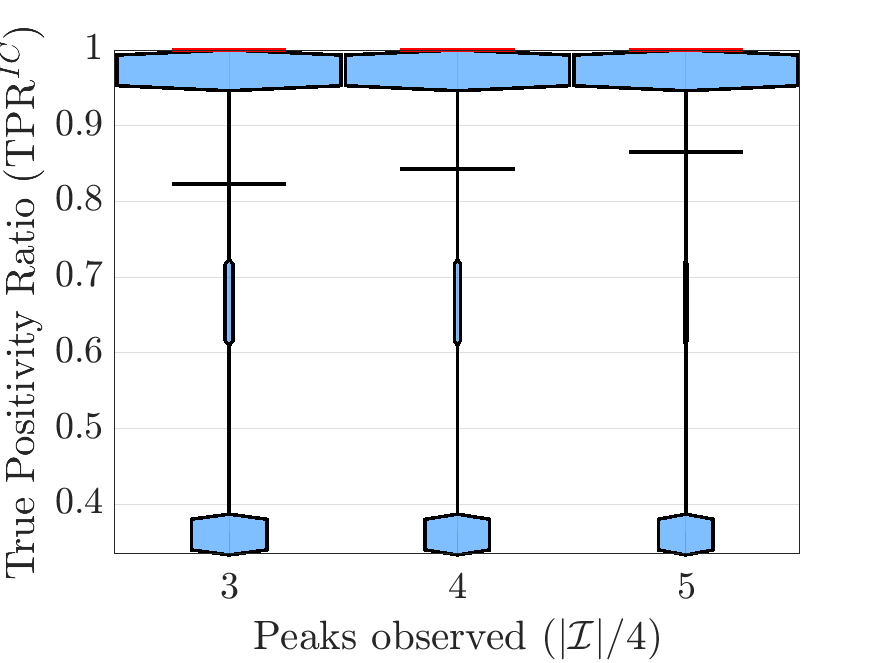}&
\includegraphics[trim={0 0 0 0},clip,width=0.25\textwidth]{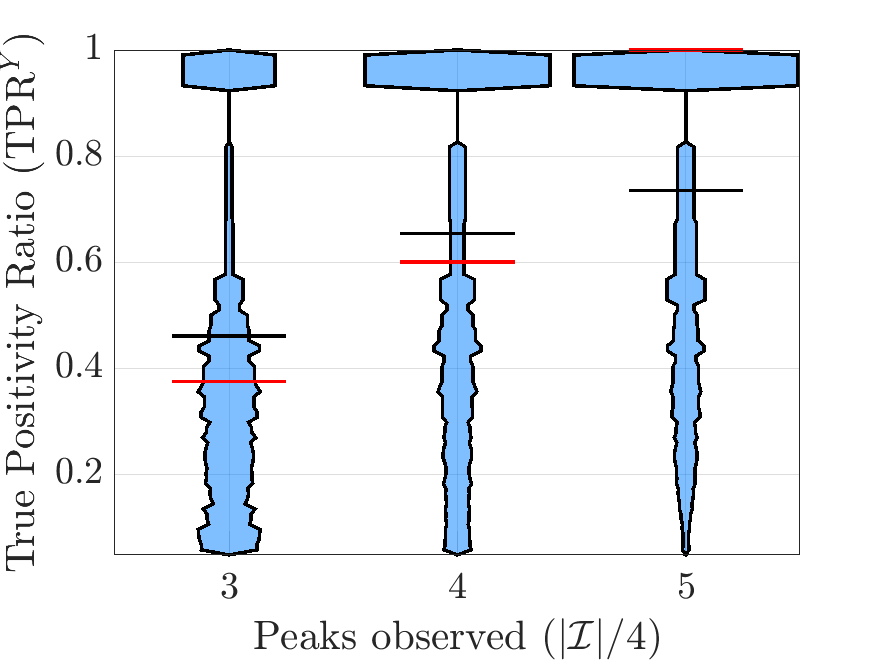}&
\includegraphics[trim={0 0 0 0},clip,width=0.25\textwidth]{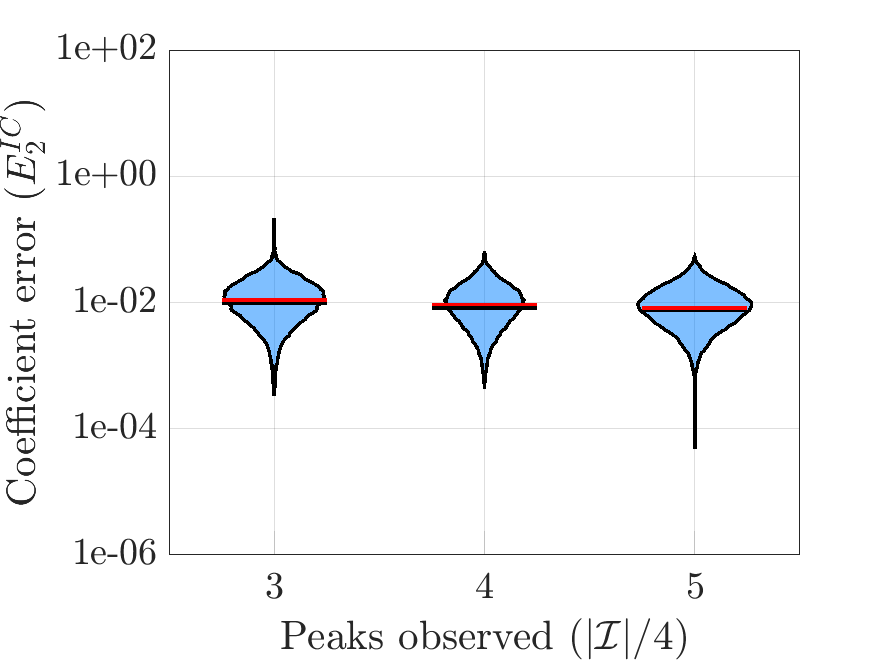}&
\includegraphics[trim={0 0 0 0},clip,width=0.25\textwidth]{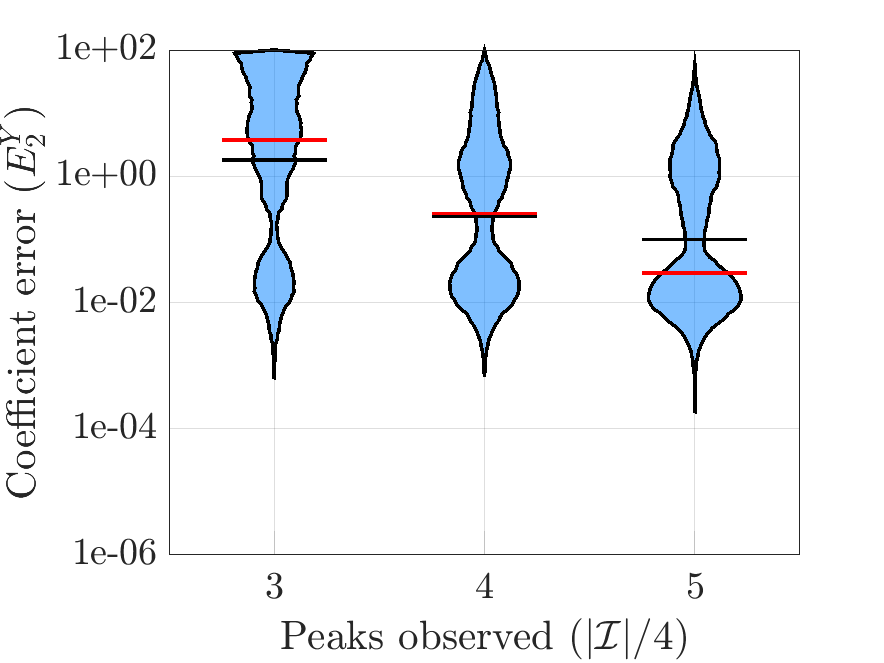}
\end{tabular}
\end{center}
\caption{{\bf Violin plots comparing aggregate performance for random and peak sampling strategies (Fig. \ref{fig:sampling_strategies} cont.)}. Left to right: TPR$^{IC}$ (model recovery accuracy for the initial conditions map), TPR$^{Y}$ (model recovery accuracy for the continuous dynamics), $E^{IC}_2$ (coefficient accuracy for initial conditions map), $E^{Y}_2$ (coefficient accuracy for the continuous dynamics). Top and bottom rows display results for random and peak sampling, respectively.}
\label{fig:sampling_strategies_app}
\end{figure*}

\begin{figure*}
\begin{center}
\begin{tabular}{@{}c@{}c@{}c@{}c@{}}
\includegraphics[trim={0 0 0 0},clip,width=0.25\textwidth]{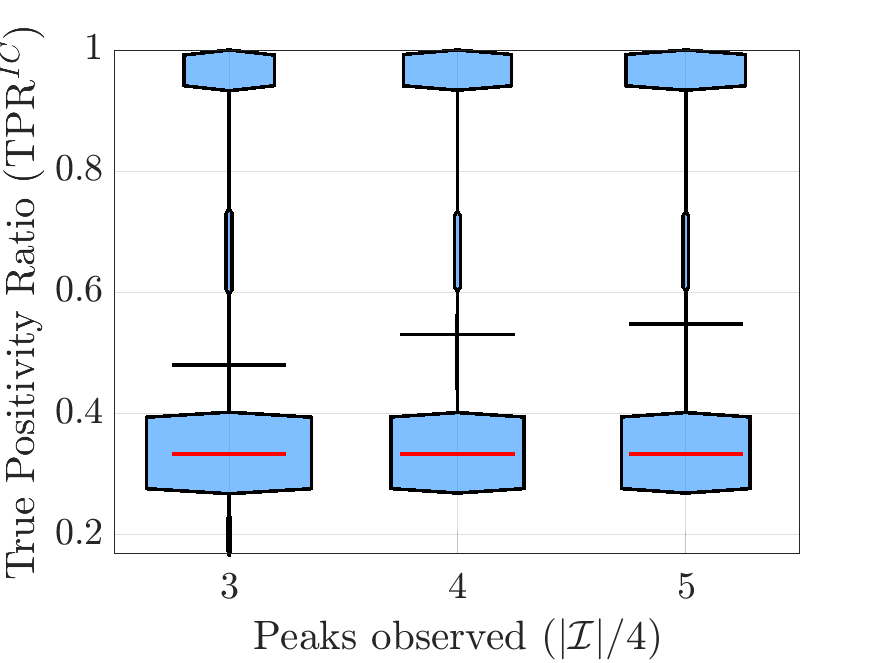}&
\includegraphics[trim={0 0 0 0},clip,width=0.25\textwidth]{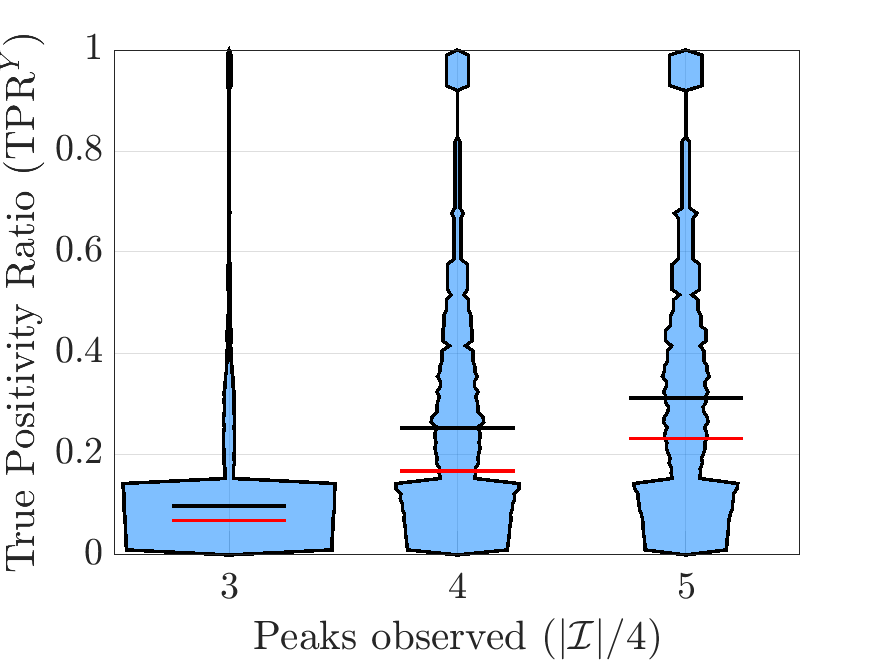}&
\includegraphics[trim={0 0 0 0},clip,width=0.25\textwidth]{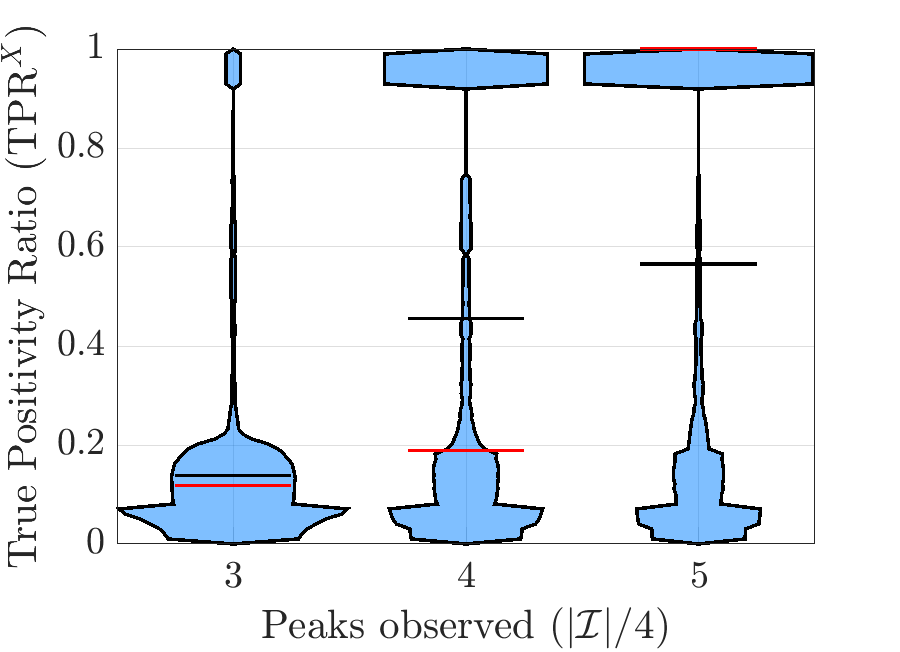}&
\includegraphics[trim={0 0 0 0},clip,width=0.25\textwidth]{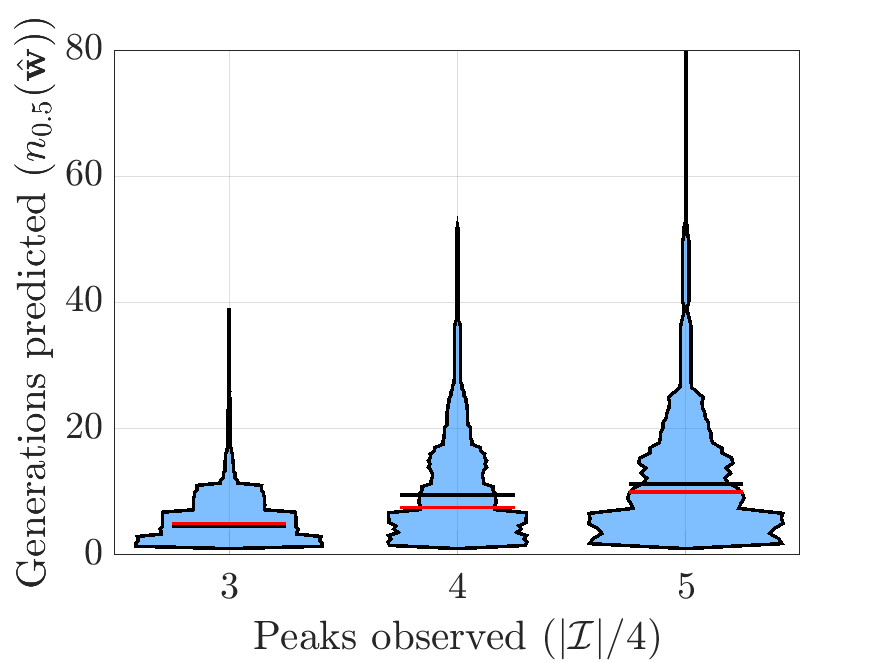}
\end{tabular}
\end{center}
\caption{{\bf Violin plots comparing aggregate performance for peak sampling at 10\% noise}. Left to right: TPR$^{IC}$ (model recovery accuracy for the initial conditions map), TPR$^{Y}$ (model recovery accuracy for the continuous dynamics), TPR$^{X}$ (model recovery accuracy for the discrete dynamics). At 10\% noise, it is clear that more samples are needed to achieve good long-term prediction. Sampling over 5 peaks yields correct identification of the discrete map in the majority of cases (3rd plot), which does not necessarily correspond to correct identification of the initial conditions map or the continuous dynamics (1st and 2nd plots). However, the predictive capability is still sufficient to accurately predict  5 generations ahead on average, even with only 3 peaks sampled (4th plot), which may be sufficient for many applications. When 5 peaks are observed, 10 generations on average may be predicted.}
\label{fig:sampling_strategies_10percent}
\end{figure*}

Figure \ref{fig:sampling_strategies_app} complements Figure \ref{fig:sampling_strategies} with additional performance metrics including information about the initial conditions map and the continuous dynamics. The trends are analogous, with peak sampling offering a clear advantage, and requiring fewer samples to achieve accurate model recovery. In Figure \ref{fig:sampling_strategies_10percent}, we report results for peak sampling under 10\% noise, showing a breakdown in the long-term predictive capabilities of the learned model. However, at this high noise level, learned models are still able to predict on average 5-10 generations in the future, which may be sufficient for many applications including the spongy moth one where insect outbreaks tend to occur every 9-12 years.

\dataccess{All software used to generate the results in this work is available at this repository: \url{https://github.com/dm973/WSINDy_Eco.git}}

\aucontribute{All authors conceived of the overall concept of the paper, designed numerical experiments, and wrote the manuscript. DM and VD developed the algorithms, and DM performed all computer programming, data collection, and data analysis. GD provided ecological motivation and field expertise.}


\funding{This research was supported in part by the NSF Division Of Environmental Biology (EEID grant DEB-2109774), and NIFA Biological Sciences (grant 2019-67014-29919).}

\ack{The authors wish to thank Prof. David Bortz (Department of Applied Mathematics, University of Colorado, Boulder) for insightful comments about the methodology of this work as well as Dr. Katie Dixon (University of Chicago, Department of Ecology \& Evolution) for guidance in its ecological applications.}


 

\normalem
\printbibliography

@article{ABC2009,
  title = {Approximate {{Bayesian}} Computation Scheme for Parameter Inference and Model Selection in Dynamical Systems},
  author = {Tina Toni, David Welch, Natalja Strelkowa},
  year = {2009},
  journal = {Journal of the Royal Society, Interface},
  pages = {187--202}
}

@article{Akaike1974IEEETransAutomControl,
  title = {A New Look at the Statistical Model Identification},
  author = {Akaike, H.},
  year = {1974},
  month = dec,
  journal = {IEEE Transactions on Automatic Control},
  volume = {19},
  number = {6},
  pages = {716--723},
  issn = {0018-9286},
  doi = {10.1109/TAC.1974.1100705},
  urldate = {2014-11-12},
  langid = {english}
}

@incollection{Akaike1977Applicationsofstatistics,
  title = {On Entropy Maximization Principle},
  booktitle = {Applications of Statistics},
  author = {Akaike, Hirotugu},
  editor = {Krishnaiah, P. R.},
  year = {1977},
  pages = {27--41},
  publisher = {North Holland},
  address = {Amsterdam, Netherlands}
}

@article{AltizerDobsonHosseiniEtAl2006EcolLett,
  title = {Seasonality and the Dynamics of Infectious Diseases},
  author = {Altizer, Sonia and Dobson, Andrew and Hosseini, Parviez and Hudson, Peter and Pascual, Mercedes and Rohani, Pejman},
  year = {2006},
  journal = {Ecology letters},
  volume = {9},
  number = {4},
  pages = {467--484},
  publisher = {Wiley Online Library}
}

@article{AndreasenDwyer2023TheAmericanNaturalist,
  title = {Seasonality and the {{Coexistence}} of {{Pathogen Strains}}},
  author = {Andreasen, Viggo and Dwyer, Greg},
  year = {2023},
  month = may,
  journal = {The American Naturalist},
  volume = {201},
  number = {5},
  pages = {639--658},
  issn = {0003-0147, 1537-5323},
  doi = {10.1086/723490},
  urldate = {2023-12-15},
  langid = {english}
}

@misc{BertsimasGurnee2022arXiv220600176article,
  title = {Learning {{Sparse Nonlinear Dynamics}} via {{Mixed-Integer Optimization}}},
  author = {Bertsimas, Dimitris and Gurnee, Wes},
  year = {2022},
  month = may,
  number = {arXiv:2206.00176},
  eprint = {2206.00176},
  primaryclass = {cs, eess, math},
  institution = {arXiv},
  urldate = {2022-06-15},
  archiveprefix = {arxiv},
  langid = {english}
}

@book{bolker2008ecoModelsNData,
  title = {Ecological Models and Data in {{R}}},
  author = {Bolker, Benjamin M},
  year = {2008},
  publisher = {Princeton University Press}
}

@article{BollerslevWooldridge1992EconometricReviews,
  title = {Quasi-Maximum Likelihood Estimation and Inference in Dynamic Models with Time-Varying Covariances},
  author = {Bollerslev, Tim and Wooldridge, Jeffrey M.},
  year = {1992},
  month = jan,
  journal = {Econometric Reviews},
  volume = {11},
  number = {2},
  pages = {143--172},
  issn = {0747-4938, 1532-4168},
  doi = {10.1080/07474939208800229},
  urldate = {2024-05-15},
  langid = {english}
}

@article{BortzMessengerDukic2023article,
  title = {Direct {{Estimation}} of {{Parameters}} in {{ODE Models Using WENDy}}: {{Weak-form Estimation}} of {{Nonlinear Dynamics}}},
  shorttitle = {Direct {{Estimation}} of {{Parameters}} in {{ODE Models Using WENDy}}},
  author = {Bortz, David M. and Messenger, Daniel A. and Dukic, Vanja},
  year = {2023},
  month = feb,
  eprint = {2302.13271},
  primaryclass = {cs, math, q-bio, stat},
  urldate = {2023-02-28},
  archiveprefix = {arxiv},
  copyright = {All rights reserved},
  langid = {english}
}

@article{BriggsGodfray1996TheoreticalPopulationBiology,
  title = {The {{Dynamics}} of {{Insect}}--{{Pathogen Interactions}} in {{Seasonal Environments}}},
  author = {Briggs, C.J. and Godfray, H.C.J.},
  year = {1996},
  month = oct,
  journal = {Theoretical Population Biology},
  volume = {50},
  number = {2},
  pages = {149--177},
  issn = {00405809},
  doi = {10.1006/tpbi.1996.0027},
  urldate = {2024-03-21},
  langid = {english}
}

@article{BruntonProctorKutz2016ProcNatlAcadSciUSA,
  title = {Discovering Governing Equations from Data by Sparse Identification of Nonlinear Dynamical Systems},
  author = {Brunton, Steven L. and Proctor, Joshua L. and Kutz, J. Nathan},
  year = {2016},
  month = apr,
  journal = {Proceedings of the National Academy of Sciences},
  volume = {113},
  number = {15},
  pages = {3932--3937},
  issn = {0027-8424, 1091-6490},
  doi = {10.1073/pnas.1517384113},
  urldate = {2021-07-23},
  langid = {english}
}

@inproceedings{BrusaferriMatteucciPortolaniEtAl202020207thIntConfControlDecisInfTechnolCoDIT,
  title = {Hybrid System Identification Using a Mixture of {{NARX}} Experts with {{LASSO-based}} Feature Selection},
  booktitle = {2020 7th {{International Conference}} on {{Control}}, {{Decision}} and {{Information Technologies}} ({{CoDIT}})},
  author = {Brusaferri, Alessandro and Matteucci, Matteo and Portolani, Pietro and Spinelli, Stefano and Vitali, Andrea},
  year = {2020},
  month = jun,
  pages = {545--550},
  publisher = {IEEE},
  address = {Prague, Czech Republic},
  doi = {10.1109/CoDIT49905.2020.9263962},
  urldate = {2024-03-08},
  isbn = {978-1-72815-953-9},
  langid = {english}
}

@book{BurnhamAnderson2007,
  title = {Model Selection and Multimodel Inference: A Practical Information-Theoretic Approach},
  author = {Burnham, K.P. and Anderson, D.R.},
  year = {2007},
  publisher = {Springer New York},
  isbn = {978-0-387-22456-5},
  lccn = {2001057677}
}

@article{BurnhamAndersonHuyvaert2011BehavEcolSociobiol,
  title = {{{AIC}} Model Selection and Multimodel Inference in Behavioral Ecology: Some Background, Observations, and Comparisons},
  author = {Burnham, Kenneth P and Anderson, David R and Huyvaert, Kathryn P},
  year = {2011},
  journal = {Behavioral ecology and sociobiology},
  volume = {65},
  pages = {23--35},
  publisher = {Springer}
}

@article{conrad2016,
  title = {Accelerating Asymptotically Exact {{MCMC}} for Computationally Intensive Models via Local Approximations},
  author = {Patrick R. Conrad, Youssef M. Marzouk, Natesh S. Pillai and Smith, Aaron},
  year = {2016},
  journal = {Journal of the American Statistical Association},
  volume = {111},
  number = {516},
  eprint = {https://doi.org/10.1080/01621459.2015.1096787},
  pages = {1591--1607},
  publisher = {Taylor \& Francis},
  doi = {10.1080/01621459.2015.1096787}
}

@article{dukic2012,
  title = {Tracking Epidemics with Google Flu Trends Data and a State-Space {{SEIR}} Model},
  author = {Vanja Dukic, Hedibert F. Lopes and Polson, Nicholas G.},
  year = {2012},
  journal = {Journal of the American Statistical Association},
  volume = {107},
  number = {500},
  eprint = {https://doi.org/10.1080/01621459.2012.713876},
  pages = {1410--1426},
  publisher = {Taylor \& Francis},
  doi = {10.1080/01621459.2012.713876}
}

@article{Dwyer1991Ecology,
  title = {The Roles of Density, Stage, and Patchiness in the Transmission of an Insect Virus},
  author = {Dwyer, Greg},
  year = {1991},
  journal = {Ecology},
  volume = {72},
  number = {2},
  pages = {559--574},
  publisher = {Wiley Online Library}
}

@article{DwyerDushoffElkintonEtAl2000,
  title = {Pathogen-{{Driven Outbreaks}} in {{Forest Defoliators Revisited}}: {{Building Models}} from {{Experimental Data}}.},
  author = {Dwyer, Greg and Dushoff, Jonathan and Elkinton, Joseph S and Levin, Simon A},
  year = {2000},
  langid = {english}
}

@article{DwyerDushoffYee2004Nature,
  title = {The Combined Effects of Pathogens and Predators on Insect Outbreaks},
  author = {Dwyer, Greg and Dushoff, Jonathan and Yee, Susan Harrell},
  year = {2004},
  journal = {Nature},
  volume = {430},
  number = {6997},
  pages = {341--345},
  publisher = {Nature Publishing Group UK London}
}

@article{DwyerMihaljevicDukic2022AmNat,
  title = {Can Eco-Evo Theory Explain Population Cycles in the Field?},
  author = {Dwyer, Greg and Mihaljevic, Joseph R and Dukic, Vanja},
  year = {2022},
  journal = {The American Naturalist},
  volume = {199},
  number = {1},
  pages = {108--125},
  publisher = {The University of Chicago Press Chicago, IL}
}

@article{ElderdDukicDwyer2006ProcNatlAcadSci,
  title = {Uncertainty in Predictions of Disease Spread and Public Health Responses to Bioterrorism and Emerging Diseases},
  author = {Elderd, Bret D and Dukic, Vanja M and Dwyer, Greg},
  year = {2006},
  journal = {Proceedings of the National Academy of Sciences},
  volume = {103},
  number = {42},
  pages = {15693--15697},
  publisher = {National Academy of Sciences}
}

@article{ElderdDushoffDwyer2008TheAmericanNaturalist,
  title = {Host-{{Pathogen Interactions}}, {{Insect Outbreaks}}, and {{Natural Selection}} for {{Disease Resistance}}},
  author = {Elderd, Bret~D. and Dushoff, Jonathan and Dwyer, Greg},
  year = {2008},
  month = dec,
  journal = {The American Naturalist},
  volume = {172},
  number = {6},
  pages = {829--842},
  issn = {0003-0147, 1537-5323},
  doi = {10.1086/592403},
  urldate = {2023-11-06},
  langid = {english}
}

@article{elderdDwyer2019popStructure,
  title = {Using Insect Baculoviruses to Understand How Population Structure Affects Disease Spread},
  author = {Elderd, Bret D and Dwyer, {\relax GREG}},
  year = {2019},
  journal = {Wildlife disease ecology: linking theory to data and application. Cambridge University Press, Cambridge, UK},
  pages = {225--261}
}

@article{ElderdDwyerDukic2013Epidemics,
  title = {Population-Level Differences in Disease Transmission: {{A Bayesian}} Analysis of Multiple Smallpox Epidemics},
  author = {Elderd, Bret D and Dwyer, Greg and Dukic, Vanja},
  year = {2013},
  journal = {Epidemics},
  volume = {5},
  number = {3},
  pages = {146--156},
  publisher = {Elsevier}
}

@article{EmerickSingh2016MathematicalBiosciences,
  title = {The Effects of Host-Feeding on Stability of Discrete-Time Host--Parasitoid Population Dynamic Models},
  author = {Emerick, Brooks and Singh, Abhyudai},
  year = {2016},
  month = feb,
  journal = {Mathematical Biosciences},
  volume = {272},
  pages = {54--63},
  issn = {00255564},
  doi = {10.1016/j.mbs.2015.11.011},
  urldate = {2024-03-21},
  langid = {english}
}

@article{EmerickSinghChhetri2020MathematicalBiosciences,
  title = {Global Redistribution and Local Migration in Semi-Discrete Host--Parasitoid Population Dynamic Models},
  author = {Emerick, Brooks and Singh, Abhyudai and Chhetri, Safal Raut},
  year = {2020},
  month = sep,
  journal = {Mathematical Biosciences},
  volume = {327},
  pages = {108409},
  issn = {00255564},
  doi = {10.1016/j.mbs.2020.108409},
  urldate = {2024-03-21},
  langid = {english}
}

@article{EskolaGeritz2007BullMathBiol,
  title = {On the {{Mechanistic Derivation}} of {{Various Discrete-Time Population Models}}},
  author = {Eskola, Hanna T. M. and Geritz, Stefan A. H.},
  year = {2007},
  month = jan,
  journal = {Bulletin of Mathematical Biology},
  volume = {69},
  number = {1},
  pages = {329--346},
  issn = {0092-8240, 1522-9602},
  doi = {10.1007/s11538-006-9126-4},
  urldate = {2024-03-21},
  langid = {english}
}

@article{FaselKutzBruntonEtAl2022ProcRSocA,
  title = {Ensemble-{{SINDy}}: {{Robust}} Sparse Model Discovery in the Low-Data, High-Noise Limit, with Active Learning and Control},
  shorttitle = {Ensemble-{{SINDy}}},
  author = {Fasel, U. and Kutz, J. N. and Brunton, B. W. and Brunton, S. L.},
  year = {2022},
  month = apr,
  journal = {Proceedings of the Royal Society A: Mathematical, Physical and Engineering Sciences},
  volume = {478},
  number = {2260},
  pages = {20210904},
  issn = {1364-5021, 1471-2946},
  doi = {10.1098/rspa.2021.0904},
  urldate = {2022-10-20},
  langid = {english}
}

@article{fleming2015effects,
  title = {Effects of Host Heterogeneity on Pathogen Diversity and Evolution},
  author = {{Fleming-Davies}, Arietta E and Dukic, Vanja and Andreasen, Viggo and Dwyer, Greg},
  year = {2015},
  journal = {Ecology letters},
  volume = {18},
  number = {11},
  pages = {1252--1261}
}

@article{FullerElderdDwyer2012TheAmericanNaturalist,
  title = {Pathogen {{Persistence}} in the {{Environment}} and {{Insect-Baculovirus Interactions}}: {{Disease-Density Thresholds}}, {{Epidemic Burnout}}, and {{Insect Outbreaks}}},
  shorttitle = {Pathogen {{Persistence}} in the {{Environment}} and {{Insect-Baculovirus Interactions}}},
  author = {Fuller, Emma and Elderd, Bret D. and Dwyer, Greg},
  year = {2012},
  month = mar,
  journal = {The American Naturalist},
  volume = {179},
  number = {3},
  pages = {E70-E96},
  issn = {0003-0147, 1537-5323},
  doi = {10.1086/664488},
  urldate = {2024-01-26},
  langid = {english}
}

@book{GoebelSanfeliceTeel2012,
  title = {Hybrid Dynamical Systems: {{Modeling}}, Stability, and Robustness},
  author = {Goebel, R. and Sanfelice, R.G. and Teel, A.R.},
  year = {2012},
  publisher = {Princeton University Press},
  isbn = {978-0-691-15389-6},
  lccn = {2011941674}
}

@article{GourierouxMonfortTrognon1984Econometrica,
  title = {Pseudo {{Maximum Likelihood Methods}}: {{Theory}}},
  shorttitle = {Pseudo {{Maximum Likelihood Methods}}},
  author = {Gourieroux, C. and Monfort, A. and Trognon, A.},
  year = {1984},
  month = may,
  journal = {Econometrica},
  volume = {52},
  number = {3},
  eprint = {1913471},
  eprinttype = {jstor},
  pages = {681},
  issn = {00129682},
  doi = {10.2307/1913471},
  urldate = {2024-05-29},
  langid = {english}
}

@article{GurevichReinboldGrigoriev2019Chaos,
  title = {Robust and Optimal Sparse Regression for Nonlinear {{PDE}} Models},
  author = {Gurevich, Daniel R. and Reinbold, Patrick A. K. and Grigoriev, Roman O.},
  year = {2019},
  month = oct,
  journal = {Chaos: An Interdisciplinary Journal of Nonlinear Science},
  volume = {29},
  number = {10},
  pages = {103113},
  issn = {1054-1500, 1089-7682},
  doi = {10.1063/1.5120861},
  urldate = {2021-09-27},
  langid = {english}
}

@book{hilborn2013ecoDetective,
  title = {The Ecological Detective: Confronting Models with Data ({{MPB-28}})},
  author = {Hilborn, Ray and Mangel, Marc},
  year = {2013},
  publisher = {Princeton University Press}
}

@article{HootenHobbs2015EcolMonogr,
  title = {A Guide to {{Bayesian}} Model Selection for Ecologists},
  author = {Hooten, M. B. and Hobbs, N. T.},
  year = {2015},
  journal = {Ecological Monographs},
  volume = {85},
  number = {1},
  eprint = {https://esajournals.onlinelibrary.wiley.com/doi/pdf/10.1890/14-0661.1},
  pages = {3--28},
  doi = {10.1890/14-0661.1}
}

@article{Hunter1995PopulDynNewApproachesSynth,
  title = {Ecology, Life History, and Phylogeny of Outbreak and Nonoutbreak Species},
  author = {Hunter, Alison F},
  year = {1995},
  journal = {Population dynamics: new approaches and synthesis},
  pages = {41--64},
  publisher = {Academic Press San Diego}
}

@book{keeling2011modelingInfDiseases,
  title = {Modeling Infectious Diseases in Humans and Animals},
  author = {Keeling, Matt J and Rohani, Pejman},
  year = {2011},
  publisher = {Princeton university press}
}

@article{KennedyDukicDwyer2014AmNat,
  title = {Pathogen Growth in Insect Hosts: Inferring the Importance of Different Mechanisms Using Stochastic Models and Response Time Data},
  author = {Kennedy, David and Dukic, Vanja and Dwyer, Greg},
  year = {2014},
  journal = {The American Naturalist},
  volume = {184},
  number = {3},
  publisher = {American Society of Naturalists}
}

@article{KennedyDukicDwyer2015EnvironEcolStat,
  title = {Combining Principal Component Analysis with Parameter Line-Searches to Improve the Efficacy of {{Metropolis}}--{{Hastings MCMC}}},
  author = {Kennedy, David A and Dukic, Vanja and Dwyer, Greg},
  year = {2015},
  journal = {Environmental and ecological statistics},
  volume = {22},
  pages = {247--274},
  publisher = {Springer US}
}

@article{KingIonidesPascualEtAl2008Nature,
  title = {Inapparent Infections and Cholera Dynamics},
  author = {King, Aaron A and Ionides, Edward L and Pascual, Mercedes and Bouma, Menno J},
  year = {2008},
  journal = {Nature},
  volume = {454},
  number = {7206},
  pages = {877--880},
  publisher = {Nature Publishing Group UK London}
}

@article{kyle2020stochasticity,
  title = {Stochasticity and Infectious-Disease Dynamics: {{Density}} and Weather Effects on a Fungal Insect Pathogen},
  author = {Kyle, Colin Hector and Liu, Jiawei and Gallagher, Molly Elizabether and Dukic, Vanja and Dwyer, Greg},
  year = {2020},
  journal = {The American Naturalist}
}

@article{LagergrenNardiniLavigneEtAl2020ProcRSocA,
  title = {Learning Partial Differential Equations for Biological Transport Models from Noisy Spatiotemporal Data},
  author = {Lagergren, John and Nardini, John T. and Lavigne, G. Michael and Rutter, Erica M. and Flores, Kevin B.},
  year = {2020},
  month = feb,
  journal = {Proceedings of the Royal Society A: Mathematical, Physical and Engineering Sciences},
  volume = {476},
  number = {2234},
  eprint = {1902.04733},
  pages = {20190800},
  issn = {1364-5021, 1471-2946},
  doi = {10.1098/rspa.2019.0800},
  urldate = {2021-07-09},
  archiveprefix = {arxiv},
  langid = {english}
}

@incollection{LaxMilgram1954Contributionstothetheoryofpartialdifferentialequations,
  title = {Parabolic Equations},
  booktitle = {Contributions to the Theory of Partial Differential Equations},
  author = {Lax, Peter D. and Milgram, Arthur N.},
  year = {1954},
  series = {Annals of {{Mathematics Studies}}},
  volume = {33},
  publisher = {Princeton University Press}
}

@article{LiWuLiu2023PhysRevResearch,
  title = {Discover Governing Differential Equations from Evolving Systems},
  author = {Li, Yuanyuan and Wu, Kai and Liu, Jing},
  year = {2023},
  month = may,
  journal = {Physical Review Research},
  volume = {5},
  number = {2},
  pages = {023126},
  issn = {2643-1564},
  doi = {10.1103/PhysRevResearch.5.023126},
  urldate = {2024-03-08},
  langid = {english}
}

@article{LiXuDuanEtAl2023ChaosInterdiscipJNonlinearSci,
  title = {A Data-Driven Framework for Learning Hybrid Dynamical Systems},
  author = {Li, Yang and Xu, Shengyuan and Duan, Jinqiao and Huang, Yong and Liu, Xianbin},
  year = {2023},
  month = jun,
  journal = {Chaos: An Interdisciplinary Journal of Nonlinear Science},
  volume = {33},
  number = {6},
  pages = {061104},
  issn = {1054-1500, 1089-7682},
  doi = {10.1063/5.0157669},
  urldate = {2024-03-08},
  langid = {english}
}

@article{maclauchlan2009DFTMinBC,
  title = {An Integrated Management System for the {{Douglas-fir}} Tussock Moth in Southern {{British Columbia}}},
  author = {Maclauchlan, Lorraine E and Hall, Peter M and Otvos, Imre S and Brooks, Julie E},
  year = {2009},
  journal = {Journal of Ecosystems and Management}
}

@article{ManganAskhamBruntonEtAl2019ProcRSocA,
  title = {Model Selection for Hybrid Dynamical Systems via Sparse Regression},
  author = {Mangan, N. M. and Askham, T. and Brunton, S. L. and Kutz, J. N. and Proctor, J. L.},
  year = {2019},
  month = mar,
  journal = {Proceedings of the Royal Society A: Mathematical, Physical and Engineering Sciences},
  volume = {475},
  number = {2223},
  pages = {20180534},
  issn = {1364-5021, 1471-2946},
  doi = {10.1098/rspa.2018.0534},
  urldate = {2024-03-04},
  langid = {english}
}

@article{McGoffMukherjeePillai2015StatSurv,
  title = {Statistical Inference for Dynamical Systems: {{A}} Review},
  author = {McGoff, Kevin and Mukherjee, Sayan and Pillai, Natesh},
  year = {2015},
  journal = {Statistics Surveys},
  volume = {9},
  number = {none},
  pages = {209--252},
  publisher = {{Amer. Statist. Assoc., the Bernoulli Soc., the Inst. Math. Statist., and the Statist. Soc. Canada}},
  doi = {10.1214/15-SS111}
}

@article{messenger2022asymptotic,
  title = {Asymptotic Consistency of the {{WSINDy}} Algorithm in the Limit of Continuum Data},
  author = {Messenger, Daniel A and Bortz, David M},
  year = {2022},
  journal = {arXiv preprint arXiv:2211.16000},
  eprint = {2211.16000},
  archiveprefix = {arxiv}
}

@article{MessengerBortz2021JComputPhys,
  title = {Weak {{SINDy For Partial Differential Equations}}},
  author = {Messenger, Daniel A. and Bortz, David M.},
  year = {2021},
  month = oct,
  journal = {Journal of Computational Physics},
  volume = {443},
  eprint = {2007.02848},
  pages = {110525},
  doi = {10.1016/j.jcp.2021.110525},
  urldate = {2020-07-07},
  archiveprefix = {arxiv},
  copyright = {All rights reserved}
}

@article{MessengerBortz2021SIAMMultiscaleModelSimul,
  title = {Weak {{SINDy}}: {{Galerkin-Based Data-Driven Model Selection}}},
  shorttitle = {Weak {{SINDy}}},
  author = {Messenger, Daniel A. and Bortz, David M.},
  year = {2021},
  journal = {SIAM Multiscale Modeling \& Simulation},
  volume = {19},
  number = {3},
  eprint = {2005.04339},
  pages = {1474--1497},
  urldate = {2020-07-07},
  archiveprefix = {arxiv},
  copyright = {All rights reserved}
}

@article{MessengerBortz2022PhysicaDNonlinearPhenomena,
  title = {Learning Mean-Field Equations from Particle Data Using {{WSINDy}}},
  author = {Messenger, Daniel A. and Bortz, David M.},
  year = {2022},
  month = nov,
  journal = {Physica D: Nonlinear Phenomena},
  volume = {439},
  pages = {133406},
  issn = {01672789},
  doi = {10.1016/j.physd.2022.133406},
  urldate = {2022-08-19},
  langid = {english}
}

@misc{MessengerBurbyBortz2023,
  title = {Coarse-{{Graining Hamiltonian Systems Using WSINDy}}},
  author = {Messenger, Daniel A. and Burby, Joshua W. and Bortz, David M.},
  year = {2023},
  month = dec,
  doi = {10.21203/rs.3.rs-3645802/v1},
  urldate = {2024-04-22},
  copyright = {https://creativecommons.org/licenses/by/4.0/},
  langid = {english}
}

@article{MessengerWheelerLiuEtAl2022JRSocInterface,
  title = {Learning {{Anisotropic Interaction Rules}} from {{Individual Trajectories}} in a {{Heterogeneous Cellular Population}}},
  author = {Messenger, Daniel A. and Wheeler, Graycen E. and Liu, Xuedong and Bortz, David M.},
  year = {2022},
  month = oct,
  journal = {Journal of The Royal Society Interface},
  volume = {19},
  number = {195},
  pages = {20220412},
  doi = {10.1098/rsif.2022.0412}
}

@article{MihaljevicPolivkaMehmelEtAl2019AmNat,
  title = {Using Biological Control Data to Understand Host-Pathogen Dynamics},
  author = {Mihaljevic, Joseph R and Polivka, Karl M and Mehmel, Constance J and Li, Chentong and Dukic, Vanja and Dwyer, Greg},
  year = {2019},
  journal = {The American Naturalist},
  publisher = {The University of Chicago Press}
}

@article{MihaljevicPolivkaMehmelEtAl2020AmNat,
  title = {An Empirical Test of the Role of Small-Scale Transmission in Large-Scale Disease Dynamics},
  author = {Mihaljevic, Joseph R and Polivka, Carlos M and Mehmel, Constance J and Li, Chentong and Dukic, Vanja and Dwyer, Greg},
  year = {2020},
  journal = {The American Naturalist},
  pages = {https--doi},
  publisher = {The University of Chicago Press}
}

@book{miller1997baculovirusesBook,
  title = {The Baculoviruses},
  author = {Miller, Lois},
  year = {1997},
  publisher = {Springer Science \& Business Media}
}

@article{Myers1993AmSci,
  title = {Population Outbreaks in Forest {{Lepidoptera}}},
  author = {Myers, Judith H},
  year = {1993},
  journal = {American Scientist},
  volume = {81},
  number = {3},
  pages = {240--251},
  publisher = {JSTOR}
}

@article{NicolaouHuoChenEtAl2023PhysRevResearch,
  title = {Data-Driven Discovery and Extrapolation of Parameterized Pattern-Forming Dynamics},
  author = {Nicolaou, Zachary G. and Huo, Guanyu and Chen, Yihui and Brunton, Steven L. and Kutz, J. Nathan},
  year = {2023},
  month = nov,
  journal = {Physical Review Research},
  volume = {5},
  number = {4},
  pages = {L042017},
  issn = {2643-1564},
  doi = {10.1103/PhysRevResearch.5.L042017},
  urldate = {2023-11-15},
  langid = {english}
}

@article{NovelliLenciBelardinelli2022JComputNonlinearDyn,
  title = {Boosting the {{Model Discovery}} of {{Hybrid Dynamical Systems}} in an {{Informed Sparse Regression Approach}}},
  author = {Novelli, Nico and Lenci, Stefano and Belardinelli, Pierpaolo},
  year = {2022},
  month = may,
  journal = {Journal of Computational and Nonlinear Dynamics},
  volume = {17},
  number = {5},
  pages = {051007},
  issn = {1555-1415, 1555-1423},
  doi = {10.1115/1.4053324},
  urldate = {2024-03-08},
  langid = {english}
}

@article{ostfeld1996densityGM,
  title = {Of Mice and Mast},
  author = {Ostfeld, Richard S and Jones, Clive G and Wolff, Jerry O},
  year = {1996},
  journal = {BioScience},
  volume = {46},
  number = {5},
  pages = {323--330},
  publisher = {JSTOR}
}

@article{OtvosCunninghamFriskie1987CanEntomol,
  title = {Aerial Application of Nuclear Polyhedrosis Virus against {{Douglas-fir}} Tussock Moth, {{Orgyia}} Pseudotsugata ({{McDunnough}})({{Lepidoptera}}: {{Lymantriidae}}): {{I}}. {{Impact}} in the Year of Application},
  author = {Otvos, {\relax IS} and Cunningham, {\relax JC} and Friskie, {\relax LM}},
  year = {1987},
  journal = {The Canadian Entomologist},
  volume = {119},
  number = {7-8},
  pages = {697--706},
  publisher = {Cambridge University Press}
}

@article{PachepskyNisbetMurdoch2008Ecology,
  title = {{{BETWEEN DISCRETE AND CONTINUOUS}}: {{CONSUMER}}--{{RESOURCE DYNAMICS WITH SYNCHRONIZED REPRODUCTION}}},
  shorttitle = {{{BETWEEN DISCRETE AND CONTINUOUS}}},
  author = {Pachepsky, E. and Nisbet, R. M. and Murdoch, W. W.},
  year = {2008},
  month = jan,
  journal = {Ecology},
  volume = {89},
  number = {1},
  pages = {280--288},
  issn = {0012-9658},
  doi = {10.1890/07-0641.1},
  urldate = {2024-03-21},
  langid = {english}
}

@article{PaezDukicDushoffEtAl2017AmNat,
  title = {Eco-Evolutionary Theory and Insect Outbreaks},
  author = {P{\'a}ez, David J and Dukic, Vanja and Dushoff, Jonathan and {Fleming-Davies}, Arietta and Dwyer, Greg},
  year = {2017},
  journal = {The American Naturalist},
  volume = {189},
  number = {6},
  pages = {616--629},
  publisher = {University of Chicago Press Chicago, IL}
}

@article{PreisigRippin1993ComputChemEngb,
  title = {Theory and Application of the Modulating Function Method---{{I}}. {{Review}} and Theory of the Method and Theory of the Spline-Type Modulating Functions},
  author = {Preisig, H.A. and Rippin, D.W.T.},
  year = {1993},
  month = jan,
  journal = {Computers \& Chemical Engineering},
  volume = {17},
  number = {1},
  pages = {1--16},
  issn = {00981354},
  doi = {10.1016/0098-1354(93)80001-4},
  urldate = {2023-02-20},
  langid = {english}
}

@article{ramsay2007,
  title = {Parameter Estimation for Differential Equations: A Generalized Smoothing Approach},
  author = {Ramsay, Jim O and Hooker, Giles and Campbell, David and Cao, Jiguo},
  year = {2007},
  journal = {Journal of the Royal Statistical Society Series B: Statistical Methodology},
  volume = {69},
  number = {5},
  pages = {741--796},
  publisher = {Oxford University Press}
}

@misc{RussoLaiu2022arXiv220915573article,
  title = {Convergence of Weak-{{SINDy Surrogate Models}}},
  author = {Russo, Benjamin and Laiu, M. Paul},
  year = {2022},
  month = oct,
  number = {arXiv:2209.15573},
  eprint = {2209.15573},
  primaryclass = {cs, math},
  institution = {arXiv},
  urldate = {2022-10-20},
  archiveprefix = {arxiv},
  langid = {english}
}

@article{SchaefferMcCalla2017PhysRevE,
  title = {Sparse Model Selection via Integral Terms},
  author = {Schaeffer, Hayden and McCalla, Scott G.},
  year = {2017},
  month = aug,
  journal = {Physical Review E},
  volume = {96},
  number = {2},
  pages = {023302},
  issn = {2470-0045, 2470-0053},
  doi = {10.1103/PhysRevE.96.023302},
  urldate = {2021-09-27},
  langid = {english}
}

@article{Schwarz1978AnnStatist,
  title = {Estimating the {{Dimension}} of a {{Model}}},
  author = {Schwarz, Gideon},
  year = {1978},
  month = mar,
  journal = {The Annals of Statistics},
  volume = {6},
  number = {2},
  issn = {0090-5364},
  doi = {10.1214/aos/1176344136},
  urldate = {2024-04-15}
}

@techreport{Shinbrot1954NACATN3288,
  title = {On the Analysis of Linear and Nonlinear Dynamical Systems for Transient-Response Data},
  author = {Shinbrot, Marvin},
  year = {1954},
  month = dec,
  number = {NACA TN 3288},
  address = {Moffett Field, CA},
  institution = {Ames Aeronautical Laboratory}
}

@techreport{SinghEmerick2020preprint,
  type = {Preprint},
  title = {Hybrid Systems Modeling of Ecological Population Dynamics},
  author = {Singh, Abhyudai and Emerick, Brooks},
  year = {2020},
  month = mar,
  institution = {Ecology},
  doi = {10.1101/2020.03.28.013524},
  urldate = {2024-03-21},
  langid = {english}
}

@article{skaller1985patternsGM11yrs,
  title = {Patterns in the Distribution of Gypsy Moth ({{{\emph{Lymantria}}}}{\emph{ Dispar}}) ({{Lepidoptera}}: {{Lymantriidae}}) Egg Masses over an 11-Year Population Cycle},
  author = {Skaller, P Michael},
  year = {1985},
  journal = {Environmental entomology},
  volume = {14},
  number = {2},
  pages = {106--117},
  publisher = {Oxford University Press Oxford, UK}
}

@article{splineODE2018,
  title = {Prediction of Dynamical Time Series Using Kernel Based Regression and Smooth Splines},
  author = {Navarrete, Raymundo and Viswanath, Divakar},
  year = {2018},
  journal = {Electronic Journal of Statistics},
  volume = {12},
  number = {2},
  pages = {2217--2237},
  publisher = {{Institute of Mathematical Statistics and Bernoulli Society}},
  doi = {10.1214/18-EJS1429}
}

@article{TangLiaoKuskeEtAl2023JComputPhys,
  title = {{{WeakIdent}}: {{Weak}} Formulation for {{Identifying Differential Equation}} Using {{Narrow-fit}} and {{Trimming}}},
  shorttitle = {{{WeakIdent}}},
  author = {Tang, Mengyi and Liao, Wenjing and Kuske, Rachel and Kang, Sung Ha},
  year = {2023},
  month = mar,
  journal = {Journal of Computational Physics},
  pages = {112069},
  issn = {00219991},
  doi = {10.1016/j.jcp.2023.112069},
  urldate = {2023-03-20},
  langid = {english}
}

@book{van2007introduction,
  title = {An Introduction to Hybrid Dynamical Systems},
  author = {{van der Schaft}, A.J. and Schumacher, H.},
  year = {2007},
  series = {Lecture Notes in Control and Information Sciences},
  publisher = {Springer London},
  isbn = {978-1-84628-542-4}
}

@article{WangHuanGarikipati2019ComputerMethodsinAppliedMechanicsandEngineering,
  title = {Variational System Identification of the Partial Differential Equations Governing the Physics of Pattern-Formation: {{Inference}} under Varying Fidelity and Noise},
  shorttitle = {Variational System Identification of the Partial Differential Equations Governing the Physics of Pattern-Formation},
  author = {Wang, Z. and Huan, X. and Garikipati, K.},
  year = {2019},
  month = nov,
  journal = {Computer Methods in Applied Mechanics and Engineering},
  volume = {356},
  pages = {44--74},
  issn = {00457825},
  doi = {10.1016/j.cma.2019.07.007},
  urldate = {2021-09-27},
  langid = {english}
}

@article{Watanabe2013JMachLearnRes,
  title = {A Widely Applicable {{Bayesian}} Information Criterion},
  author = {Watanabe, Sumio},
  year = {2013},
  month = mar,
  journal = {Journal of Machine Learning Research},
  volume = {14},
  number = {1},
  pages = {867--897},
  publisher = {JMLR.org},
  issn = {1532-4435},
  issue_date = {January 2013}
}

@article{williams1991densityGM,
  title = {Oak Defoliation and Population Density Relationships for the Gypsy Moth ({{Lepidoptera}}: {{Lymantriidae}}},
  author = {Williams, {\relax DW} and Fuester, {\relax RW} and Metterhouse, {\relax WW} and Balaam, {\relax RJ} and Bullock, {\relax RH} and Chianesei, {\relax RJ}},
  year = {1991},
  journal = {Journal of Economic Entomology},
  volume = {84},
  number = {5},
  pages = {1508--1514},
  publisher = {Oxford University Press Oxford, UK}
}

@techreport{WoodallEsparzaGutovaEtAl2023preprinta,
  type = {Preprint},
  title = {Model Discovery Approach Enables Non-Invasive Measurement of Intra-Tumoral Fluid Transport in Dynamic {{MRI}}},
  author = {Woodall, Ryan T. and Esparza, Cora C. and Gutova, Margarita and Wang, Maosen and Cunningham, Jessica J. and Brummer, Alexander B. and Stine, Caleb A. and Brown, Christine C. and Munson, Jennifer M. and Rockne, Russell C.},
  year = {2023},
  month = aug,
  institution = {Bioengineering},
  doi = {10.1101/2023.08.28.554919},
  urldate = {2024-01-05},
  langid = {english}
}

@article{WoodsElkinton1987JInvertebrPathol,
  title = {Biomodal Patterns of Mortality from Nuclear Polyhedrosis Virus in Gypsy Moth ({{Lymantria}} Dispar) Populations},
  author = {Woods, {\relax SA} and Elkinton, {\relax JS}},
  year = {1987},
  journal = {Journal of Invertebrate Pathology},
  volume = {50},
  number = {2},
  pages = {151--157},
  publisher = {Elsevier}
}

@article{Wooldridge2024,
  title = {A {{Unified Approach}} to {{Robust}}, {{Regression-Based Specification Tests}}},
  author = {Wooldridge, Jeffrey M},
  year = {2024},
  langid = {english}
}

@article{YangWongKou2021ProcNatlAcadSci,
  title = {Inference of Dynamic Systems from Noisy and Sparse Data via Manifold-Constrained {{Gaussian}} Processes},
  author = {Yang, Shihao and Wong, Samuel W. K. and Kou, S. C.},
  year = {2021},
  journal = {Proceedings of the National Academy of Sciences},
  volume = {118},
  number = {15},
  eprint = {https://www.pnas.org/doi/pdf/10.1073/pnas.2020397118},
  pages = {e2020397118},
  doi = {10.1073/pnas.2020397118}
}

@article{ZhangSchaeffer2019MultiscaleModelSimul,
  title = {On the Convergence of the Sindy Algorithm},
  author = {Zhang, Linan and Schaeffer, Hayden},
  year = {2019},
  journal = {Multiscale Modeling \& Simulation},
  volume = {17},
  number = {3},
  eprint = {https://doi.org/10.1137/18M1189828},
  pages = {948--972},
  doi = {10.1137/18M1189828}
}



\end{document}